\documentclass[lettersize,journal]{IEEEtran}  

\usepackage{cite}
\usepackage{amsmath,amssymb,amsfonts}
\usepackage{algorithmic}
\usepackage{graphicx}
\usepackage{textcomp}
\usepackage{bm}
\usepackage{hyperref}
\usepackage{url} 
\usepackage{subcaption}
\usepackage[table,xcdraw]{xcolor}

\IEEEoverridecommandlockouts                              

\title{
Online Identification using Adaptive Laws and Neural Networks for Multi-Quadrotor Centralized Transportation System
}

\author{Tianhua Gao$^{1}$, Kohji Tomita$^{2}$, and Akiya Kamimura$^{3}$ 
\thanks{The authors are with the Intelligent Systems Research Institute, National
Institute of Advanced Industrial Science and Technology (AIST), Japan (\{$^{1}$kou.tenka, $^{2}$k.tomita, $^{3}$kamimura.a\}@aist.go.jp). $^{1}$T. Gao is also with 
the Graduate School of
Systems and Information Engineering, University of Tsukuba, Japan ($^{1}$gao.tianhua.tkb\_gb@u.tsukuba.ac.jp). 
Corresponding author: Akiya Kamimura.
Contact address:
Tsukuba Central2, 1-1-1 Umezono, Tsukuba, Ibaraki 305-8568, Japan. Phone:
+81 80-2309-1517.
For simulation videos in Section 5, refer to:\url{https://staff.aist.go.jp/kamimura.a/ES2025/video.mp4}}%
}

\begin{document}

\maketitle
\thispagestyle{empty}
\pagestyle{empty}

\begin{abstract}
This paper introduces an adaptive-neuro identification method that enhances the robustness of a centralized multi-quadrotor transportation system. This method leverages online tuning and learning on decomposed error subspaces, enabling efficient real-time compensation to time-varying disturbances and model uncertainties acting on the payload. The strategy is to decompose the high-dimensional error space into a set of low-dimensional subspaces. In this way, the identification problem for unseen features is naturally transformed into submappings (``slices'’) addressed by multiple adaptive laws and shallow neural networks, which are updated online via Lyapunov-based adaptation without requiring persistent excitation (PE) and offline training.  Due to the model-free nature of neural networks, this approach can be well adapted to highly coupled and nonlinear centralized transportation systems. It serves as a feedforward compensator for the payload controller without explicitly relying on the dynamics coupled with the payload, such as cables and quadrotors.
The proposed control system has been proven to be stable in the sense of Lyapunov, and its enhanced robustness under time-varying disturbances and model uncertainties was demonstrated by numerical simulations.

\end{abstract}

\noindent\textbf{1. Introduction}

Among various aerial manipulation techniques \cite{2022 Past Present and Future of Aerial Robotic Manipulators}, \cite{2024 Autonomous aerial robotics for package delivery: a technical review}, \cite{2024 A Review of Real-Time Implementable Cooperative Aerial Manipulation Systems}, cable-suspended payload transportation using multi-quadrotor systems has gained significant attention due to its structural simplicity, enhanced operational flexibility, and energy efficiency. This transportation paradigm introduces distinct control challenges that have motivated the development of various control architectures in the recent literature. These efforts can be categorized into two primary orientations: the decentralized methods (e.g.,\cite{2011 Cooperative manipulation and transportation with aerial robots}-\cite{2024 Decentralized adaptive controller for multi-drone cooperative transport with offset and moving center of gravity}) and the centralized methods (e.g., \cite{2013 Dynamics Control and Planning for Cooperative
Manipulation of Payloads Suspended by Cables
from Multiple Quadrotor Robots}-\cite{2024 Efficient Optimization-Based Cable Force Allocation for Geometric Control of a Multirotor Team Transporting a Payload}).

Decentralized methods are generally more economical and simpler to implement, as they eliminate the need for full-state estimation of the cable–payload dynamics. A common strategy in these decentralized frameworks is to simplify coordination by avoiding complex feedback from the payload. Several studies\cite{2017 Collaborative transportation using MAVs via passive force control}, \cite{2022 Controller Design and Disturbance Rejection of
Multi-Quadcopters for Cable Suspended Payload
Transportation Using Virtual Structure} have completely discarded payload feedback, instead adopting leader–follower schemes to coordinate the formation of multirotor systems. These algorithms exhibit robustness since they do not rely on the knowledge of the payload, but leave the payload pose uncontrolled during transportation. Consequently, in recent state-of-the-art research, some research has increasingly focused on addressing the payload pose manipulation problem.  Studies such as \cite{2020 Full-Pose Manipulation Control of a Cable-Suspended Load With Multiple UAVs Under Uncertainties}, \cite{2022 Indirect Force Control of a Cable-Suspended Aerial Multi-Robot Manipulator} have achieved full-pose manipulation of cable-suspended platforms under quasi-static conditions. Furthermore, pose regulation for cable-suspended beams using incomplete system dynamics has been proposed in recent studies \cite{2023 Force-Based Pose Regulation of a Cable-Suspended Load Using UAVs with Force Bias}, \cite{2023 Equilibria Stability and Sensitivity for the Aerial
Suspended Beam Robotic System Subject to
Parameter Uncertainty}. Despite these recent advances, achieving precise and dynamic pose tracking with decentralized methods remains inherently challenging, motivating further exploration of centralized control approaches.

Centralized methods focus on the dynamical tracking of the payload trajectory and orientation. These methods typically rely on the complete dynamics of the system, including the cables, which has traditionally been considered a major challenge. However, advances in sensor technology have enabled recent studies (e.g., \cite{2024 Aerial Transportation of Cable-Suspended Loads With an Event Camera}) to estimate the full state of the cables. Moreover, the feasibility of modeling the cable as a rigid and taut link has been validated through real-world experiments \cite{2021 Cooperative Transportation of Cable Suspended Payloads With MAVs Using Monocular Vision and Inertial Sensing}. Furthermore, the collision issue, previously considered a drawback of centralized methods, has been addressed using optimization-based strategies, as demonstrated in \cite{2024 Efficient Optimization-Based Cable Force Allocation for Geometric Control of a Multirotor Team Transporting a Payload}. As a result, centralized methods are becoming increasingly promising for achieving dynamic transportation with stable payload pose. However, despite these recent advancements, significant challenges remain due to the highly coupled system dynamics \cite{2023 Nonlinear Model Predictive Control for Cooperative Transportation and Manipulation of Cable Suspended Payloads with Multiple Quadrotors}. In such highly nonlinear and indirect actuation systems, designing a model-based disturbance observer becomes particularly challenging, as the propagation of disturbances and their interactions across subsystems are hard to isolate and compensate for accurately. These complexities also pose limitations on the effectiveness of traditional robust control strategies, highlighting the need for alternative frameworks that are model-free in nature and capable of simplifying controller design while enhancing robustness.

Given that centralized approaches generally adopt geometric tracking control \cite{2018 Geometric Control of Quadrotor UAVs Transporting
a Cable-Suspended Rigid Body}-\cite{2022 Geometric control of two quadrotors carrying a rigid rod with elastic cables}, \cite{2023 Nonlinear Model Predictive Control for Cooperative Transportation and Manipulation of Cable Suspended Payloads with Multiple Quadrotors}-\cite{2024 Efficient Optimization-Based Cable Force Allocation for Geometric Control of a Multirotor Team Transporting a Payload} for payload manipulation, we turn our attention to the methodologies in learning-based geometric control using neural networks as a potential source of insight. In \cite{2021 Geometric Adaptive Control With Neural Networks
for a Quadrotor in Wind Fields}, researchers proposed a geometric adaptive controller enhanced by neural networks to compensate for wind disturbances, achieving uniformly ultimately bounded (UUB) tracking errors on the special Euclidean group SE(3). Another recent study \cite{2024 Neural Moving Horizon Estimation for Robust Flight Control} presented a NeuroMHE-augmented geometric controller using a multilayer perceptron (MLP), with performance validated through experiments. However, due to the absence of a Lyapunov-based design, the approach does not ensure theoretical stability guarantees. In addition, studies have shown that the expressive power of feedforward neural networks increases significantly with depth compared to width \cite{2016 The Power of Depth for Feedforward Neural Networks}. Motivated by this, several studies \cite{2024 Meta-Learning Augmented MPC for Disturbance-Aware Motion Planning and Control of Quadrotors}, \cite{2025 High Maneuverability and Efficiency Control for Hybrid Quadrotor With All-Moving Wings in SE(3) Based on Deep Reinforcement Learning} employed deep neural networks (DNN) to improve the representational capacity for better control performance. However, these DNN methods are black-box in nature, which raises concerns in safety-critical scenarios \cite{2019 Stop explaining black box machine learning models for high stakes decisions and use interpretable models instead}. These limitations indicate the importance of developing learning-based methods with theoretical guarantees and sufficient representational capacity.

In this work, we leverage multiple adaptive laws and neural networks to develop an online identification framework that bypasses the need for complex coupled models of cables and quadrotor systems, enabling real-time compensation for time-varying disturbances and model uncertainties. A novel feature of our method is that the entire online learning process is decomposed into multiple parallel processes that operate simultaneously. Specifically, we decompose the high-dimensional error space of the payload into several low-dimensional subspaces, and construct adaptive laws and neural network submappings (``slices'') within these subspaces. This idea was inspired by emerging neuroscience studies \cite{2019 Cortical Areas Interact through a Communication Subspace}-\cite{2025 Multiplexed subspaces route neural activity across brain-wide networks} which uncovered the ``shared subspace'' property underlying cognition control.  For clarity and future reference, we refer to our ``subspace learning'' method as Dimension-Decomposed Learning (DiD-L) based on the Sliced Adaptive-Neuro Mapping (SANM) module in this paper, and our contributions are summarized as follows:
\begin{itemize}
\item Proposed a novel adaptive-neuron geometric control using an online identification method (i.e., SANM) to enhance the robustness of the centralized multi-quadrotor transportation system. 
\item Proved the stability of the proposed control strategy for tracking the desired motion of a cable-suspended payload, in the presence of parametric uncertainties and time-varying disturbances.  
\item Demonstrated the effectiveness and robustness of the proposed controller through numerical simulations and analysis.
\end{itemize}

The remainder of this paper is organized as follows. Section 2 presents the problem formulation. Section 3 elaborates on the control strategy. Section 4 analyzes the stability of the proposed control system. Section 5 demonstrates the simulation results. The conclusions and future work are discussed in Section 6. Beyond our previous work \cite{2025 Robustness Enhancement for Multi-Quadrotor Centralized Transportation System via Online Tuning and Learning}, this article improves the algorithm, provides a more rigorous stability proof, and presents more comprehensive technical details and simulation results.

\vspace{1em} 
\noindent\textbf{2. Problem Formulation}

\noindent\textbf{2.1 Kinematics and Dynamics with Disturbance}

\begin{figure}[t]
      \centering
      \includegraphics[scale=0.23]{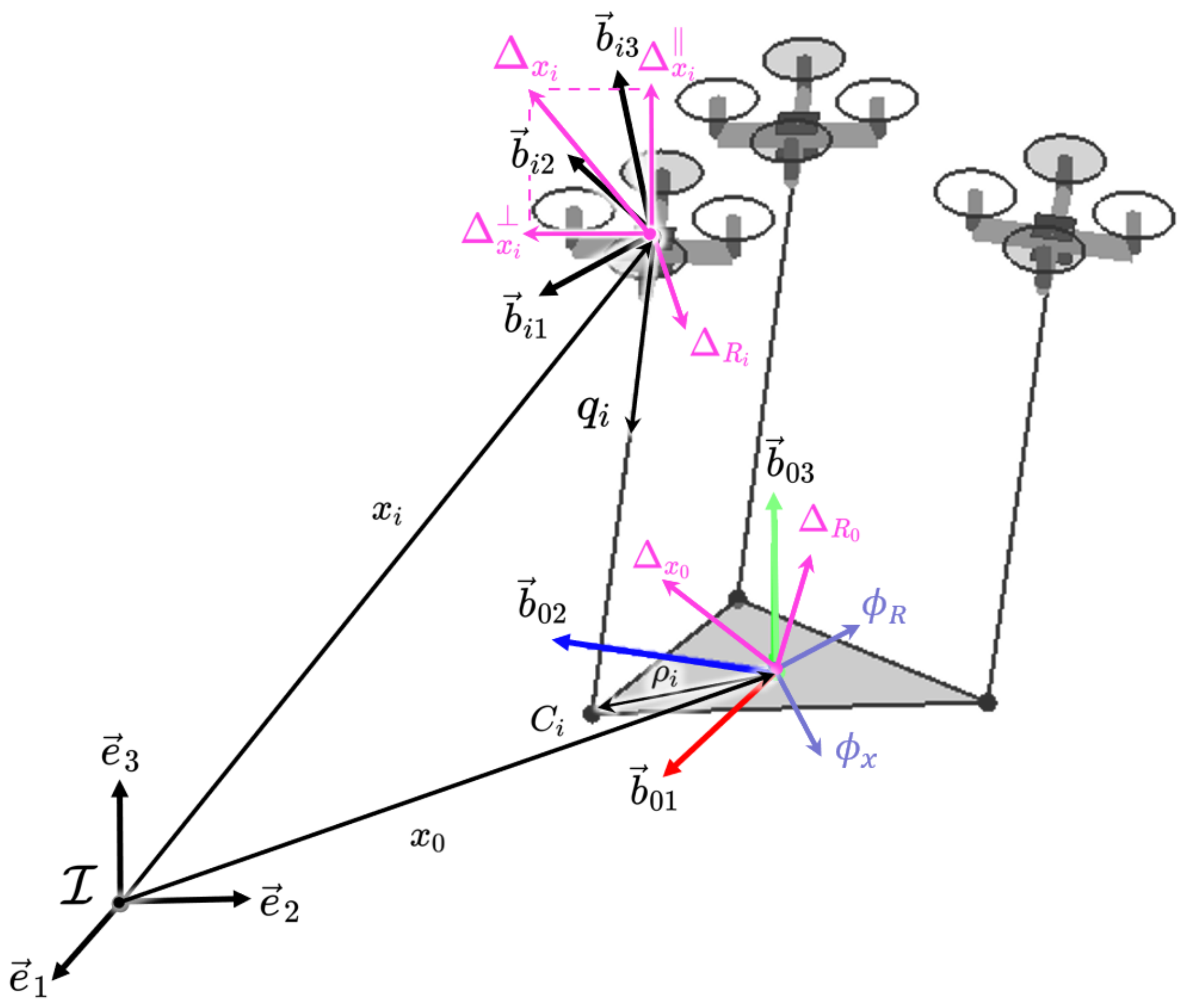}
      \caption{\footnotesize Model of the multi-quadrotor centralized
transportation system with disturbances: $\Delta_{\bm{x}_{i}}$, $\Delta_{\bm{R}_i}$, $\Delta_{\bm{x}_{i}}^{\parallel}$, $\Delta_{\bm{x}_{i}}^{\bot}$, $\Delta_{\bm{x}_0}$, $\Delta_{\bm{R}_0}\in\mathbb{R}^3$ and augmented time-varying disturbance dynamics: $\bm{\phi}_{\bm{\textit{x}}}$, $\bm{\phi}_{\bm{\textit{R}}}\in \mathbb{R}^3$. The $\Delta_{\bm{x}_{i}}$  and $\Delta_{\bm{R}_i}$ denote the disturbance force  and  moment 
 exerted on the $i^{th}$ quadrotor, respectively.  The disturbance force $\Delta_{\bm{x}_{i}}$ can be decomposed into parallel and normal components along the cable, denoted as $\Delta_{\bm{x}_{i}}^{\parallel}$ and  $\Delta_{\bm{x}_{i}}^{\bot}$.  Similarly, the payload experiences disturbance force  $\Delta_{\bm{x}_0}$ and moment $\Delta_{\bm{R}_0}$. The time-varying features of the disturbance on the payload are captured by the additionally introduced terms $\bm{\phi}_{\bm{\textit{x}}}$ and $\bm{\phi}_{\bm{\textit{R}}}$. For the symbol reference, refer to TABLE \uppercase\expandafter{\romannumeral 1}.}
      \label{Gao1}
   \end{figure}

\begin{table*}[thbp]
\label{table1}
  \centering
  \caption{List of Symbol References}
  \begin{tabular}{>{\columncolor{gray!10}}c|l|>{\columncolor{gray!10}}c|l}
    \hline
    \hline
      $C_i\in{E}^3$&Connection point between payload and $i^{th}$ cable& $\bm{a}_i\in\mathbb{R}^3$& \,\,Accelaration of the $i^{th}$ connection point $C_i$ \,\,\\
    $g\in\mathbb{R}$ &Gravitational acceleration & $\bm{\omega}_i\in\mathbb{R}^3$ &\,\, Angular velocity of $i^{th}$ cable\,\,\\
    $l_i\in\mathbb{R}$ &Length of $i^{th}$ cable& $\bm{\rho}_i\in\mathbb{R}^3$ & \,\,Position of  $C_i$ in $\mathcal{B}_0$\,\,\\
    $m_0,m_i\in\mathbb{R}$ & Mass of payload and $i^{th}$ quadrotor& $\bm{J}_0,\bm{J}_i\in\mathbb{R}^{3\times3}$ & \,\,Inertia tensor of payload, $i^{th}$ quadrotor\,\,\\
    $\bm{x}_0,\bm{x}_i\in\mathbb{R}^3$ & Position of payload, $i^{th}$ quadrotor in $\mathcal{I}$ &$\bm{R}_0\in \mathbf{SO}(3)$ & \,\,Rotation Matrix of $\mathcal{B}_0$ relative to $\mathcal{I}$\,\,\\
    $\bm{\dot{x}}_0,\bm{\dot{x}}_i\in\mathbb{R}^3$ & Linear velocity of payload, $i^{th}$ quadrotor in $\mathcal{I}$ &$\bm{R}_{i}\in \mathbf{SO}(3)$ & \,\,Rotation Matrix of $\mathcal{B}_i$ relative to $\mathcal{I}$\,\,\\
    $\bm{\ddot{x}}_0,\bm{\ddot{x}}_i\in\mathbb{R}^3$ & Linear acceleration of payload, $i^{th}$ quadrotor in $\mathcal{I}$ &$\bm{q}_i\in \mathbf{S}^2$ &\,\,Unit vector from $i^{th}$ quadrotor to $C_i$ in $\mathcal{I}$\,\, \\
    $\bm{\Omega}_0,\bm{\dot{\Omega}}_0\in\mathbb{R}^3$ & Angular velocity, acceleration of payload in $\mathcal{I}$ &  $\mathbf{u}_i\in\mathbb{R}^3$ & \,\,Control force at $i^{th}$ quadrotor\,\,\\
    $\bm{\Omega}_i\in\mathbb{R}^3$& Angular velocity of $i^{th}$ quadrotor in $\mathcal{B}_i$ & $\mathbf{M}_i\in\mathbb{R}^3$ & \,\,Control moment at $i^{th}$ quadrotor \,\,\\ 
    \hline
    \hline
  \end{tabular}
   \label{table1}
\end{table*}

\begin{table*}[!t]
  \centering
  \caption{List of Notations: Maps, Operators, Subscripts and Superscripts}
  \begin{tabular}{>{\columncolor{gray!10}}c|l|>{\columncolor{gray!10}}c|l}
    \hline
    \hline
    $[\,\bullet\,]_{\times}$ & Skew-symmetric map: $\mathbb{R}^3\to\mathfrak{so}(3)$ &$\otimes$&Tensor Product \\
   $\bullet^{[\cdot]}$ &  Element extraction map: $ (\mathbb{R}^3\cup\mathbb{R}^{3\times 3})\times\mathbb{N}\to\mathbb{R}$& $\bullet^{\dagger}$& Pseudoinverse of a matrix\\
   $\bullet^\vee$& Vee map: $\mathfrak{so}(3)\to\mathbb{R}^3$ &$\bullet_i$& Quantities associated with the $i^{th}$ quadrotor for $i\in[1, n]$\\
   $\bullet_{\bm{d}}$& Desired value&$\bullet_0$& Quantities associated with the payload\\
   $\bullet^{\text{vec}}$ & Feature vector formed by diagonal elements of a matrix&$\bullet^{\parallel}_i$& Quantities parallel to $\bm{q}_i$\\
   $\bar{\bullet}$ & Estimation value&$\bullet^{\bot}_i$& Quantities perpendicular to $\bm{q}_i$\\
   $\widetilde{\bullet}$ & Estimation error value & $\overset{\tiny \text{max}}{\bullet}$ & Maximum value of a quantity\\
    $\lambda_{\min}(\bullet)$ &Minimum eigenvalue of a matrix & $\overset{\tiny \text{min}}{\bullet}$ & Minimum value of a quantity \\
  $\lambda_{\max}(\bullet)$ &Maximum eigenvalue of a matrix&$\bullet'$&Reference value of a quantity\\
  
    \hline
    \hline
  \end{tabular}
  \label{table2}
\end{table*}

This subsection presents the disturbance-augmented complete dynamics for the multi-quadrotor centralized transportation system. 
Recent studies have extensively explored the centralized geometric control strategy originally proposed in \cite{2018 Geometric Control of Quadrotor UAVs Transporting
a Cable-Suspended Rigid Body}, demonstrating its feasibility through experimental validations \cite{2021 Cooperative Transportation of Cable Suspended Payloads With MAVs Using Monocular Vision and Inertial Sensing}, \cite{2023 Nonlinear Model Predictive Control for Cooperative Transportation and Manipulation of Cable Suspended Payloads with Multiple Quadrotors}, as well as open-source simulation platforms \cite{2024 RotorTM: A Flexible Simulator for Aerial Transportation and Manipulation}. To maintain consistency and facilitate reference to the modeling framework in \cite{2018 Geometric Control of Quadrotor UAVs Transporting
a Cable-Suspended Rigid Body}, we adopt its original notation throughout this paper.
 As illustrated in Fig. \ref{Gao1}, in the $n$-quadrotor scenario, an East-North-Up (ENU) inertial frame $\mathcal{I}:=\{\bm{\vec{e}}_j\}_{j\in\{1,2,3\}}$, a North-East-Up (NEU) payload body-fixed frame $\mathcal{B}_{0}:=\{\bm{\vec{b}}_{0j}\}_{j\in\{1,2,3\}}$  and $n$ North-East-Up (NEU) quadrotor body-fixed frames $\mathcal{B}_{i}:=\{\bm{\vec{b}}_{ij}\}_{j\in\{1,2,3\},i\in[1, n]} $ are defined for modeling. 
 Building upon this, we further augment the dynamics in \cite{2018 Geometric Control of Quadrotor UAVs Transporting
a Cable-Suspended Rigid Body} with unknown time-varying dynamics term of translational and rotational disturbance $\bm{\phi}_{\bm{\textit{x}}}\in \mathbb{R}^3$, $\bm{\phi}_{\bm{\textit{R}}}\in \mathbb{R}^3$ at the acceleration-level, and then reformulate the complete dynamics as follows:
\begin{equation}
{
\begin{aligned}
\bm{\ddot{x}}_0=\frac{1}{m_0}\left(\bm{\mathrm{F}_d}+\Delta_{\bm{x}_0}+\bm{\displaystyle \sum}_{i=1}^n\Delta_{\bm{x}_{i}}^{\parallel}\right)-g\bm{\vec{e}}_3+\bm{Y_{\textit{x}}}+\bm{\phi}_{\bm{\textit{x}}},
\end{aligned}
}
\label{Payload Translational Dynamics}
\end{equation}
\begin{equation}
{
\begin{aligned}
\bm{\dot{\Omega}}_0=&\bm{J}_0^{-1}\left(\bm{\mathrm{M}_d}+\Delta_{\bm{R}_0}+\bm{\displaystyle \sum}_{i=1}^n[\bm{\rho}_i]_{\times}\bm{R}_0^{\top}\Delta_{\bm{x}_{i}}^{\parallel}\right)+\bm{Y_{\textit{R}}}\\
&+\bm{\phi}_{\bm{\textit{R}}}(\bm{J}_0, \bm{\Omega}_0),
\end{aligned}
}
\label{Payload Rotational Dynamics}
\end{equation}
\begin{equation}
{
\begin{aligned}
 \bm{\dot{R}}_0=\bm{R}_0[\bm{\Omega}_0]_{\times},
\end{aligned}
}
\label{Payload Rotational Kinetatics}
\end{equation}
\begin{equation}
{
\begin{aligned}
 \bm{\ddot{q}}_i=\frac{1}{m_il_i}[\bm{q}_i]_{\times}^2(\mathbf{u}_i+\Delta_{\bm{x}_{i}}-m_i\bm{a}_i)-\lVert \bm{\dot{q}}_i \rVert^2_2\bm{q}_i,
\end{aligned}
}
\label{Cable Orientation Dynamics}
\end{equation}
\begin{equation}
{
\begin{aligned}
\bm{\dot{\omega}}_i=\frac{1}{l_i}[\bm{q}_i]_{\times}\bm{a}_i-\frac{1}{m_il_i}[\bm{q}_i]_{\times}(\mathbf{u}_i^{\bot}+\Delta_{\bm{x}_{i}}^{\bot}),
\end{aligned}
}
\label{Cable Angular Velocity Dynamics}
\end{equation}
\begin{equation}
{
\begin{aligned}
\bm{\dot{\Omega}}_i=\bm{J}_i^{-1}(\mathbf{M}_i-\bm{\Omega}_i\times\bm{J}_i\bm{\Omega}_i+\Delta_{\bm{R}_i}),
\end{aligned}
}
\label{quadrotor rotational dynamics}
\end{equation}
\begin{equation}
{
\begin{aligned}
\bm{\dot{R}}_i=\bm{R}_i[\bm{\Omega}_i]_{\times},
\end{aligned}
}
\label{quadrotor rotational kinematics}
\end{equation}
where Eqs.~\eqref{Payload Translational Dynamics}, \eqref{Payload Rotational Dynamics} and \eqref{Payload Rotational Kinetatics} are the translational dynamics, rotational dynamics and rotational kinematics of the payload, respectively. Throughout this paper, symbols with subscript $\bullet_0$ are associated with the payload. The $\bm{x}_0\in\mathbb{R}^3$ denotes the position of the payload in the inertia frame $\mathcal{I}$. The attitude of the payload is represented by a rotation matrix $\bm{R}_0\in \mathbf{SO}(3) = \{\bm{R}_0\in\mathbb{R}^{3\times3}\mid\bm{R}_0^{\top}\bm{R}_0 = \mathbf{I}^{3\times3}, \mathrm{det}(\bm{R}_0) = 1\}$, which describes the rotation of the payload body-fixed frame $\mathcal{B}_0$ relative to the inertial reference $\mathcal{I}$. The $\bm{\Omega}_0\in\mathbb{R}^3$ denotes the angular velocity of the payload in the body-fixed frame $\mathcal{B}_0$. The inertia tensor of the payload is denoted by $\bm{J}_0\in\mathbb{R}^{3\times3}$. The constants $m_0\in\mathbb{R}$ and $g\in\mathbb{R}$ are the mass of the payload and the gravitational acceleration, respectively. The $\bm{\mathrm{F}_d}$ and $\bm{\mathrm{M}_d}\in\mathbb{R}^3$ are the desired resultant control force and moment acting on the payload, which will be designed as first-level wrench control signals in Section 3.1. The vector $\bm{\rho}_i\in\mathbb{R}^3$ denotes the position of $C_i$ (connection point between payload and $i^{th}$ cable) in the payload body-fixed frame $\mathcal{B}_0$. The $\Delta_{\bm{x}_{0}}\in\mathbb{R}^3$ and  $\Delta_{\bm{R}_0}\in\mathbb{R}^3$ are the disturbance force and moment exerted on the payload.

\noindent\textbf{\textit{Notation 1:}}
The symbol $[\,\bullet\,]_{\times}\!:\!\mathbb{R}^3\!\to\!\mathfrak{so}(3)$ denotes the skew-symmetric map defined by the condition that $[\mathfrak{a}]_{\times}\mathfrak{b}=\mathfrak{a}\times\mathfrak{b}, \forall\mathfrak{a},\mathfrak{b}\in\mathbb{R}^3$.

\noindent\textbf{\textit{Remark 1:}} The term $\bm{\phi}_{\bm{\textit{R}}}(\bm{J}_0, \bm{\Omega}_0)$ in Eq.~\eqref{Payload Rotational Dynamics} denotes the rotational disturbance dynamics term $\bm{\phi}_{\bm{\textit{R}}}$ that absorbs a nonlinear gyroscopic term $-\bm{J}_0^{-1}[\bm{\Omega}_0]_{\times}\bm{J}_0\bm{\Omega}_0$. 
In cases where the inertial dynamics is not fully known and thus absorbed into such disturbance terms, recent state-of-the-art (e.g., \cite{2025 Kinematics-informed neural network control on SO(3)}) has employed neural networks to implicitly capture dynamic information. In line with this approach, we also employ neural networks to approximate $\bm{\phi}_{\bm{\textit{R}}}(\bm{J}_0, \bm{\Omega}_0)$ in this work. 

The $\bm{Y_{\textit{x}}}\in\mathbb{R}^3$ and $\bm{Y_{\textit{R}}}\in\mathbb{R}^3$ are errors caused by tracking deviations of cables, respectively:
\begin{equation}
{
\begin{aligned}
\bm{Y_{\textit{x}}}=\frac{1}{m_0}\bm{\sum}_{i=1}^n\left(\bm{\mathrm{\mu}}_{i}-\bm{\mathrm{\mu}}_{i_{\bm{d}}}\right), 
\end{aligned}
}
\label{Yx}
\end{equation}

\begin{equation}
{
\begin{aligned}
\bm{Y_{\textit{R}}}=\bm{J}_0^{-1}\bm{\sum}_{i=1}^n[\bm{\rho}_i]_{\times}\bm{R}_0^{\top}\left(\bm{\mathrm{\mu}}_{i}-\bm{\mathrm{\mu}}_{i_{\bm{d}}}\right),
\end{aligned}
}
\label{YR}
\end{equation}
where $\bm{\mathrm{\mu}}_{i}\in \mathbb{R}^3$ denotes the internal tension along the $i^{th}$  cable. Its actual value is derived by projecting its desired value $\bm{\mathrm{\mu}}_{i_{\bm{d}}}\in\mathbb{R}^3$ onto the cable direction:
\begin{equation}
{
\begin{aligned}
\bm{\mathrm{\mu}}_{i}=(\bm{q}_{i}\otimes\bm{q}_i)\bm{\mathrm{\mu}}_{i_{\bm{d}}}.
\end{aligned}
}
\end{equation}

\noindent\textbf{\textit{Remark 2:}} The value of $\bm{\mathrm{\mu}}_{i_{\bm{d}}}$ is obtained using a tension planner, which will be explained in Section 3.5.

The Eqs.~\eqref{Cable Orientation Dynamics} and \eqref{Cable Angular Velocity Dynamics} describe the directional dynamics and rotational dynamics of the $i^{th}$ cable, respectively. The direction of each cable is represented by a unit vector $\bm{q}_i\in \mathbf{S}^2$ from the $i^{th}$ quadrotor to the $i^{th}$ connection point $C_i$. The $\Delta_{\bm{x}_{i}}\in\mathbb{R}^3$ denotes the disturbance force exerted on the $i^{th}$ quadrotor, which can be decomposed into parallel and normal components along the cable, denoted as $\Delta_{\bm{x}_{i}}^{\parallel}\in\mathbb{R}^3$ and $\Delta_{\bm{x}_{i}}^{\bot}\in\mathbb{R}^3$. These quadrotor disturbance terms appear here because the cable dynamics are inherently coupled with the quadrotor dynamics. The $\bm{\omega}_i\in\mathbb{R}^3$ denotes the angular velocity of the $i^{th}$ cable. The constants $m_i\in\mathbb{R}$ and $l_i\in\mathbb{R}$ are the mass of the $i^{th}$ quadrotor and the length of the $i^{th}$ cable, respectively.  Here, $\mathbf{u}_i\in\mathbb{R}^3$ denotes the control force at the $i^{th}$ quadrotor, which is composed of a parallel component $\mathbf{u}_i^{\parallel}\in\mathbb{R}^3$ and a normal component $\mathbf{u}_i^{\bot}\in\mathbb{R}^3$. Their relationship can be represented by:
\begin{equation}
{
\begin{aligned}
\mathbf{u}_i=\mathbf{u}_i^{\parallel}+\mathbf{u}_i^{\bot},
\end{aligned}
}
\end{equation}
\begin{equation}
{
\begin{aligned}
 \mathbf{u}_i^{\parallel}=(\bm{q}_{i}\otimes\bm{q}_i)\mathbf{u}_i,
\end{aligned}
}
\end{equation}
\begin{equation}
{
\begin{aligned}
 \mathbf{u}_i^{\bot}=(\mathbf{I}^{3\times3}-\bm{q}_{i}\otimes\bm{q}_i)\mathbf{u}_i,
\end{aligned}
}
\end{equation}
where the value of the normal component $\mathbf{u}_i^{\bot}$ will be given by the normal controller designed in Section 3.6, and the parallel component $\mathbf{u}_i^{\parallel}$ is determined by the following dynamically coupled constraint:
\begin{equation}
{
\begin{aligned}
\mathbf{u}_i^{\parallel} =\bm{\mathrm{\mu}}_i+m_il_i\lVert \bm{\omega}_i \rVert^2_2\bm{q}_i-m_i(\bm{q}_{i}\otimes \bm{q}_i) \bm{a}_i.
\end{aligned}
}
\label{parallel component}
\end{equation}
The acceleration $\bm{a}_i\in\mathbb{R}^3$ of the $i^{th}$ connection point $C_i$ is expressed as:
\begin{equation}
{
\begin{aligned}
\bm{a}_i=\bm{\ddot{x}}_0+g\bm{\vec{e}}_3+\bm{R}_0[\bm{\Omega}_0]_{\times}^2\bm{\rho}_{i}-\bm{R}_0[\bm{\rho}_i]_{\times}\bm{\dot{\Omega}}_0,
\end{aligned}
}
\end{equation}
where the quantities with the subscript $\bullet_0$ are associated with the payload. These terms reveal the coupling between the payload and the cable, which has been elaborated in Eqs.~\eqref{Payload Translational Dynamics}, \eqref{Payload Rotational Dynamics} and \eqref{Payload Rotational Kinetatics}.

The Eqs.~\eqref{quadrotor rotational dynamics} and \eqref{quadrotor rotational kinematics} are the rotational dynamics and kinematics of the $i^{th}$ quadrotor, respectively. The $\bm{x}_i\in\mathbb{R}^3$ denotes the position of the $i^{th}$ quadrotor in the inertia frame $\mathcal{I}$. The attitude of the $i^{th}$ quadrotor is represented by a rotation matrix $\bm{R}_i\in \mathbf{SO}(3) = \{\bm{R}_i\in\mathbb{R}^{3\times3}\mid\bm{R}_i^{\top}\bm{R}_i = \mathbf{I}^{3\times3}, \mathrm{det}(\bm{R}_i) = 1\}$, which describes the rotation of the $i^{th}$ quadrotor body-fixed frame $\mathcal{B}_i$ relative to the inertial reference $\mathcal{I}$. The $\bm{\Omega}_i\in\mathbb{R}^3$ denotes the angular velocity of each quadrotor in the body-fixed frame $\mathcal{B}_i$. The $\bm{J}_i\in\mathbb{R}^{3\times3}$ represents the inertia tensor of the $i^{th}$ quadrotor. The control moment at the $i^{th}$ quadrotor is denoted by $\mathbf{M}_i\in\mathbb{R}^3$. The $\Delta_{\bm{R}_i}\in\mathbb{R}^3$ represents the disturbance moment exerted on each quadrotor.

\noindent\textbf{\textit{Remark 3:}} Through the foregoing  rearrangement in Section 2.1, we condense and refine the existing centralized transportation model. Compared to the formulation presented in \cite{2018 Geometric Control of Quadrotor UAVs Transporting
a Cable-Suspended Rigid Body}, our formulation explicitly includes the desired control input of the payload wrench $\{\bm{\mathrm{F}_d}, \bm{\mathrm{M}_d}\}$ instead of the control input thrusts and moments of the quadrotors $\{f_i\in\mathbb{R},\mathbf{M}_i\in\mathbb{R}^3\}_{1\leq i\leq n}$, facilitating the design of the feedforward compensator for the payload controller. This formulation also makes it more convenient for the reader to construct the coupled control architecture in the \textit{MATLAB Simulink} following a proper control flow sequence.

\vspace{1em} 

\noindent\textbf{2.2 Control Problem Formulation}

\begin{figure*}[!t] 
 \centering
 \includegraphics[scale=0.15]{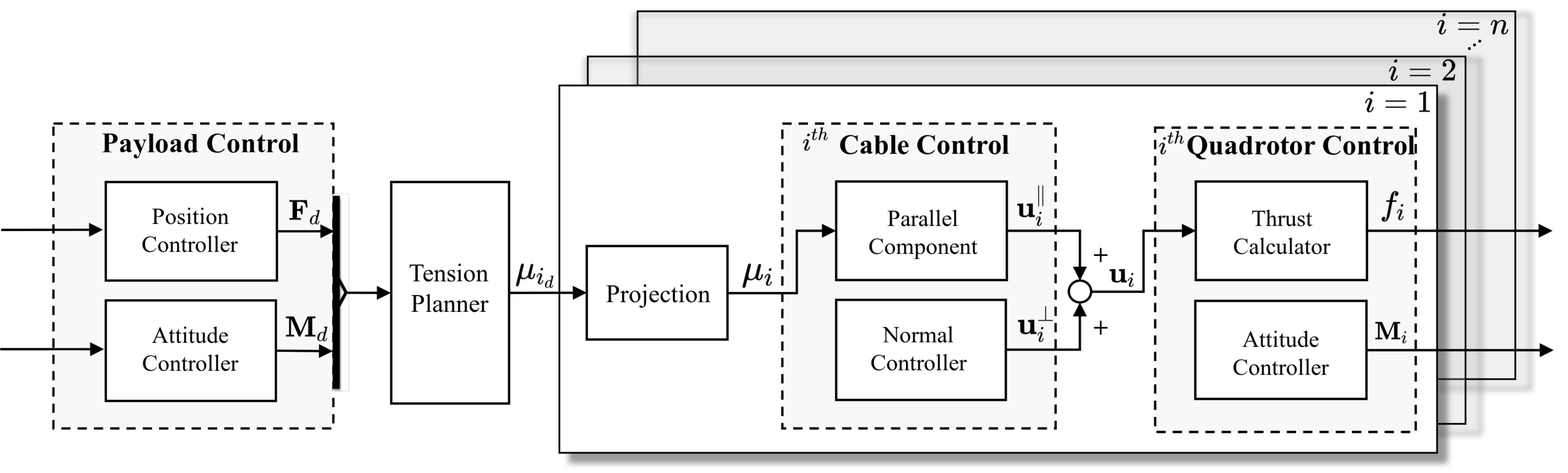}
 \caption{\footnotesize Main control flow of the existing  control strategy for the multi-quadrotor centralized transportation system: $ \{\bm{\mathrm{F}_d}, \bm{\mathrm{M}_d}\}\to\{\bm{\mathrm{\mu}}_{i_{\bm{d}}}\}\to\{\bm{\mathrm{\mu}}_{i}\}\to\{\mathbf{u}_i^{\parallel}, \mathbf{u}_i^{\bot}\}\to\{f_i,\mathbf{M}_i\}$. The $\bm{\mathrm{F}_d}$ and $\bm{\mathrm{M}_d}\in\mathbb{R}^3$ are the desired resultant control force and moment that act on the payload. Their combination, $\{\bm{\mathrm{F}_d}, \bm{\mathrm{M}_d}\}$, represents the desired control wrench. The desired internal tension of the $i^{th}$ cable is denoted by $\bm{\mathrm{\mu}}_{i_{\bm{d}}}\in\mathbb{R}^3$. The actual tension along the $i^{th}$ cable $\bm{\mathrm{\mu}}_{i}\in \mathbb{R}^3$ is then obtained by projecting $\bm{\mathrm{\mu}}_{i_{\bm{d}}}$ onto the cable direction vector $\bm{q}_i\in \mathbf{S}^2$. The control force at the $i^{th}$ quadrotor consists of a parallel component $\mathbf{u}_i^{\parallel}\in\mathbb{R}^3$ and a normal component $\mathbf{u}_i^{\bot}\in\mathbb{R}^3$. Their sum defines the control force $\mathbf{u}_i\in\mathbb{R}^3$ of the $i^{th}$ quadrotor. Eventually, the thrust of each quadrotor $f_i\in\mathbb{R}$ is calculated from this control force, while its orientation is regulated through geometric attitude control input $\mathbf{M}_i\in\mathbb{R}^3$. It can be seen that this control strategy is a payload-centric, multi-level and coupled control architecture, with the desired control wrench $\{\bm{\mathrm{F}_d}, \bm{\mathrm{M}_d}\}$ serving as the first-level command. }
      \label{ControFlow}
\end{figure*}

The existing geometric control for the multi-quadrotor centralized transportation \cite{2018 Geometric Control of Quadrotor UAVs Transporting
a Cable-Suspended Rigid Body} adopts a multi-level control flow:  
\begin{equation}
{
\begin{aligned}
 \{\bm{\mathrm{F}_d}, \bm{\mathrm{M}_d}\}\to\{\bm{\mathrm{\mu}}_{i_{\bm{d}}}\}\to\{\bm{\mathrm{\mu}}_{i}\}\to\{\mathbf{u}_i^{\parallel}, \mathbf{u}_i^{\bot}\}\to\{f_i,\mathbf{M}_i\},\notag
\end{aligned}
}
\end{equation}
for $1\leq i\leq n$. A schematic illustration of this control flow is shown in Fig.~\ref{ControFlow}. The desired payload control wrench $\{\bm{\mathrm{F}_d}, \bm{\mathrm{M}_d}\}$ in \cite{2018 Geometric Control of Quadrotor UAVs Transporting
a Cable-Suspended Rigid Body} employs a Proportional-Integral-Derivative (PID)-driven design. The PID control is inherently a model-free method and is therefore well suited as a high-level controller for the highly coupled and complex dynamics of the centralized transportation system. However, during transportation, the payload is typically subject to highly nonlinear and time-varying disturbances, such as wind fields and external torques. Moreover, its mass and inertia tensor remain uncertain parameters in different missions. Consequently, the effectiveness of PID control in such a centralized transportation framework is limited. This motivates the need for an identification approach to compensate for model uncertainties and unknown time-varying disturbances.

In this study, our objective is to design the desired control wrench for the payload $\{\bm{\mathrm{F}_d}, \bm{\mathrm{M}_d}\}$, given the desired position $\bm{x}_{0_{\bm{d}}}(t)\in\mathbb{R}^3$ and the desired attitude $\bm{R}_{0_{\bm{d}}}(t)\in \mathbf{SO}(3)$ of the payload.
We aim to improve the robustness of payload tracking by introducing a feedforward compensator to the PID controller. In recent studies such as \cite{2023 Quadrotor Neural Network Adaptive Control: Design
and Experimental Validation},  artificial neural networks (ANNs) have been employed to compensate for disturbances in quadrotor systems. Since multilayer neural networks have been proven to be universal approximators \cite{1989 Multilayer feedforward networks are universal approximators}, they serve as an effective tool to identify payload disturbance features without relying on the explicit model of cables or quadrotors. 
To integrate into the control architecture, we define payload tracking errors, which are inputs for both the PID controllers and the neural network-based compensator.

The translational tracking errors of the payload, including the position error $\bm{e}_{\bm{x}_0}\in\mathbb{R}^3$ and the velocity error $\bm{e}_{\bm{v}}\in\mathbb{R}^3 $, are defined as follows:
\begin{equation}
{
\begin{aligned}
    \bm{e}_{\bm{x}_0}:=\bm{x}_0-\bm{x}_{0_{\bm{d}}}, 
    \label{e_x0}
\end{aligned}
}
\end{equation}
\begin{equation}
{
\begin{aligned}
    \bm{e}_{\bm{v}_0}:=\bm{v}_0-\bm{\dot{x}}_{0_{\bm{d}}}=\bm{\dot{e}}_{\bm{x}_0}, 
    \label{e_v0}
\end{aligned}
}
\end{equation}
The rotational tracking errors of the payload, including the attitude error  $\bm{e}_{\bm{R}_0}\in\mathbb{R}^3$ and the angular velocity error $\bm{e}_{\bm{\Omega}_0}\in\mathbb{R}^3$, are defined in the Lie algebra-induced representations as follows:
\begin{equation}
{
\begin{aligned}
     \bm{e}_{\bm{R}_0}:=\frac{1}{2}(\bm{R}_{0_{\bm{d}}}^{\top}\bm{R}_0-\bm{R}_0^{\top}\bm{R}_{0_{\bm{d}}})^\vee, 
    \label{e_R0}
\end{aligned}
}
\end{equation}
\begin{equation}
{
\begin{aligned}
     \bm{e}_{\bm{\Omega}_0}:=\bm{\Omega}_0-\bm{R}_0^{\top}\bm{R}_{0_{\bm{d}}}\bm{\Omega}_{0_{\bm{d}}},
    \label{e_Omega0}
\end{aligned}
}
\end{equation}
where the desired angular velocity of payload $\bm{\Omega}_{0_{\bm{d}}}\in\mathbb{R}^3$ is obtained by:
\begin{equation}
   \bm{\Omega}_{0_{\bm{d}}}:=(\bm{R}^{\top}_{0_{\bm{d}}}\dot{\bm{R}}_{0_{\bm{d}}})^{\vee}.
    \label{Omega0}
\end{equation}
\textbf{\textit{Notation 2: }}
Vee map $\bullet^\vee: \mathfrak{so}(3)\to\mathbb{R}^3$ denotes the inverse of skew-symmetric map $[\,\bullet\,]_{\times}$. 
For any skew-symmetric matrix $S \in \mathfrak{so}(3)$, we have:
\begin{equation}
    S \triangleq
\begin{bmatrix}
0 & -s_3 & s_2 \\
s_3 & 0 & -s_1 \\
-s_2 & s_1 & 0
\end{bmatrix},
\quad
S^\vee \triangleq
\begin{bmatrix}
s_1 \\
s_2 \\
s_3
\end{bmatrix}.
\end{equation}
This mapping enables a Euclidean vector representation of rotational errors in $\mathbb{R}^3$ by using Lie algebra $\mathfrak{so}(3)$.

\vspace{1em} 
\noindent\textbf{3. Contol System Design}

The internal tension along the $i^{th}$ cable $\bm{\mathrm{\mu}}_{i}$ and the parallel component of the control force on the $i^{th}$ quadrotor $\mathbf{u}_i^{\parallel}$ have been introduced in Section 2.1, as they are determined by the coupling relations imposed by the system dynamics rather than by the controller design. 
This section further presents the design of the remaining control system, including the desired payload control wrench $\{\bm{\mathrm{F}_d}, \bm{\mathrm{M}_d}\}$, the online identification module: Sliced Adaptive-Neuro Mapping (SANM), the adaptive law ``slices'', the neural network ``slices'', the desired tension of the $i^{th}$ cable calculated from the tension planner $\bm{\mathrm{\mu}}_{i_{\bm{d}}}$, the normal control force of the $i^{th}$ quadrotor $\mathbf{u}_i^{\bot}$, and the quadrotor control $\{f_i,\mathbf{M}_i\}$. 

\vspace{1em} 

\noindent\textbf{3.1 Payload Control :} $\{\bm{\mathrm{F}_d}, \bm{\mathrm{M}_d}\}$

\begin{figure*}[!t] 
 \centering
 \includegraphics[scale=0.13]{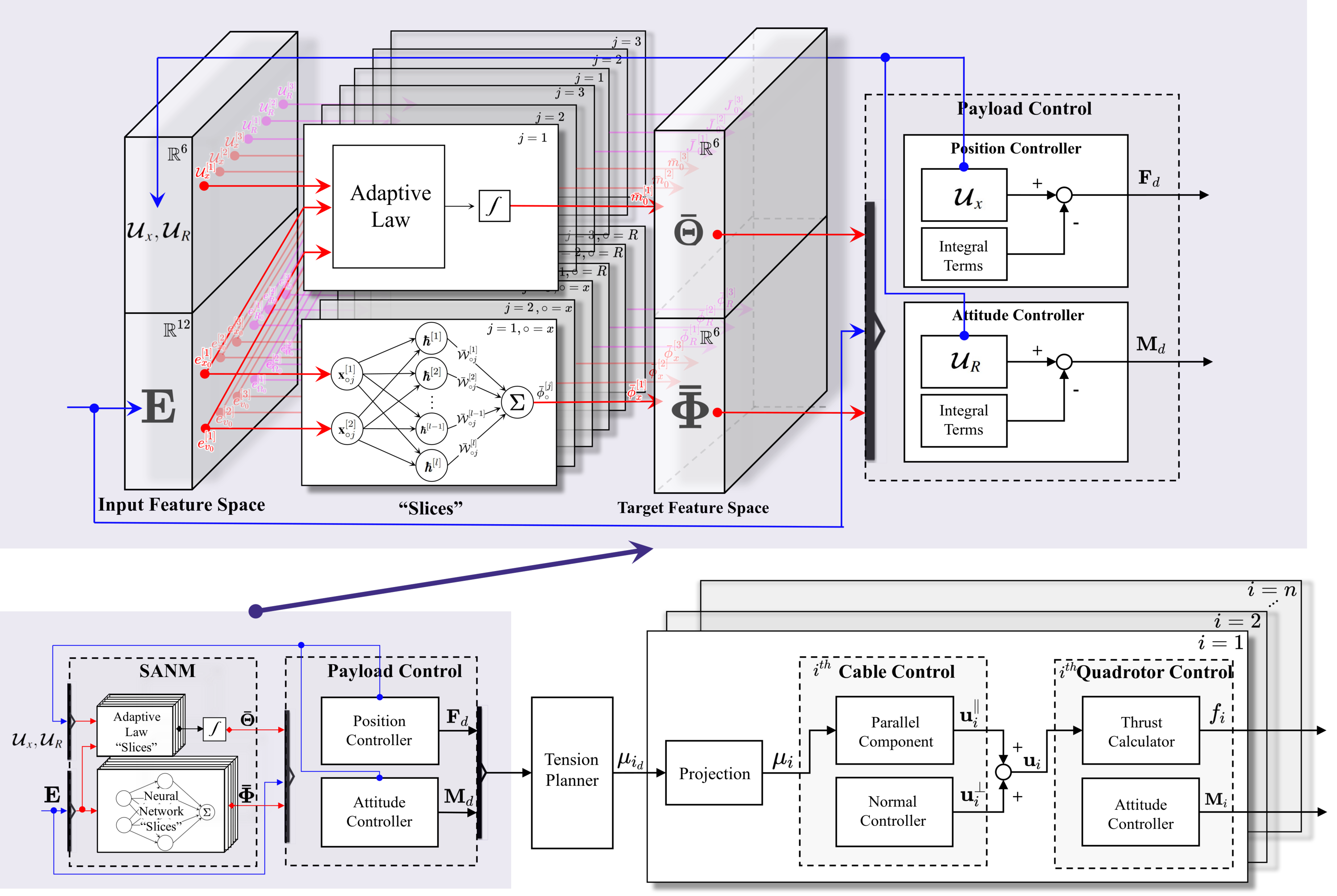}
 \caption{\footnotesize Main control flow of our control strategy for enhancing the robustness of the multi-quadrotor centralized transportation system. Compared with the existing strategy, we adopt an online identification module termed the Sliced Adaptive-Neuro Mapping (SANM) as a feedforward compensator of the payload control. The structure of SANM is illustrated in the upper part of this schematic. The design of this strategy is detailed in Section 3. The payload control is first described in Section 3.1. The SANM module is then introduced in Section 3.2 and its ``slices'' are designed in Section 3.3 and 3.4. The tension planner is developed in Section 3.5. The cable control is presented in Section 3.6. The quadrotors adopt geometric control, which is introduced in Section 3.7.}
      \label{ControFlow_SANM}
\end{figure*}

This subsection describes the design of the desired payload control wrench $\{\bm{\mathrm{F}_d}, \bm{\mathrm{M}_d}\}$.
First, we define the estimated values of the model parameters and the disturbance dynamics of the payload $\{\bm{\bar{m}}_0,\bm{\bar{J}}_0,\bm{\bar{\phi}}_{\bm{\textit{x}}},\bm{\bar{\phi}}_{\bm{\textit{R}}}\}$, which are intended to be identified online by the SANM module. The $\bm{\bar{m}}_0\in \mathbb{R}^{3}$ denotes the estimated payload mass vector and the $\bm{\bar{J}}_0\in\mathbb{R}^{3\times3}$ denoets the estimated diagonalized  inertia tensor. The $\bm{\bar{\phi}_\bm{\textit{x}}}\in\mathbb{R}^3$ and $\bm{\bar{\phi}_\bm{\textit{R}}}\in\mathbb{R}^3$ represents the estimated dynamics of the translational and rotational disturbance, respectively. 
These estimated values are then incorporated into the Proportional-Derivative (PD) control laws for translational payload tracking. Specifically,
$\bm{\mathcal{U}_{\bm{\textit{x}}}}^{[j]}\in\mathbb{R}$ and $\bm{\mathcal{U}_{\bm{\textit{R}}}}^{[j]}\in\mathbb{R}$ denote the $j^{\text{th}}$ component of the translational control law $\bm{\mathcal{U}_{\bm{\textit{x}}}}\in\mathbb{R}^3$ and the  rotational control law $\bm{\mathcal{U}_{\bm{\textit{R}}}}\in\mathbb{R}^3$  of the payload along $\bm{\vec{e}}_{j}$-axis and $\bm{\vec{b}}_{0j}$-axis, respectively:
   \begin{equation}
{
\begin{aligned}
\bm{\mathcal{U}_{\bm{\textit{x}}}}^{[j]}:=\bm{\bar{m}}_0^{[j]}
\left(\bm{-\mathcal{K}}_{\bm{\textit{x}} j}^{\top}\bm{\mathcal{E}}_{\bm{\textit{x}} j}+\bm{\ddot{x}}_{0_{\bm{d}}}^{[j]}+g\bm{\delta}_{j3} -\bm{\bar{\phi}}_{\bm{\textit{x}}}^{[j]}\right),
\end{aligned}
}
\label{Translational Adaptive-Neuro Control}
\end{equation}
\begin{equation}
{
\begin{aligned}
\bm{\mathcal{U}_{\bm{\textit{R}}}}^{[j]}:=&\bm{\bar{J}}_0^{[j]}
\bigg{\{}-k_{R_0}\bm{e}^{[j]}_{\bm{R}_0}-k_{\Omega_0}\bm{e}^{[j]}_{\bm{\Omega}
_0}-\left([\bm{\Omega}_0]_{\times}\bm{R}_0^{\top}\bm{R}_{0_{\bm{d}}}\bm{\Omega}_{0_{\bm{d}}}\right)^{[j]}\\
&+\left(\bm{R}_0^{\top}\bm{R}_{0_{\bm{d}}}\bm{\dot{\Omega}}_{0_{\bm{d}}}\right)^{[j]}
-\bm{\bar{\phi}}_{\bm{\textit{R}}}^{[j]}\bigg{\}},
\end{aligned}
}
\label{Rotational Adaptive-Neuro Control}
\end{equation}
where $\{\bm{\bar{m}}_0^{[j]}\in\mathbb{R}\}_{j\in\{1,2,3\}}$ are the $j^{\text{th}}$ elements of $\bm{\bar{m}}_0$, i.e., the estimated values of the
mass along the $\bm{\vec{e}}_{j}$-axis.  The $\{\bm{\bar{\phi}}_{\bm{\textit{x}}}^{[j]}\in\mathbb{R}\}_{j\in\{1,2,3\}}$ are the $j^{\text{th}}$ elements of $\bm{\bar{\phi}_\bm{\textit{x}}}$, i.e., the estimated dynamics of translational  disturbance decomposed along  $\bm{\vec{e}}_{j}$-axis. The $\{\bm{\bar{J}}^{[j]}\in\mathbb{R}\}_{j\in\{1,2,3\}}$ are the $j^{th}$ elements of the main diagonal of $\bm{\bar{J}}_0$, i.e., the moment of inertia along $\bm{\vec{b}}_{0j}$-axis.  The $\{\bm{\bar{\phi}}_{\bm{\textit{R}}}^{[j]}\in\mathbb{R}\}_{j\in\{1,2,3\}}$ are the $j^{\text{th}}$ elements of $\bm{\bar{\phi}_\bm{\textit{R}}}$, i.e., the estimated dynamics of the rotational disturbance decomposed along the $\bm{\vec{b}}_{0j}$-axis. In addition,  $\bm{\mathcal{K}}_{\bm{\textit{x}} j}:=\left(\bm{k_{\text{p}}}^{[j]},\bm{k_{\text{d}}}^{[j]}\right)^{\top}$ is the gain vector for translational PD control along $\bm{\vec{e}}_j$-axis  with positive gain vectors $\bm{k_{\text{p}}}$, $\bm{k_{\text{d}}}\in\mathbb{R}^{3}$. The $\bm{\mathcal{E}}_{\bm{\textit{x}} j}:=\left(\bm{e}_{\bm{x}_0}^{[j]}, \bm{\dot{e}}_{\bm{x}_0}^{[j]}\right)^{\top}$ denotes the translation error vector along the $\bm{\vec{e}}_j$-axis. The symbol $\bm{\delta}_{j3}$ denotes a Kronecker delta where $\bm{\delta}_{j3}=1$ when $j=3$ and $\bm{\delta}_{j3}=0$ otherwise. The $k_{R_0}$ and $k_{\Omega_0}\in \mathbb{R}^{+}$ are positive gains for rotational PD control. 

\noindent\textbf{\textit{Notation 3:}} The superscript $\bullet^{[j]}$  denotes an element extraction map $\bullet^{[j]}: (\mathbb{R}^3\cup\mathbb{R}^{3\times 3})\times\{1,2,3\}\to\mathbb{R}$ which extracts the $j^{th}$ element from either a vector or the main diagonal of a matrix for $j\in\{1,2,3\}$. 

\noindent\textbf{\textit{Remark 4:}} The reason we design each component of the payload control laws separately is to align with the ``dimension-decomposition'' characteristic of our proposed online learning framework. 
As will be discussed later, the proposed SANM module performs online identification by decomposing the high-dimensional input space along each axis, thereby enabling parallel online learning on low-dimensional subspaces. 

Then, the integral terms are appended to give the $j^{th}$ element of the desired payload control wrench $\{\bm{\mathrm{F}_d}, \bm{\mathrm{M}_d}\}$:
\begin{equation}
{
\begin{aligned}
\!\!\!\!\!\!\!\!\!\!\!\!\!\!\!\!\!\!\!\!\!\!\!\!\!\!\!\bm{\mathrm{F}_d}^{[j]}:=\,\, &\bm{\mathcal{U}_{\bm{\textit{x}}}}^{[j]}-\bar{\Delta}_{\bm{x}_0}^{[j]}-\bm{\sum}^n_{i=1}\bar{\Delta}_{\bm{x}_i}^{\parallel[j]},\\[-5pt]
\end{aligned}
}
\label{Payload Translational Control}
\end{equation}
\begin{equation}
{
\begin{aligned}
\bm{\mathrm{M}_d}^{[j]}:=\bm{\mathcal{U}_{\bm{\textit{R}}}}^{[j]}-\bar{\Delta}^{[j]}_{\bm{R}_0}-\bm{\sum}_{i=1}^n\left([\bm{\rho}_i]_{\times}\bm{R}_0^{\top}\bar{\Delta}_{\bm{x}_i}^{\parallel}\right)^{[j]},\\[-5pt]
\end{aligned}
}
\label{Payload Rotational Control}
\end{equation}
where  $\bar{\Delta}_{\bm{x}_0}$, $\bar{\Delta}_{\bm{R}_0}$  and $\bar{\Delta}_{\bm{x}_i}^{\parallel}\in\mathbb{R}^3$ are estimated disturbances of the payload and $i^{th}$ quadrotor, which are updated through the following equations derived from the Lyapunov analysis:
\begin{equation}
{
\begin{aligned}
\dot{\bar{\Delta}}^{[j]}_{\bm{x}_0}:=\frac{\textit{h}_{x_0}}{m'_0}\bm{\mathcal{E}}_{\bm{\textit{x}} j}^{\top}\bm{P}_j\bm{B}, \,\,\,\,\, \dot{\bar{\Delta}}^{[j]}_{\bm{R}_0}:=\frac{\textit{h}_{R_0}}{\bm{J}_0'^{[j]}}\bm{e}_{\bm{\Omega}_0}^{[j]},
\end{aligned}
     }
     \label{integral compensations0}
\end{equation}
\begin{equation}
{
\begin{aligned}
    \dot{\bar{\Delta}}_{\bm{x}_i}:=&\textit{h}_{x_i}(\bm{q}_i \otimes \bm{q}_i)\bigg{\{}\bm{\sum}_{j=1}^3\frac{1}{m'_0}\mathfrak{u}_j\bm{\mathcal{E}}_{\bm{\textit{x}} j}^{\top}\bm{P}_j\bm{B}\\[-5pt]
    &-\bm{J}'^{-1}_0\bm{R}_0[\bm{\rho}_i]_{\times}\bm{e}_{\bm{\Omega}_0}+\frac{\textit{h}_{x_i}}{m_i l_i}[\bm{q}_i]_{\times}(\bm{e}_{\bm{\omega}_i}+\textit{c}_{q}\bm{e}_{\bm{q}_i})\bigg{\}}.
\end{aligned}
     }\label{integral compensations}
\end{equation}
Here, $m'_0\in\mathbb{R}$ and $\bm{J}_0'\in\mathbb{R}^3$ denote the reference mass and the inertia tensor of the payload. The parameters $\textit{c}_{q}$, $\textit{h}_{x_0}$, $\textit{h}_{R_0}$,  $\textit{h}_{x_i}\in\mathbb{R}^{+}$ are positive constants. The vector $\mathfrak{u}_j\in\mathbb{R}^3$ denotes a unit vector with its $j^{th}$ element equal to 1. The term $\bar{\Delta}_{\bm{x}_i}^{\parallel}$ is obtained by projecting $\bar{\Delta}_{\bm{x}_i}$ onto the direction of $\bm{q}_i$, i.e., $\bar{\Delta}_{\bm{x}_i}^{\parallel}:=(\bm{q}_i \otimes \bm{q}_i)\bar{\Delta}_{\bm{x}_i}$.  The $\{\bm{P}_j\in\mathbb{R}^{2\times2}\}_{j\in\{1,2,3\}}$ denote  Lyapunov matrices acquired from Lyapunov equations given in Eq.~\eqref{Lyapunov equations}. The $\bm{e}_{\bm{\omega}_i}$ and $\bm{e}_{\bm{q}_i}$ are tracking errors of the $i^{th}$ cable, which will be introduced in Section 3.6. The vector $\bm{B}=\left(0,1\right)^{\top}$. The estimation errors of these integral compensation terms are defined as follows: 
\begin{align}
\widetilde{\Delta}_{\bm{x}_0}\in\mathbb{R}^3\triangleq\Delta_{\bm{x}_0}-\bar{\Delta}_{\bm{x}_0},\label{Estimation errors of integral terms1}\\ 
\widetilde{\Delta}_{\bm{R}_0}\in\mathbb{R}^3\triangleq\Delta_{\bm{R}_0}-\bar{\Delta}_{\bm{R}_0},\label{Estimation errors of integral terms2}\\
\widetilde{\Delta}_{\bm{x}_i}\in\mathbb{R}^3\triangleq\Delta_{\bm{x}_i}-\bar{\Delta}_{\bm{x}_i}.\label{Estimation errors of integral terms3}
\end{align}

\vspace{1em} 

\noindent\textbf{3.2 Sliced Adaptive-Neuro Mapping (SANM)}

In this subsection,  we introduce the online identification module termed Sliced Adaptive-Neuro Mapping (SANM) to acquire the estimated values $\{\bm{\bar{m}}_0,\bm{\bar{J}}_0,\bm{\bar{\phi}}_{\bm{\textit{x}}},\bm{\bar{\phi}}_{\bm{\textit{R}}}\}$. Consider an Adaptive-Neuro mapping $\bm{\mathcal{S}}_{AN}$:
\begin{equation}
(\!\!\!\!\!\!\!\!\!\!\!\!\!\underbrace{\bm{\bar{\Theta}}_,\bm{\bar{\Phi}}}_{\textbf{Target Feature Space}}\!\!\!\!\!\!\!\!\!\!\!\!\!) \triangleq \bm{\mathcal{S}}_{AN}(\!\!\!\underbrace{(\bm{\mathcal{U}_{\bm{\textit{x}}}}, \bm{\mathcal{U}_{\bm{\textit{R}}}}),\mathbf{E}}_{\textbf{Input Feature Space}}\!\!\!):\mathbb{R}^6\times\bm{\mathcal{C}}\to\mathbb{R}^6\times\mathbb{R}^6,\notag
\end{equation}
where $\mathbf{E}\triangleq\big{(}\bm{e}_{\bm{x}_0}^{\top},\bm{e}_{\bm{v}_0}^{\top}, \bm{e}^{\top}_{\bm{R}_0},\bm{e}^{\top}_{\bm{\Omega}_0}\big{)}^{\top}\!\!\!\in\!\!\bm{\mathcal{C}}\!\!\subset\!\!\mathbb{R}^{12}$ is defined as the full-state error vector of the payload, and $\bm{\mathcal{C}}$ denotes a compact subset of that bounds the state error. The vector $\bm{\bar{\Theta}}\!\!:=\!\!\left(\bm{\bar{m}}_0^{[1]},\bm{\bar{m}}_0^{[2]},\bm{\bar{m}}_0^{[3]},\bm{\bar{J}}_0^{[1]},\bm{\bar{J}}_0^{[2]},\bm{\bar{J}}_0^{[3]}\right)^{\top}\!\!\in\!\!\mathbb{R}^{6}$ denotes the estimated model feature and vector $\bm{\bar{\Phi}} \!:=\!\!\left(\bm{\bar{\phi}}_{\bm{\textit{x}}}^{[1]},\bm{\bar{\phi}}_{\bm{\textit{x}}}^{[2]},\bm{\bar{\phi}}_{\bm{\textit{x}}}^{[3]},\bm{\bar{\phi}}_{\bm{\textit{R}}}^{[1]},\bm{\bar{\phi}}_{\bm{\textit{R}}}^{[2]},\bm{\bar{\phi}}_{\bm{\textit{R}}}^{[3]}\right)^{\top}\!\!\in\mathbb{R}^6$ represents the estimated disturbance dynamics feature.  

Subsequently,  we decompose the aforementioned high-dimensional Adaptive-Neuro mapping into a set of low-dimensional submappings (“slices”), described as follows for $j\in\{1,2,3\}$:
\begin{equation}
\small{
\begin{aligned}
&\bigoplus_{j=1}^{3}\Bigg{\{}\Big{(}\underbrace{\bm{\bar{m}}_0^{[j]},\bm{\bar{J}}_0^{[j]},\bm{\bar{\phi}}_{\bm{\textit{x}}}^{[j]},\bm{\bar{\phi}}_{\bm{\textit{R}}}^{[j]}}_{\textbf{Sliced Target Feature Space}}\Big{)}\!:=\\
&\! \bm{\mathcal{S}}^{[j]}_{AN}\Big{(}\underbrace{\bm{\mathcal{U}_{\bm{\textit{x}}}}^{[j]},\bm{\mathcal{U}_{\bm{\textit{R}}}}^{[j]},(\bm{e}_{\bm{x}_0}^{[j]}, \bm{e}_{\bm{v}_0}^{[j]}),(\bm{e}_{\bm{R}_0}^{[j]}, \bm{e}_{\bm{\Omega}_0}^{[j]})}_{\textbf{Sliced Input Feature Space}}\Big{)}\!:\!\mathbb{R}\!\times\!\!\mathbb{R}\!\times\!(\mathbb{R}^{2})^2\!\!\to\!(\mathbb{R})^4\Bigg{\}}.\\
\end{aligned}
}\notag
\end{equation}
where the symbol $\bigoplus$ denotes the direct sum. The structure of this Sliced Adaptive-Neuro Mapping (SANM) is shown in Fig.~\ref{ControFlow_SANM}. Here, the original 12-dimensional target feature is represented as a set of 12 one-dimensional features.
This approach decomposes the high-dimensional error space of the payload into several low-dimensional subspaces, in which adaptive laws and neural network submappings (``slices'') are constructed. The design of these ``slices'' is provided in the following subsections.

\vspace{1em} 
\noindent\textbf{3.3 Adaptive Law ``Slices'' : }$\{\bm{\bar{m}}_0^{[j]},\bm{\bar{J}}_0^{[j]}\}_{j\in\{1,2,3\}}$

This subsection presents the adaptive law submappings (``slices'') of the SANM module. The components of the estimated model feature introduced in Section 3.2: $\{\bm{\bar{m}}_0^{[j]},\bm{\bar{J}}_0^{[j]}\}_{j\in\{1,2,3\}}$, are designed to be updated online by the following inherently bounded adaptive laws:  
\begin{equation}
{
\footnotesize
\bm{\dot{\bar{m}}}_0^{[j]}:=
\begin{cases}
   \frac{-\bm{\bar{m}}_0^{[j]^2}}{\eta_{\textit{m}_j}}\bm{\mathcal{E}}_{\bm{\textit{x}} j}^{\top}\bm{P}_j\bm{B}\bm{\mathcal{U}_{\bm{\textit{x}}}}^{[j]}, \,\, \bm{\mathcal{E}}_{\bm{\textit{x}} j}^{\top}\bm{P}_j\bm{B}\bm{\mathcal{U}_{\bm{\textit{x}}}}^{[j]}>0\\[2pt]
 \frac{-\bm{\bar{m}}_0^{[j]^2}}{\eta_{\textit{m}_j}}\bm{\mathcal{E}}_{\bm{\textit{x}} j}^{\top}\bm{P}_j\bm{B}\bm{\mathcal{U}_{\bm{\textit{x}}}}^{[j]}, \,\, \bm{\mathcal{E}}_{\bm{\textit{x}} j}^{\top}\bm{P}_j\bm{B}\bm{\mathcal{U}_{\bm{\textit{x}}}}^{[j]}\leq 0, \,\,\bm{\bar{m}}_0^{[j]}<\overset{\tiny \text{max}}{m_0}\\[2pt]
 \mathfrak{s}_{\textit{m}_j}\frac{-\bm{\bar{m}}_0^{[j]^2}}{\eta_{\textit{m}_j}}, \,\, \,\,\,\,\,\,\,\,\,\,\,\,\,\,\,\,\,\,\,\,\,\,\,\bm{\mathcal{E}}_{\bm{\textit{x}} j}^{\top}\bm{P}_j\bm{B}\bm{\mathcal{U}_{\bm{\textit{x}}}}^{[j]}\leq 0 , \,\,\bm{\bar{m}}_0^{[j]}\geq\overset{\tiny \text{max}}{m_0}
\end{cases}
}
\label{Adaptive Law of mass}
\end{equation}
\begin{equation}
{
\footnotesize
\bm{\dot{\bar{\mathit{J}}}}_0^{[j]}\!\!:=\!\!
\begin{cases}
   \frac{-\bm{\bar{J}}_0^{[j]^2}}{\eta_{J_j}}(\bm{e}^{[j]}_{\bm{\Omega}_0}\!\!+\!\!c_R\bm{e}^{[j]}_{\bm{R}_0})\bm{\mathcal{U}_{\bm{\textit{R}}}}^{[j]}, (\bm{e}^{[j]}_{\bm{\Omega}_0}\!\!+\!\!c_R\bm{e}^{[j]}_{\bm{R}_0})\bm{\mathcal{U}_{\bm{\textit{R}}}}^{[j]}>0\\[2pt]
\!\!\frac{-\bm{\bar{J}}_0^{[j]^2}}{\eta_{J_j}}(\bm{e}^{[j]}_{\bm{\Omega}_0}\!\!+\!\!c_R\bm{e}^{[j]}_{\bm{R}_0})\bm{\mathcal{U}_{\bm{\textit{R}}}}^{[j]}, (\bm{e}^{[j]}_{\bm{\Omega}_0}\!\!+\!\!c_R\bm{e}^{[j]}_{\bm{R}_0})\bm{\mathcal{U}_{\bm{\textit{R}}}}^{[j]}\!\leq\! 0, \,\,\bm{\bar{J}}_0^{[j]}\!<\!\overset{\tiny \text{max}}{\bm{J}_0}\overset{[j]}{\rule{0pt}{1.5ex}}\\[2pt]
 \mathfrak{s}_{J_j}\frac{-\bm{\bar{J}}_0^{[j]^2}}{\eta_{J_j}}, \,\, \,\,\,\,\,\,\,\,\,\,\,\,\,\,\,\,\,\,\,\,\,\,\,\,\,\,\,\,\,\,\,\,\,\,(\bm{e}^{[j]}_{\bm{\Omega}_0}\!\!+\!\!c_R\bm{e}^{[j]}_{\bm{R}_0})\bm{\mathcal{U}_{\bm{\textit{R}}}}^{[j]}\!\leq\! 0, \,\,\bm{\bar{J}}_0^{[j]}\!\geq\!\overset{\tiny \text{max}}{\bm{J}_0}\overset{[j]}{\rule{0pt}{1.5ex}}
\end{cases}
}
\label{Adaptive Law of Inertia Tensor}
\end{equation}
where the  $c_R\in\mathbb{R}^+$ is a positive constant. The $\{1/\eta_{\textit{m}_j}\in\mathbb{R}^+\}_{j\in\{1,2,3\}}$ and $\{1/\eta_{J_j}\in\mathbb{R}^+\}_{j\in\{1,2,3\}}$ are the update rates. The $\{\mathfrak{s}_{\textit{m}_j}\in\mathbb{R}^+\}_{j\in\{1,2,3\}}$ and $\{\mathfrak{s}_{J_j}\in\mathbb{R}^+\}_{j\in\{1,2,3\}}$ are the scaling factors.  The preset maximum mass and maximum inertia tensor of the payload are denoted by  $\overset{\tiny \text{max}}{m_0}\in\mathbb{R}$ and  $\overset{\tiny \text{max}}{\bm{J}_0}\in\mathbb{R}^{3\times3}$ , respectively.  As in Eqs. \eqref{integral compensations0} and \eqref{integral compensations},  the $\{\bm{P}_j\in\mathbb{R}^{2\times2}\}_{j\in\{1,2,3\}}$ denote  Lyapunov matrices acquired from Lyapunov equations and the vector $\bm{B}=\left(0,1\right)^{\top}$.

The estimation errors of the mass feature components
$\{\widetilde{m}_j\!\in\!\mathbb{R}\}_{j\in\{1,2,3\}}$  and the inertia feature components
$\{\widetilde{J}_{j}\!\in\!\mathbb{R}\}_{j\in\{1,2,3\}}$  are defined in reciprocal forms, respectively: 
\begin{align}
     \widetilde{m}_j\triangleq\frac{1}{m_0}-\frac{1}{\bm{\bar{m}}_0^{[j]}}, \,\,\, \widetilde{J}_{j}\triangleq\frac{1}{\bm{J}_0^{[j]}}-\frac{1}{\bar{\bm{J}}_0^{[j]}}.
\label{Estimation errors of mass and inertia features}
\end{align}

\vspace{1em} 
\noindent\textbf{3.4 Neural Network ``Slices'' : } $\{\bm{\bar{\phi}}_{\bm{\textit{x}}}^{[j]},\bm{\bar{\phi}}_{\bm{\textit{R}}}^{[j]}\}_{j\in\{1,2,3\}}$

This subsection presents the neural network submappings (``slices'') of the SANM module. The universal approximation theorem \cite{1989 Multilayer feedforward networks are universal approximators} ensured that the unknown time-varying dynamics term of translational and rotational disturbance, $\bm{\phi}_{\bm{\textit{x}}}$ and $\bm{\phi}_{\bm{\textit{R}}}$, can be approximated with arbitrary accuracy on a compact set by sufficiently parameterized neural networks. Based on this theorem, the elements of $\bm{\phi}_{\bm{\textit{x}}}$ and $\bm{\phi}_{\bm{\textit{R}}}$, denoted as $\{\bm{\phi}_{\circ}^{[j]}\}_{\circ\in\{\bm{\textit{x}}, \bm{\textit{R}}\}, j\in\{1,2,3\}}$, can also be approximated by multiple neural networks. Therefore, we construct several Radial Basis Function (RBF) neural networks, each configured with a 2 inputs-$l$ hidden layer neurons-1 output (2-$l$-1) structure, as described below:
\begin{equation}
{
\begin{aligned}
\bm{\phi}_{\circ}^{[j]}=\bm{\mathcal{W}}_{\circ j}^{\top}\bm{\hbar}(\textbf{x}_{\circ j})+\epsilon_{\circ j},
\label{Phi}
\end{aligned}
}
\end{equation}
where index subscript $\bullet_{\circ\in\{\bm{\textit{x}}, \bm{\textit{R}}\}}$, refers to the symbols relating to translational and rotational dynamics. The 
 $\{\textbf{x}_{\circ j}\in \mathbb{R}^2\}_{\circ\in\{\bm{\textit{x}}, \bm{\textit{R}}\}, j\in\{1,2,3\}}$ denote the input vectors of $j^{th}$ neural network, and  $\{\bm{\mathcal{W}}_{\circ j}\!\in\!\mathbb{R}^{l}\}_{\circ\in\{\bm{\textit{x}}, \bm{\textit{R}}\}, j\in\{1,2,3\}}$, $\{\epsilon_{\circ j}\!\in\!\mathbb{R}\}_{\circ\in\{\bm{\textit{x}}, \bm{\textit{R}}\}, j\in\{1,2,3\}}$, $\{\bm{\hbar}(\textbf{x}_{\circ j})\!\in\!\mathbb{R}^l\}_{\circ\in\{\bm{\textit{x}}, \bm{\textit{R}}\}, j\in\{1,2,3\}}$  are the corresponding weight vectors, bounded intrinsic approximation errors and Gaussian activation functions, respectively. Here, the weight vectors are designed to be bounded within each compact set $\mathbb{W}_{\circ j}=\{\bm{\mathcal{W}}_{\circ j}\!\in\!\mathbb{R}^l\,|\,\|\bm{\mathcal{W}}_{\circ j}\|\leq r_{\circ j}\}_{\circ\in\{\bm{\textit{x}}, \bm{\textit{R}}\}, j\in\{1,2,3\}}$ for the corresponding positive constant $r_{\circ j}\in\mathbb{R}^+$.

The output of the $k^{th}$ hidden layer neuron in the $j^{th}$ neural network ``slice" is given as follows:
\begin{equation}
{
\begin{aligned}
&\bm{\hbar}^{[k]}(\textbf{x}_{\circ j}):=\mathrm{exp}\left(-\frac{\lVert\textbf{x}_{\circ j}-\textbf{c}_{kj}\rVert^2}{2b^{2}_{kj}}\right),
\end{aligned}
\label{Gaussian activation function}
}
\end{equation}
where $\{\textbf{c}_{kj}\in\mathbb{R}^2\}_{1\leq k \leq l, j\in\{1,2,3\}}$ denotes the center vector of the $k^{th}$ neuron and $\{b_{kj}\in\mathbb{R}\}_{1\leq k \leq l, j\in\{1,2,3\}}$ represents the width of the $k^{th}$ Gaussian function for $1\leq k \leq l$. 

To approximate Eq.~\eqref{Phi}, the estimated disturbance dynamics features $\{\bm{\bar{\phi}}_{\circ}^{[j]}\}_{\circ\in\{\bm{\textit{x}}, \bm{\textit{R}}\}, j\in\{1,2,3\}}$ are given by the following neural networks with time-varying estimated weights $\{\bm{\bar{\mathcal{W}}}_{\circ j}\in\mathbb{W}_{\circ j}\}_{\circ\in\{\bm{\textit{x}}, \bm{\textit{R}}\}, j\in\{1,2,3\}}$:
\begin{equation}
{
\begin{aligned}
\bm{\bar{\phi}}_{\circ}^{[j]}:=\bm{\bar{\mathcal{W}}}_{\circ j}^{\top}\bm{\hbar}(\textbf{x}_{\circ j}),\\[-5pt]
\end{aligned}
\label{Phi_hat}
}
\end{equation}
where input  $\textbf{x}_{\bm{\textit{x}} j}:=\bm{\mathcal{E}}_{\bm{\textit{x}} j}:=\left(\bm{e}_{\bm{x}_0}^{[j]}, \bm{\dot{e}}_{\bm{x}_0}^{[j]}\right)^{\top}$ takes  the payload translational error vector along $\bm{\vec{e}}_j$-axis, and input $\textbf{x}_{\bm{\textit{R}} j}:= \left(\bm{e}_{\bm{R}_0}^{[j]}, \bm{e}_{\bm{\Omega}_0}^{[j]}\right)^{\top}$ takes the payload rotational error vector along $\bm{\vec{b}}_{0j}$-axis.  This equation gives the mathematical formulation of each neural network ``slice", as shown in Fig.~\ref{ControFlow_SANM}.

According to the stability analysis in Section 4,  the estimated weights $\bm{\bar{\mathcal{W}}}_{\bm{\textit{x}} j}\in\mathbb{W}_{\bm{\textit{x}} j}$ and $\bm{\bar{\mathcal{W}}}_{\bm{\textit{R}} j}\in\mathbb{W}_{\bm{\textit{R}} j}$ are designed to be updated online by the following Lyapunov adaptation:
\begin{align}
\bm{\dot{\bar{\mathcal{W}}}}_{\bm{\textit{x}} 
j}:=&\gamma_{\bm{\textit{x}}j}\bm{\mathcal{E}}_{\bm{\textit{x}} j}^{\top}\bm{P}_j\bm{B}\bm{\hbar}(\textbf{x}_{\bm{\textit{x}} j}),
\label{Estimated Weights_x}\\
\bm{\dot{\bar{\mathcal{W}}}}_{\bm{\textit{R}} 
 j}:=&\gamma_{\bm{\textit{R}}j} \left(\bm{e}^{[j]}_{\bm{\Omega}_0}+c_R\bm{e}^{[j]}_{\bm{R}_0}\right)\bm{\hbar}(\textbf{x}_{\bm{\textit{R}} j}),
\label{Estimated Weights_R}
\end{align}
where $\gamma_{\bm{\textit{x}}j}\in\mathbb{R}^+$, $\gamma_{\bm{\textit{R}}j}\in\mathbb{R}^+$ and $c_R\in\mathbb{R}^+$ are the corresponding update rate constants. 
The optimal weights that can be identified by these update laws are expressed as:
\begin{equation}
    \bm{\mathcal{W}}_{\circ j}^*\triangleq\mathrm{arg}\, \underset{\bm{\mathcal{W}}_{\circ j}\in\mathbb{W}_{\circ j}}{\mathrm{min}}\left(\mathrm{sup}\big{\vert}\bm{\phi}_{\circ }^{[j]}-\bm{\bar{\phi}}_{\circ }^{[j]}\big{\vert}\right), \label{W*}
\end{equation}
where $\mathrm{arg}\,\mathrm{min}$ denotes the selection of $ \bm{\mathcal{W}}_{\circ j}$ that minimizes the supremum of the error between $\bm{\phi}_{\circ}^{[j]}$ and $\bm{\bar{\phi}}_{\circ}^{[j]}$. 

The optimal approximation error of each neural network ``slice'', defined as the residual between the actual unknown disturbance component $\bm{\phi}_{\circ}^{[j]}$ and its best neural approximation under the optimal weights $\bm{\mathcal{W}}_{\circ j}^*$, is given by:  
\begin{equation}
    {
    \begin{aligned}
\bm{\varpi}^{[j]}_{\circ}\triangleq \bm{\phi}_{\circ}^{[j]}-\bm{\bar{\phi}}_{\circ}^{[j]}(\textbf{x}_{\circ j}\vert\bm{\mathcal{W}}^*_{\circ j}).
    \label{approximation error}
    \end{aligned}
    }
\end{equation}
where $\{\|\bm{\varpi}^{[j]}_{\circ}\|\}_{\circ\in\{\bm{\textit{x}}, \bm{\textit{R}}\}, j\in\{1,2,3\}}$ are bounded according to the universal approximation theorem \cite{1989 Multilayer feedforward networks are universal approximators} and the boundedness of the neural network inputs is guaranteed by Proposition 3.  Here, $\{\bm{\varpi}^{[j]}_{\circ}\in\mathbb{R}\}_{\circ\in\{\bm{\textit{x}}, \bm{\textit{R}}\}, j\in\{1,2,3\}}$ denote the $j^{th}$ components of the optimal approximation error vectors $\{\bm{\varpi}_{\circ}\in\mathbb{R}^{3}\}_{\circ\in\{\bm{\textit{x}}, \bm{\textit{R}}\}}$.

From Eqs.~\eqref{Phi_hat}, \eqref{W*}, \eqref{approximation error}, the problem of the approximation error $\bm{\phi}_{\circ }^{[j]}-\bm{\bar{\phi}}_{\circ }^{[j]}$ can be reformulated in terms of the weight estimation error as:
\begin{equation}
   \bm{\phi}_{\circ}^{[j]}-\bm{\bar{\phi}}_{\circ }^{[j]}=\bm{\tilde{\mathcal{W}}}_{\circ j}^{\top}\bm{\hbar}(\textbf{x}_{\circ j})+\bm{\varpi}^{[j]}_{\circ},
    \label{approximation error to weight error}
\end{equation}
where the weight estimation errors $\{\bm{\tilde{\mathcal{W}}}_{\circ j}\}_{\circ\in\{\bm{\textit{x}}, \bm{\textit{R}}\}, j\in\{1,2,3\}}$ are defined as:
\begin{equation}
    \bm{\tilde{\mathcal{W}}}_{\circ j}\in\mathbb{R}^l\triangleq\bm{\mathcal{W}}^*_{\circ j}-\bm{\bar{\mathcal{W}}}_{\circ j}.
    \label{weight Estimation errors}
\end{equation}

\vspace{1em} 
\noindent\textbf{3.5 Tension Planner :} $\bm{\mathrm{\mu}}_{i_{\bm{d}}}$

This subsection develops a tension planning module that computes the desired internal tension $\bm{\mathrm{\mu}}_{i_{\bm{d}}}\in\mathbb{R}^3$ along the $i^{th}$ cable. From a mechanical perspective, the set of all desired internal tensions $\{\bm{\mathrm{\mu}}_{i_{\bm{d}}}\}_{i\in[1, n]}$ must collectively generate the desired resultant force $\bm{\mathrm{F}_d}$ and moment $\bm{\mathrm{M}_d}$. This requirement leads to the following equilibrium constraints:
\begin{equation}
\bm{\sum}_{i=1}^n\bm{\mathrm{\mu}}_{i_{\bm{d}}}=\bm{\mathrm{F}_d},
    \label{tension equation1}
\end{equation}
\begin{equation}
\bm{\sum}_{i=1}^n[\bm{\rho}_i]_{\times}\bm{R}_0^{\top}\bm{\mathrm{\mu}}_{i_{\bm{d}}}=\bm{\mathrm{M}_d}.
    \label{tension equation2}
\end{equation}
To obtain a feasible solution that satisfies the above constraints, a minimum-norm formulation is adopted as follows: 
\begin{equation}
    {
    \begin{aligned}
 &{\renewcommand{\arraystretch}{0.1}\begin{bmatrix}\bm{\mathrm{\mu}}_{1_{\bm{d}}} \\\cdot\\\cdot\\ \bm{\mathrm{\mu}}_{n_{\bm{d}}}\\\end{bmatrix}}:=\bm{\mathrm{diag}}{\renewcommand{\arraycolsep}{0.4pt}\begin{bmatrix}
         \bm{R_0} &\cdot&\cdot&\bm{R_0}
     \end{bmatrix}}{\renewcommand{\arraycolsep}{0.4pt}\begin{bmatrix}
\mathbf{I}^{3\times3}&\cdot&\cdot&\mathbf{I}^{3\times3}\\
        [\bm{\rho}_{1}]_{\times} &\cdot&\cdot& [\bm{\rho}_{n}]_{\times} 
    \end{bmatrix}}^\dagger\begin{bmatrix} \bm{R}_0^{\top}\bm{\mathrm{F}_d} \\ \bm{\mathrm{M}_d}\end{bmatrix},
\end{aligned}
    }
\end{equation}
where $\bm{\mathrm{diag}}[\bullet\bullet\bullet]$ represents a block-diagonal matrix constructed from the input blocks. The $\mathbf{I}^{3\times3}$ is the $3\times3$ identity matrix. The superscript $\bullet^\dagger$ denotes the pseudoinverse of a matrix.

\vspace{1em} 
\noindent\textbf{3.6 Cable Control :} $\{\mathbf{u}_i^{\parallel}, \mathbf{u}_i^{\bot}\}$

This subsection introduces the design of the control for the $i^{th}$ cable.  In a centralized multi-quadrotor transportation system, each cable is indirectly actuated through its connection to a quadrotor. In other words, the driving force of the cable originates from the tension generated by the motion of the quadrotor.  Specifically, the quadrotor manipulates the cable by exerting forces that translate into tension along the direction of the cable. Therefore, cable control can be decomposed into two orthogonal components, both applied at the center of mass of the quadrotor: the parallel component $\mathbf{u}_i^{\parallel}\in\mathbb{R}^3$, which aligns with the direction of the cable and somewhat determines the magnitude of tension, and the normal component $\mathbf{u}_i^{\bot}\in\mathbb{R}^3$, which governs the orientation of the cable by inducing changes in its direction.  

The parallel component $\mathbf{u}_i^{\parallel}$ has already been given through the dynamically coupled constraint in Eq.~\eqref{parallel component}. In contrast, the normal component $\mathbf{u}_i^{\bot}$ plays a key role in regulating the orientation of the cable.   Hence, a normal controller is designed as follows:
\begin{equation}
\begin{aligned}
    \mathbf{u}_i^{\bot}:=&m_i l_i[\bm{q}_{i}]_{\times}\{-k_{q}e_{\bm{q}_{i}}-k_{\omega}e_{\bm{\omega}_i}-(\bm{q}_{i}\cdot\bm{\omega}_{i_{\bm{d}}})\bm{\dot{q}}_i\\
    &-[\bm{q}_{i}]^2_{\times}\dot{\bm{\omega}}_{i_{\bm{d}}}\}-m_i[ \bm{q}_{i}]^2_{\times}\bm{a}_i-\bar{\Delta}_{\bm{x}_i}^{\bot},
    \end{aligned}
\label{ui_normal}
\end{equation}
where $k_{q}\in\mathbb{R}^+$, $k_{\omega}\in\mathbb{R}^+$ are positive constants. The $\bar{\Delta}_{\bm{x}_i}^{\bot}\in\mathbb{R}^3:=-[ \bm{q}_{i}]^2_{\times}\bar{\Delta}_{\bm{x}_i}$ denotes the normal component of the estimated disturbance of the $i^{th}$ quadrotor from the integral compensation given in Eq.~\eqref{integral compensations}. The desired angular velocity of the $i^{th}$ cable is represented by $\bm{\omega}_{i_{\bm{d}}}\in\mathbb{R}^3$:

\begin{equation}
\begin{aligned}
\bm{\omega}_{i_{\bm{d}}}:=\bm{q}_{i_{\bm{d}}}\times\bm{\dot{q}}_{i_{\bm{d}}},
    \end{aligned}
\label{omega_id}
\end{equation}
in which $\bm{q}_{i_{\bm{d}}} \in\mathbf{S}^2$ is defined as the desired direction of the $i^{th}$ cable, given by:
\begin{equation}
\begin{aligned}
\bm{q}_{i_{\bm{d}}}:=-\frac{\bm{\mathrm{\mu}}_{i_{\bm{d}}}}{\lVert\bm{\mathrm{\mu}}_{i_{\bm{d}}}\lVert}.
    \end{aligned}
\end{equation}
The $\bm{e}_{\bm{q}_{i}}\in\mathbb{R}^3$ and $\bm{e}_{\bm{\omega}_i}\in\mathbb{R}^3$ are defined as the direction and angular velocity tracking errors of the $i^{th}$ cable, respectively, and are given by:
\begin{equation}
\begin{aligned}
\bm{e}_{\bm{q}_{i}}:=\bm{q}_{i_{\bm{d}}}\times\bm{q}_{i}, \,\,\bm{e}_{\bm{\omega}_i}:=\bm{\omega}_{i}+[ \bm{q}_{i}]^2_{\times}\bm{\omega}_{i_{\bm{d}}}.
\end{aligned}
\label{e_qi and e_omegai}
\end{equation}

\vspace{1em} 
\noindent\textbf{3.7 Quadrotor Control :} $\{f_i,\mathbf{M}_i\}$

This subsection presents the geometric control for the quadrotors.
In Eq. (11), the control force at the  $i^{th}$ quadrotor $\mathbf{u}_i$ is given as the sum of the parallel (parallel to the cable) component $\mathbf{u}_i^{\parallel}$, and the normal (perpendicular to the cable) component  $\mathbf{u}_i^{\bot}$.  
For the underactuated nature of quadrotors, $\mathbf{u}_i$ is physically generated by the thrust force acting along the quadrotor body-fixed $\bm{\vec{b}}_{i3}$ axis. Therefore, the quadrotor body-fixed $\bm{\vec{b}}_{i3}$ axis is  aligned with the direction of $\mathbf{u}_i$:
\begin{equation}
\begin{aligned}
\bm{\vec{b}}_{i3} :=\frac{\mathbf{u}_i}{\lVert\mathbf{u}_i\lVert} \in \mathbf{S}^{2}.
    \end{aligned}
\end{equation}
This alignment imposes a two-degree-of-freedom constraint on the three-degree-of-freedom attitude, leaving one remaining degree of freedom to be determined. A desired facing direction, denoted as, $\bm{\vec{b}}_{i1\bm{d}}(t)\in \mathbf{S}^{2}$, is specified for the first body-fixed axis to determine the remaining rotational degree of freedom. Since $\bm{\vec{b}}_{i1\bm{d}}(t)$ is required to remain orthogonal to $\bm{\vec{b}}_{i3}$, it is projected onto the plane normal to $\bm{\vec{b}}_{i3}$ to satisfy the orthogonality constraint. Consequently, the desired attitude of the $i^{th}$ quadrotor is constructed and computed by:
\begin{equation}
\begin{aligned}
 \bm{R}_{{i_{\bm{c}}}}:={\renewcommand{\arraystretch}{1}\begin{bmatrix}-\frac{[\bm{\vec{b}}_{i3}]^2_{\times}\bm{\vec{b}}_{i1\bm{d}}}{\lVert[\bm{\vec{b}}_{i3}]^2_{\times}\bm{\vec{b}}_{i1\bm{d}}\lVert},&\frac{[\bm{\vec{b}}_{i3}]_{\times}\bm{\vec{b}}_{i1\bm{d}}}{\lVert[\bm{\vec{b}}_{i3}]_{\times}\bm{\vec{b}}_{i1\bm{d}}\lVert},&\bm{\vec{b}}_{i3}\end{bmatrix}}\in\mathbf{SO}(3).
    \end{aligned}
\end{equation}
The desired angular velocity of the $i^{th}$ quadrotor $\bm{\Omega}_{i_{\bm{c}}}\in\mathbb{R}^3$ is given by:
\begin{equation}
   \bm{\Omega}_{i_{\bm{c}}}:=(\bm{R}^{\top}_{i_{\bm{c}}}\dot{\bm{R}}_{i_{\bm{c}}})^{\vee}.
    \label{Omega_ic}
\end{equation}
Next, following the standard formulation in geometric tracking control on $\mathbf{SO}(3)$ \cite{2011 Geometric tracking control of the attitude dynamics of a rigid body on SO(3)}, the tracking errors for the attitude and angular velocity of each quadrotor are defined as, respectively:
\begin{equation}
\begin{aligned}
     \bm{e}_{\bm{R}_i}&:=\frac{1}{2}(\bm{R}^{\top}_{i_{\bm{c}}}\bm{R}_{i}-\bm{R}^{\top}_{i}\bm{R}_{i_{\bm{c}}})^\vee,\\ \bm{e}_{\bm{\Omega}_i}&:=\bm{\Omega}_{i}-\bm{R}^{\top}_{i}\bm{R}_{i_{\bm{c}}}\bm{\Omega}_{i_{\bm{c}}}.
\end{aligned}
\end{equation}
For the $i^{th}$ quadrotor, the corresponding thrust $f_i\in\mathbb{R}$ and moment $\mathbf{M}_i\in\mathbb{R}^3$ are designed as follows, respectively:
\begin{equation}
\begin{aligned}
f_i=\mathbf{u}_i\cdot \bm{R}_{i}\bm{\vec{e}}_3,
\end{aligned}
\end{equation}
\begin{equation}
\begin{aligned}
\mathbf{M}_i=&-k_{R_i}\bm{e}_{\bm{R}_i}-k_{\Omega_i}\bm{e}_{\bm{\Omega}_i}+\bm{\Omega}_{i}\times\bm{J}_i\bm{\Omega}_{i}\\
    &-\bm{J}_i([ \bm{\Omega}_{i}]_{\times}\bm{R}^{\top}_{i}\bm{R}_{i_{\bm{c}}}\bm{\Omega}_{i_{\bm{c}}}-\bm{R}^{\top}_{i}\bm{R}_{i_{\bm{c}}}\bm{\dot{\Omega}}_{i_{\bm{c}}}),
\end{aligned}
\end{equation}
where $k_{R_i}\in\mathbb{R}^+$ and $k_{\Omega_i}\in\mathbb{R}^+$ are positive constants. 

\vspace{1em} 
\noindent\textbf{\textit{As noted in}}\cite{2011 Geometric tracking control of the attitude dynamics of a rigid body on SO(3)}: Under the aforementioned geometric tracking control, the zero equilibrium of the
tracking error $\bm{e}_{\bm{R}_i}$, $\bm{e}_{\bm{\Omega}_i}$ is exponentially stable.

\vspace{1em} 
\noindent\textbf{4. Stability Analysis} 

For the cable-suspended payload system, consider the following semi-global open domain:
\begin{equation}
{
   \begin{aligned}
    \mathcal{D}=&\Big{\{}\Big{(}\bm{e}_{\bm{x}_0}, \bm{e}_{\bm{v}_0}, \bm{e}_{\bm{R}_0}, \bm{e}_{\bm{\Omega}_0}, \bm{e}_{\bm{q}_{i}}, \bm{e}_{\bm{\omega}_i},\\
    &\,\,\,\,
    \widetilde{\Delta}_{\bm{x}_0},\widetilde{\Delta}_{\bm{R}_0}, \widetilde{\Delta}_{\bm{x}_i},(\tilde{m}_j,\widetilde{J}_{j},\bm{\tilde{\mathcal{W}}}_{\bm{\textit{x}} j},\bm{\tilde{\mathcal{W}}}_{\bm{\textit{R}} j})_{j\in\{1,2,3\}}\Big{)}\\
    &\,\,\,\,\in(\mathbb{R}^3)^4\!\times\!(\mathbb{R}^3\!\times\!\mathbb{R}^3)^n\!\times\!(\mathbb{R}^3)^2\!\times\!(\mathbb{R}^3)^n\times\!\\
    &\,\,\,\,\,\,\prod^{3}_{j=1}(\mathbb{R}\!\times\!\mathbb{R}\!\times\!\mathbb{R}^{l}\!\times\!\mathbb{R}^{l})\big{|} \lVert\bm{e}_{\bm{x}_0}\lVert<e_{x_{\text{max}}},\\
    &\,\,\,\,\lVert\bm{e}_{\bm{v}_0}\lVert<e_{v_{\text{max}}}, 0<\Psi_{\textit{R}_0}<\psi_{\textit{R}_0}<1,\\
    &\,\,\,\,\,\,0<\Psi_{\textit{q}_i}<\psi_{\textit{q}_i}<1, \lVert\widetilde{\Delta}_{\bm{x}_0}\lVert<\varepsilon_{\Delta}, \lVert\widetilde{\Delta}_{\bm{R}_0}\lVert<\varepsilon_{\Delta},\\
    &\,\,\,\,\lVert\widetilde{\Delta}_{\bm{x}_i}\lVert<\varepsilon_{\Delta},  \lVert\tilde{m}_j\lVert<\varepsilon_{m}, \lVert\widetilde{J}_{j}\lVert<\varepsilon_{J}, \lVert\bm{\tilde{\mathcal{W}}}_{\bm{\textit{x}} j}\lVert<\varepsilon_{\mathcal{W}}  \\
    &\,\,\,\,\lVert\bm{\tilde{\mathcal{W}}}_{\bm{\textit{R}} j}\lVert<\varepsilon_{\mathcal{W}}. \Big{\}}, 
\end{aligned} 
}
\label{D}
\end{equation}
where the state errors in the first line have been defined in Eqs.~\eqref{e_x0}, \eqref{e_v0}, \eqref{e_R0}, \eqref{e_Omega0}, \eqref{e_qi and e_omegai}. The estimation errors in the second line have been defined in Eqs.~\eqref{Estimation errors of integral terms1}, \eqref{Estimation errors of integral terms2}, \eqref{Estimation errors of integral terms3}, \eqref{Estimation errors of mass and inertia features}, \eqref{weight Estimation errors}. The $e_{x_{\text{max}}}$, $e_{v_{\text{max}}}$, $\psi_{\textit{R}_0}$, $\psi_{\textit{q}_i}$, $\varepsilon_{\Delta}$, $\varepsilon_{m}$, $\varepsilon_{J}$ and $\varepsilon_{\mathcal{W}}\in\mathbb{R}^+$ are positive constants. The $\Psi_{\textit{R}_0}\!:\!\mathbf{SO}(3)\times\mathbf{SO}(3)\!\to\!\mathbb{R}$ denotes an attitude configuration error scalar function for the payload:
\begin{equation}
    \Psi_{\textit{R}_0}(\bm{R}_0,\bm{R}_{0_{\bm{d}}})\triangleq\frac{1}{2}\mathrm{tr}\Big{[}\mathrm{I}^{3\times3}-\bm{R}_{0_{\bm{d}}}^{\top}\bm{R}_0\Big{]}.
    \label{Psi_R0}
\end{equation}
The $\Psi_{\textit{q}_i}\!:\!\mathbf{S}^2 \times\mathbf{S}^2\!\to\!\mathbb{R}$ denotes a configuration error scalar function for the $i^{th}$ cable:
\begin{equation}
   \Psi_{\textit{q}_i}(\bm{q}_{i},\bm{q}_{i_{\bm{d}}})\triangleq1-\bm{q}_{i}\cdot\bm{q}_{i_{\bm{d}}}.
    \label{Psi_qi}
\end{equation}
 In this domain, $\|\bm{e}_{\bm{R}_0}\|$ is also bounded by $\|\bm{e}_{\bm{R}_0}\|=\sqrt{\Psi_{\textit{R}_0}(2-\Psi_{\textit{R}_0})}\leq\sqrt{\psi_{\textit{R}_0}(2-\psi_{\textit{R}_0})}<1$ with a positive scalar $0<\psi_{\textit{R}_0}<1$. Similarly, $\|\bm{e}_{\bm{q}_{i}}\|$ is bounded by $\|\bm{e}_{\bm{q}_{i}}\|=\sqrt{\Psi_{\textit{q}_i}(2-\Psi_{\textit{q}_i})}\leq\sqrt{\psi_{\textit{q}_i}(2-\psi_{\textit{q}_i})}<1$ with a positive scalar $0<\psi_{\textit{q}_i}<1$. Based upon these, the configuration error functions admit the following quadratic bounds with respect to their corresponding error vectors:
\begin{equation}
    \frac{1}{2}\|\bm{e}_{\bm{R}_0}\|^{2}\leq\Psi_{\textit{R}_0}\leq\frac{1}{2-\psi_{\textit{R}_0}}\|\bm{e}_{\bm{R}_0}\|^{2},
    \label{Psi_R0_bound}
\end{equation}
\begin{equation}
    \frac{1}{2}\|\bm{e}_{\bm{q}_{i}}\|^{2}\leq\Psi_{\textit{q}_i}\leq\frac{1}{2-\psi_{\textit{q}_i}}\|\bm{e}_{\bm{q}_{i}}\|^{2}.
    \label{Psi_qi_bound}
\end{equation}
Within the aforementioned domain $\mathcal{D}$, the following results are presented:

\noindent\textbf{\textit{Proposition 1: (Asymptotic Stability under Known Model Parameters and Constant Disturbances)}} When the mass and inertia tensor of the payload are exactly known and the payload is subject to constant external disturbances, the SANM module can be disabled since online identification is redundant. Under these conditions, the state solution of payload error dynamics $\bm{z}_{0}(t)=\left(\|\bm{e}_{\bm{x}_0}\|,\|\bm{e}_{\bm{v}_0}\|,\|\bm{e}_{\bm{R}_0}\|,\|\bm{e}_{\bm{\Omega}_0}\|\right)^{\top}\!\!\in\mathbb{R}^{4}$
and the state solution of cable direction error dynamics $\bm{z}_{\bm{q}_i}(t)=\left(\lVert\bm{e}_{\bm{q}_i}\lVert, \lVert\bm{e}_{\bm{\omega}_i}\lVert\right)^{\top}\!\!\in\mathbb{R}^{2}$ asymptotically converge to zero. Furthermore, all estimation errors $\lVert\widetilde{\Delta}_{\bm{x}_0}\lVert$, $ \lVert\widetilde{\Delta}_{\bm{R}_0}\lVert$, $ \lVert\widetilde{\Delta}_{\bm{x}_i}\lVert$ are uniformly bounded.

\noindent\textbf{\textit{Proof:}} See \cite{2018 Geometric Control of Quadrotor UAVs Transporting a Cable-Suspended Rigid Body}.

\noindent\textbf{\textit{Proposition 2: (Practical Stability under Unknown Model Parameters and Time-Varying Disturbances)}} When the mass and inertia tensor of the payload are unknown and the payload is subject to time-varying external disturbances, the SANM module can compensate for these effects. With the SANM module activated, the state solution of the payload error dynamics $\bm{z}_{0}(t)=\left(\|\bm{e}_{\bm{x}_0}\|,\|\bm{e}_{\bm{v}_0}\|,\|\bm{e}_{\bm{R}_0}\|,\|\bm{e}_{\bm{\Omega}_0}\|\right)^{\top}\!\!\in\mathbb{R}^{4}$, and the state solution of the cable direction error dynamics $\bm{z}_{\bm{q}_i}(t)=\left(\lVert\bm{e}_{\bm{q}_i}\lVert, \lVert\bm{e}_{\bm{\omega}_i}\lVert\right)^{\top}\!\!\in\mathbb{R}^{2}$ are uniformly ultimately bounded. Furthermore, all estimation errors $\lVert\widetilde{\Delta}_{\bm{x}_0}\lVert$, $ \lVert\widetilde{\Delta}_{\bm{R}_0}\lVert$, $ \lVert\widetilde{\Delta}_{\bm{x}_i}\lVert$, $ \lVert\tilde{m}_j\lVert$, $\lVert\widetilde{J}_{j}\lVert$, $\lVert\bm{\tilde{\mathcal{W}}}_{\bm{\textit{x}} j}\lVert$, $\lVert\bm{\tilde{\mathcal{W}}}_{\bm{\textit{R}} j}\lVert$  are uniformly bounded.

\noindent\textbf{\textit{Proof:}} See Sections 4.1 and 4.2.

\noindent\textbf{\textit{Proposition 3: (Compact Set Constraint on Neural Network Inputs)}} The full-state error vector of the payload $\mathbf{E}$ is bounded by a compact set: $ \bm{\mathcal{C}}\!=\!\Big{\{}\mathbf{E}\in\mathbb{R}^{12}| \,\|\mathbf{E}\|\leq\!\|\bm{e}_{\bm{x}_0}\|+\|\bm{e}_{\bm{v}_0}\|+\|\bm{e}_{\bm{R}_0}\|+\|\bm{e}_{\bm{\Omega}_0}\|\leq r_{c}\Big{\}}$
for a positive constant $r_{c}$. This implies that all inputs of the neural networks$\{\textbf{x}_{\circ j}\in \mathbb{R}^2\}_{\circ\in\{\bm{\textit{x}}, \bm{\textit{R}}\}, j\in\{1,2,3\}}$ are also bounded within their respective compact sets. 
This compactness condition ensures the prerequisite of the universal approximation theorem (UAT) \cite{1989 Multilayer feedforward networks are universal approximators}.

\noindent\textbf{\textit{Proof:}} See Sections 4.1 and 4.2.

\vspace{1em} 
\noindent\textbf{4.1 Error Dynamics} 

To prove \textbf{\textit{Propositions 2 and 3}}, we first derive the error dynamics of the cable-suspended payload system.  

\textit{(1) Payload Translational Error Dynamics:}
Combining Eqs.~\eqref{Payload Translational Dynamics}, \eqref{Translational Adaptive-Neuro Control}, \eqref{Payload Translational Control} yields the payload translational error dynamics along the $\bm{\vec{e}}_j$-axis:
 \begin{equation}
 {
\begin{aligned}
      \bm{\dot{\mathcal{E}}}_{\bm{\textit{x}} j}={\renewcommand{\arraystretch}{1.5}\begin{bmatrix}\bm{\dot{e}}_{\bm{x}_0}^{[j]}\\ \bm{\ddot{e}}_{\bm{x}_0}^{[j]}\\\end{bmatrix}}=&\bm{\Lambda}_{\bm{\textit{x}} j}\bm{\mathcal{E}}_{\bm{\textit{x}} j}+\bm{B}\bigg{\{} \widetilde{m}_{j}\bm{\mathcal{U}_{\bm{\textit{x}}}}^{[j]}+\left(\bm{\phi}_{\bm{\textit{x}}}^{[j]}-\bm{\bar{\phi}}_{\bm{\textit{x}}}^{[j]}\right)
      \\&+\frac{1}{m_0}\left(\widetilde{\Delta}_{\bm{x}_0}^{[j]}+\bm{\sum}_{i=1}^n\widetilde{\Delta}_{\bm{x}_i}^{\parallel[j]}\right)+\bm{Y_{\bm{\textit{x}}}}^{[j]}\bigg{\}},
\end{aligned}
}
\label{translational error dynamics1}
\end{equation}
 where $\widetilde{m}_j$, $\widetilde{\Delta}_{\bm{x}_0}$ and $\widetilde{\Delta}_{\bm{x}_i}$ are estimation errors defined in Eqs.~\eqref{Estimation errors of mass and inertia features}, \eqref{Estimation errors of integral terms1} and \eqref{Estimation errors of integral terms3}, respectively. The $\widetilde{\Delta}_{\bm{x}_i}^{\parallel}\in\mathbb{R}^3\triangleq\Delta_{\bm{x}_i}^{\parallel}-\bar{\Delta}_{\bm{x}_i}^{\parallel}$ denotes the parallel component of $\widetilde{\Delta}_{\bm{x}_i}$. The vector $\bm{B}=\left(0,1\right)^{\top}$. The PD gain matrix $\bm{\Lambda}_{\bm{\textit{x}} j}\in\mathbb{R}^{2\times2}$ is given by:
\begin{equation}
{
\begin{aligned}
     \bm{\Lambda}_{\bm{\textit{x}} j}={\renewcommand{\arraystretch}{1}\begin{bmatrix}0&1\\-\bm{k_{\text{p}}}^{[j]}&-\bm{k_{\text{d}}}^{[j]}\end{bmatrix}}.
\end{aligned}
    }
    \label{Lambda_xj}
\end{equation}
By substituting Eqs.~ \eqref{Phi_hat},~\eqref{W*} and~\eqref{approximation error} with subscripts $\bullet_{\circ:=\bm{\textit{x}}}$,  Eq.~\eqref{translational error dynamics1} can be further expressed as:
\begin{equation}
{
\begin{aligned}
     \bm{\dot{\mathcal{E}}}_{\bm{\textit{x}} j}=&\bm{\Lambda}_{\bm{\textit{x}} j}\bm{\mathcal{E}}_{\bm{\textit{x}} j}+\bm{B}\bigg{\{} \widetilde{m}_j\bm{\mathcal{U}_{\bm{\textit{x}}}}^{[j]}+\left(\bm{\mathcal{W}}^*_{\bm{\textit{x}} j}-\bm{\bar{\mathcal{W}}}_{\bm{\textit{x}} j}\right)^{\top}\bm{\hbar}(\textbf{x}_{\bm{\textit{x}} j})+\bm{\varpi}^{[j]}_{\bm{\textit{x}}}\\
     &+\frac{1}{m_0}\left(\widetilde{\Delta}_{\bm{x}_0}^{[j]}+\bm{\sum}_{i=1}^n\widetilde{\Delta}_{\bm{x}_{i}}^{\parallel[j]}\right)+\bm{Y_{\bm{\textit{x}}}}^{[j]}\bigg{\}}.
\end{aligned}
    }\label{translational error dynamics2}
\end{equation}

\textit{(2) Payload Rotational Error Dynamics:} As noted in \cite{2010 Geometric tracking control of a quadrotor UAV on SE(3)} and \cite{2011 Geometric tracking control of the attitude dynamics of a rigid body on SO(3)}, the payload rotational error dynamics is given by:
\begin{equation}
{
\begin{aligned}
    \bm{\dot{e}}_{\bm{R}_0}&=\frac{1}{2}\left(\bm{R}_{0_{\bm{d}}}^{\top}\bm{R}_0[\bm{e}_{\bm{\Omega}_0}]_{\times}+[\bm{e}_{\bm{\Omega}_0}]_{\times}\bm{R}_0^{\top}\bm{R}_{0_{\bm{d}}}\right)^{\vee}\\
&=\frac{1}{2}\left(\mathrm{tr}\big{[}\bm{R}_{0}^{\top}\bm{R}_{0_{\bm{d}}}\big{]}\mathrm{I}^{3\times3}-\bm{R}_{0}^{\top}\bm{R}_{0_{\bm{d}}}\right)\bm{e}_{\bm{\Omega}_0},\\
&\equiv H(\bm{R}_{0_{\bm{d}}}^{\top}\bm{R}_0)\bm{e}_{\bm{\Omega}_0},\\
\end{aligned}
    }\label{rotational error dynamics0}
\end{equation}
\begin{equation}
{
\begin{aligned}
  \bm{\dot{e}}_{\bm{\Omega}_0}=\bm{\dot{\Omega}}_0+[\bm{\Omega}_0]_{\times}\bm{R}_0^{\top}\bm{R}_{0_{\bm{d}}}\bm{\Omega}_{0_{\bm{d}}}-\bm{R}_0^{\top}\bm{R}_{0_{\bm{d}}}\bm{\dot{\Omega}}_{0_{\bm{d}}},
\end{aligned}
    }\label{rotational error dynamics1}
\end{equation}
where $\|H(\bm{R}_{0_{\bm{d}}}^{\top}\bm{R}_0)\|\leq1$ for any $\bm{R}_{0_{\bm{d}}}^{\top}\bm{R}_0\in\mathbf{SO}(3)$.

By substituting Eqs.~\eqref{Payload Rotational Dynamics}, \eqref{Rotational Adaptive-Neuro Control}, \eqref{Payload Rotational Control} and Eqs.~\eqref{W*}, \eqref{approximation error} with subscripts $\bullet_{\circ:=\bm{\textit{R}}}$, the Eq.~\eqref{rotational error dynamics1} can be further derived in the form of the $j^{th}$ element along the $\bm{\vec{b}}_{0j}$-axis as:
\begin{equation}
{\small
\begin{aligned}
 \bm{\dot{e}}^{[j]}_{\bm{\Omega}_0}\!=\!&-k_{R_0}\bm{e}^{[j]}_{\bm{R}_0}\!-\!k_{\Omega_0}\bm{e}^{[j]}_{\bm{\Omega}_0}\!+\!\widetilde{J}_{j}\bm{\mathcal{U}_{\bm{\textit{R}}}}^{[j]}\!+\!\left(\bm{\mathcal{W}}^*_{\bm{\textit{R}} j}\!-\!\bm{\bar{\mathcal{W}}}_{\bm{\textit{R}} j}^{\top}\right)\bm{\hbar}(\textbf{x}_{\bm{\textit{R}} j})\\
&+\bm{\varpi}^{[j]}_{\bm{\textit{R}}}+\bm{J}_0^{-1[j]}\bigg{\{}\widetilde{\Delta}^{[j]}_{\bm{R}_0}+\bm{\sum}_{i=1}^n\left([\bm{\rho}_i]_{\times}\bm{R}_0^{\top}\widetilde{\Delta}_{\bm{x}_i}^{\parallel}\right)^{[j]}\bigg{\}}+\bm{Y}^{[j]}_{\textit{R}},
\end{aligned}
    }\label{rotational error dynamics2}
\end{equation}
where $\widetilde{J}_{j}$ and $\widetilde{\Delta}_{\bm{R}_0}$ are estimation errors defined in Eqs.~\eqref{Estimation errors of mass and inertia features} and \eqref{Estimation errors of integral terms2}.

\textit{(3) Cable Direction Error Dynamics:} From Eqs.~\eqref{Cable Angular Velocity Dynamics}, \eqref{ui_normal} and \eqref{e_qi and e_omegai}, the cable direction error dynamics is given by:
\begin{equation}
\begin{aligned}
{
    -[\bm{q}_i]_{\times}^2\bm{\dot{e}}_{\bm{\omega}_i}=-k_{q}\bm{e}_{\bm{q}_i}-k_{\omega}\bm{e}_{\bm{\omega}_i}-\frac{1}{m_{i}l_{i}}[\bm{q}_i]_{\times}\widetilde{\Delta}_{\bm{x}_{i}}^{\bot}.
}
\end{aligned}
\end{equation}

\vspace{1em} 
\noindent\textbf{4.2 Proof of Propositions 2 and 3} 

Define the Lyapunov candidate function for complete cable-suspended payload dynamics as $\bm{\mathcal{V}}=\bm{\mathcal{V}}_{\bm{\textit{x}}}+\bm{\mathcal{V}}_{\bm{\textit{R}}}+\bm{\mathcal{V}}_{\bm{q}}+\bm{\mathcal{V}}_{\Delta}$ with the following  Lyapunov candidate terms:
\begin{equation}
{
    \begin{aligned}
        \bm{\mathcal{V}}_{\bm{\textit{x}}}=&\bm{\sum}_{j=1}^3\frac{1}{2}\bm{\mathcal{E}}_{\bm{\textit{x}} j}^{\top}\bm{P}_j\bm{\mathcal{E}}_{\bm{\textit{x}} j}+\frac{1}{2}\eta_{\textit{m}_j}\widetilde{m}_j^2 + \frac{1}{2\gamma_{\bm{\textit{x}}j}}\bm{\tilde{\mathcal{W}}}_{\bm{\textit{x}} j}^{\top}\bm{\tilde{\mathcal{W}}}_{\bm{\textit{x}} j},
    \end{aligned}
    }
\end{equation}
\begin{equation}
{
    \begin{aligned}
\bm{\mathcal{V}}_{\bm{\textit{R}}}=&k_{R_0}\Psi_{\textit{R}_0}+\bm{\sum}_{j=1}^3\frac{1}{2}\lVert\bm{e}^{[j]}_{\bm{\Omega}_0}\lVert^2+c_R\bm{e}^{[j]}_{\bm{R}}\bm{e}^{[j]}_{\bm{\Omega}}\\
&+\frac{1}{2}\eta_{J_j}\widetilde{J}_j^2+\frac{1}{2\gamma_{\bm{\textit{R}}j}}\bm{\tilde{\mathcal{W}}}_{\bm{\textit{R}} j}^{\top}\bm{\tilde{\mathcal{W}}}_{\bm{\textit{R}} j},
\end{aligned}
    }
\end{equation}

\begin{equation}
{
    \begin{aligned}
\bm{\mathcal{V}}_{\bm{q}}=\bm{\sum}_{i=1}^{n}\frac{1}{2}\lVert\bm{e}_{\bm{\omega}_i}\lVert^2+k_{q}\Psi_{\textit{q}_i}+\textit{c}_{q}\bm{e}_{\bm{q}_i}\cdot\bm{e}_{\bm{\omega}_i},
\end{aligned}
    }
\end{equation}
\begin{equation}
{
    \begin{aligned}
    \bm{\mathcal{V}}_{\Delta}=\frac{1}{2\textit{h}_{x_0}}\lVert\widetilde{\Delta}_{\bm{x}_0}\lVert^2+\frac{1}{2\textit{h}_{R_0}}\lVert\widetilde{\Delta}_{\bm{R}_0}\lVert^2+\bm{\sum}_{i=1}^{n}\frac{1}{2\textit{h}_{x_i}}\lVert\widetilde{\Delta}_{\bm{x}_i}\lVert^2,
\end{aligned}
}
\end{equation}
where $\bm{\mathcal{V}}_{\bm{\textit{x}}}$, $\bm{\mathcal{V}}_{\bm{\textit{R}}}$, $\bm{\mathcal{V}}_{\bm{q}}$ and $\bm{\mathcal{V}}_{\Delta}$ denote the Lyapunov candidate functions for payload translational error dynamics, payload rotational error dynamics, cable direction error dynamics and integral estimation error dynamics, respectively. The $\bm{\tilde{\mathcal{W}}}_{\bm{\textit{x}} j}$ and $\bm{\tilde{\mathcal{W}}}_{\bm{\textit{R}} j}$ are weight estimation errors defined in Eq.~\eqref{weight Estimation errors}.  The $\eta_{\textit{m}_j}$, $\eta_{J_j}$, $\gamma_{\bm{\textit{x}}j}$ and $\gamma_{\bm{\textit{R}}j}\in\mathbb{R}^+$  are positive constants that determine the update rates of the adaptive law and neural network ``slices". The $\textit{h}_{x_0}$, $\textit{h}_{R_0}$,  $\textit{h}_{x_i}\in\mathbb{R}^{+}$ are positive constants that determine the integral compensation gains.  Here, if the positive constants $\textit{c}_{R}\in\mathbb{R}^+$ and $\textit{c}_{q}\in\mathbb{R}^+$ are sufficiently small, $\bm{\mathcal{V}}_{\bm{\textit{R}}}$ and $\bm{\mathcal{V}}_{\bm{q}}$ are  positive-definite. Since $\bm{\mathcal{V}}_{\textit{x}}\geq0$ and $\bm{\mathcal{V}}_{\Delta}\geq0$,  it follows that $\bm{\mathcal{V}}$ is also positive-definite. The Lyapunov matrices $\{\bm{P}_j\in\mathbb{R}^{2\times2}\}_{j\in\{1,2,3\}}$ are symmetric positive-definite  that follows the following Lyapunov equations with PD gain matrices $\bm{\Lambda}_{\bm{\textit{x}} j}$ given in Eq.~\eqref{Lambda_xj} and matrix $\bm{Q}_j>0$ for $j\in\{1,2,3\}$:
\begin{equation}
\begin{aligned}
{
\bm{\Lambda}_{\textit{x}j}^{\top}\bm{P}_j+\bm{P}_j\bm{\Lambda}_{\textit{x}j}=-\bm{Q}_j.
}
\end{aligned}
\label{Lyapunov equations}
\end{equation}

With the fact that $\dot{\Psi}_{\textit{R}_0}=\bm{e}_{\bm{R}_0}\cdot\bm{e}_{\bm{\Omega}_0}$ \cite{2010 Geometric tracking control of a quadrotor UAV on SE(3)}, the time-derivative of the complete Lyapunov function can be then given by $\bm{\dot{\mathcal{V}}}=\bm{\dot{\mathcal{V}}}_{\bm{\textit{x}}}+\bm{\dot{\mathcal{V}}}_{\bm{\textit{R}}}+\bm{\dot{\mathcal{V}}}_{\bm{q}}+\bm{\dot{\mathcal{V}}}_{\Delta}$ with:
\begin{equation}
    {\footnotesize
    \begin{aligned}
\bm{\dot{\mathcal{V}}}_{\bm{\textit{x}}}=&\bm{\sum}_{j=1}^3-\frac{1}{2}\bm{\mathcal{E}}_{\bm{\textit{x}} j}^{\top}\bm{Q}_j\bm{\mathcal{E}}_{\bm{\textit{x}} j}
      +\widetilde{m}_j\left(\bm{\mathcal{E}}_{\bm{\textit{x}} j}^{\top}\bm{P}_j\bm{B} \bm{\mathcal{U}}_{\textit{x}}^{[j]}+\eta_{\textit{m}_j}\frac{\bm{\dot{\bar{m}}}_0^{[j]}}{\bm{\bar{m}}_0^{[j]^2}}\right)\\
       &+\frac{1}{\gamma_{\textit{x}j}}\left(\bm{\mathcal{W}}^*_{\bm{\textit{x}} j}-\bm{\bar{\mathcal{W}}}_{\bm{\textit{x}} j}\right)^{\top}\left(\gamma_{\textit{x}j}\bm{\mathcal{E}}_{\bm{\textit{x}} j}^{\top}\bm{P}_j\bm{B}\bm{\hbar}(\textbf{x}_{\textit{x}j})-\bm{\dot{\bar{\mathcal{W}}}}_{\textit{x}j}\right)\\
       &+\bm{\mathcal{E}}_{\bm{\textit{x}} j}^{\top}\bm{P}_j\bm{B}\Bigg{\{}\bm{\varpi}^{[j]}_{\textit{x}}+\frac{1}{m_0}\left(\widetilde{\Delta}_{\bm{x}_0}^{[j]}+\bm{\sum}_{i=1}^n\widetilde{\Delta}_{\bm{x}_i}^{\parallel[j]}\right)+\bm{Y}_{\textit{x}}^{[j]}\Bigg{\}}, 
    \end{aligned}
    }
\end{equation}
\begin{equation}
    {\footnotesize
    \begin{aligned}
\!\!\!\!\!\!\!\!\!\!\bm{\dot{\mathcal{V}}}_{\bm{\textit{R}}}=&\bm{\sum}_{j=1}^3\!\Bigg{\{}\!-k_{R_0}c_{R}\|\bm{e}_{\bm{R}_0}^{[j]}\|^{2}\!-\!k_{\Omega_0}\|\bm{e}_{\bm{\Omega}_0}^{[j]}\|^{2}\\&\!-\!k_{\Omega_0}c_{R}\bm{e}_{\bm{\Omega}_0}^{[j]}\bm{e}_{\bm{R}_0}^{[j]}+c_R\bm{e}_{\bm{\Omega}_0}^{[j]}\bm{\dot{e}}_{\bm{R}_0}^{[j]}\\
       &+\frac{1}{\gamma_{\bm{\textit{R}}j}}\left(\bm{\mathcal{W}}^*_{\bm{\textit{R}} j}-\bm{\bar{\mathcal{W}}}_{\bm{\textit{R}} j}\right)^{\top}\bigg{\{}\gamma_{\bm{\textit{R}}j}\left(\bm{e}^{[j]}_{\bm{\Omega}_0}+c_{R}\bm{e}_{\bm{R}_0}^{[j]}\right)\bm{\hbar}(\textbf{x}_{\textit{R}j})-\bm{\dot{\bar{\mathcal{W}}}}_{\textit{R}j}\bigg{\}}\\[-10pt]
    \end{aligned}\notag
    }
\end{equation}

\begin{equation}
    {\footnotesize
    \begin{aligned}
       &\!\!\!+\widetilde{J}_j
\bigg{\{}\left(\bm{e}^{[j]}_{\bm{\Omega}_0}+c_{R}\bm{e}_{\bm{R}_0}^{[j]}\right)\bm{\mathcal{U}}_{\textit{R}}^{[j]}+\eta_{J_j}\frac{\bm{\dot{\bar{\mathit{J}}}}_0^{[j]}}{\bm{\bar{J}}_0^{[j]^2}}\bigg{\}}+\left(\bm{e}^{[j]}_{\bm{\Omega}_0}+c_{R}\bm{e}_{\bm{R}_0}^{[j]}\right)\bigg{\{}\!\!\!\\
&\bm{\varpi}^{[j]}_{\textit{R}}+\frac{1}{\bm{J}_0^{[j]}}\Big{\{}\widetilde{\Delta}^{[j]}_{\bm{R}_0}+\bm{\sum}_{i=1}^n\left([\bm{\rho}_i]_{\times}\bm{R}_0^{\top}\widetilde{\Delta}_{\bm{x}_i}^{\parallel}\right)^{[j]}\Big{\}}+\bm{Y}_{\textit{R}}^{[j]}\bigg{\}}\!\Bigg{\}}.
    \end{aligned}
    }
\end{equation}
\begin{equation}
    {\small
    \begin{aligned}
      \bm{\dot{\mathcal{V}}}_{\bm{q}}=&\bm{\sum}_{i=1}^{n}-(k_{\omega}-\textit{c}_{q})\lVert\bm{e}_{\bm{\omega}_i}\lVert^2-\textit{c}_{q}k_{q}\lVert\bm{e}_{\bm{q}_i}\lVert^2\\&-\textit{c}_{q}k_{\omega}\bm{e}_{\bm{q}_i}\cdot\bm{e}_{\bm{\omega}_i}-\left(\bm{e}_{\bm{\omega}_i}+\textit{c}_{q}\bm{e}_{\bm{q}_i}\right)\cdot\frac{[\bm{q}_i]_{\times}}{m_{i}l_{i}}\widetilde{\Delta}_{\bm{x}_{i}}^{\bot},
    \end{aligned}
    }
\end{equation}
\begin{equation}
    {\small
    \begin{aligned}
    \bm{\dot{\mathcal{V}}}_{\Delta}=&-\frac{1}{\textit{h}_{x_0}}\widetilde{\Delta}_{\bm{x}_0}\cdot\dot{\bar{\Delta}}_{\bm{x}_0}-\frac{1}{\textit{h}_{R_0}}\widetilde{\Delta}_{\bm{R}_0}\cdot\dot{\bar{\Delta}}_{\bm{R}_0}\\
     &-\bm{\sum}_{i=1}^{n}\frac{1}{\textit{h}_{x_i}}\left(\widetilde{\Delta}^{\parallel}_{\bm{x}_{i}}\cdot\dot{\bar{\Delta}}^{\parallel}_{\bm{x}_{i}}+\widetilde{\Delta}^{\bot}_{\bm{x}_{i}}\cdot\dot{\bar{\Delta}}^{\bot}_{\bm{x}_{i}}\right).
    \end{aligned}
    }
\end{equation}
If the terms $\bm{\dot{\bar{m}}}_0^{[j]}$, $\bm{\dot{\bar{\mathit{J}}}}_0^{[j]}$, $\bm{\dot{\bar{\mathcal{W}}}}_{\bm{\textit{x}} 
j}$, $\bm{\dot{\bar{\mathcal{W}}}}_{\bm{\textit{R}} 
j}$, $\dot{\bar{\Delta}}_{\bm{x}_0}$, $\dot{\bar{\Delta}}_{\bm{R}_0}$ and $\dot{\bar{\Delta}}_{\bm{x}_{i}}$, are designed as given in Eqs.~\eqref{Adaptive Law of mass}, \eqref{Adaptive Law of Inertia Tensor}, \eqref{Estimated Weights_x}, \eqref{Estimated Weights_R}, \eqref{integral compensations0} and \eqref{integral compensations}, then the time-derivative of the complete Lyapunov function $\bm{\dot{\mathcal{V}}}$ reduces to:
\begin{equation}
{\small
    \begin{aligned}
\bm{\dot{\mathcal{V}}}=&\bm{\sum}_{j=1}^3\Bigg{\{}\!\!-\frac{1}{2}\bm{\mathcal{E}}_{\bm{\textit{x}} j}^{\top}\bm{Q}_j\bm{\mathcal{E}}_{\bm{\textit{x}} j}\!+\!\bm{\mathcal{E}}_{\bm{\textit{x}}j}^{\top}\bm{P}_j\bm{B}\bm{\Xi}^{[j]}_{\bm{\textit{x}}}\!\!-\!k_{R_0}c_{R}\|\bm{e}_{\bm{R}_0}^{[j]}\|^{2}\\
&-k_{\Omega_0}\lVert\bm{e}^{[j]}_{\bm{\Omega}_0}\lVert^2+k_{\Omega_0}c_{R}\bm{e}_{\bm{\Omega}_0}^{[j]}\bm{e}_{\bm{R}_0}^{[j]}+c_R\bm{e}_{\bm{\Omega}_0}^{[j]}\bm{\dot{e}}_{\bm{R}_0}^{[j]}\\&+\left(\bm{e}^{[j]}_{\bm{\Omega}_0}+c_{R}\bm{e}_{\bm{R}_0}^{[j]}\right)\bm{\Xi}^{[j]}_{\bm{\textit{R}}}\Bigg{\}}-\bm{\sum}_{i=1}^n\bm{z}_{\bm{q}_i}^{\top}\bm{\mathcal{Z}}\bm{z}_{\bm{q}_i},
\end{aligned}
}
\label{dot V}
\end{equation}
where $\bm{z}_{\bm{q}_i}=\left(\lVert\bm{e}_{\bm{q}_i}\lVert, \lVert\bm{e}_{\bm{\omega}_i}\lVert\right)^{\top}$, $\bm{\mathcal{Z}}={\begin{bmatrix}
         \textit{c}_{q}k_{q} &\frac{\textit{c}_{q}k_{\omega}}{2}\\
\frac{\textit{c}_{q}k_{\omega}}{2}&k_{\omega}\text{-}\textit{c}_{q}
\end{bmatrix}}$ and
{\footnotesize
\begin{align}
&\begin{aligned}
&\bm{\Xi}^{[j]}_{\bm{\textit{x}}}=\bm{\varpi}^{[j]}_{\textit{x}}+\left(\frac{1}{m_0}-\frac{1}{m'_0}\right)\left(\widetilde{\Delta}_{\bm{x}_0}^{[j]}+\bm{\sum}_{i=1}^n\widetilde{\Delta}_{\bm{x}_i}^{\parallel[j]}\right)+\bm{Y}_{\textit{x}}^{[j]},\\
&\bm{\Xi}^{[j]}_{\bm{\textit{R}}}=\bm{\varpi}^{[j]}_{\textit{R}}\!+\!\left(\frac{1}{\bm{J}_0^{[j]}}-\frac{1}{\bm{J}_0'^{[j]}}\right)\bigg{\{}\widetilde{\Delta}^{[j]}_{\bm{R}_0}\!\!+\!\!\bm{\sum}_{i=1}^n\left([\bm{\rho}_i]_{\times}\bm{R}_0^{\top}\widetilde{\Delta}_{\bm{x}_i}^{\parallel}\right)^{[j]}\!\!\!\bigg{\}}\!+\!\bm{Y}_{\textit{R}}^{[j]}.
\end{aligned}\notag
\end{align}}

\begin{figure*}[!t] 
 \centering
 \includegraphics[scale=0.095]{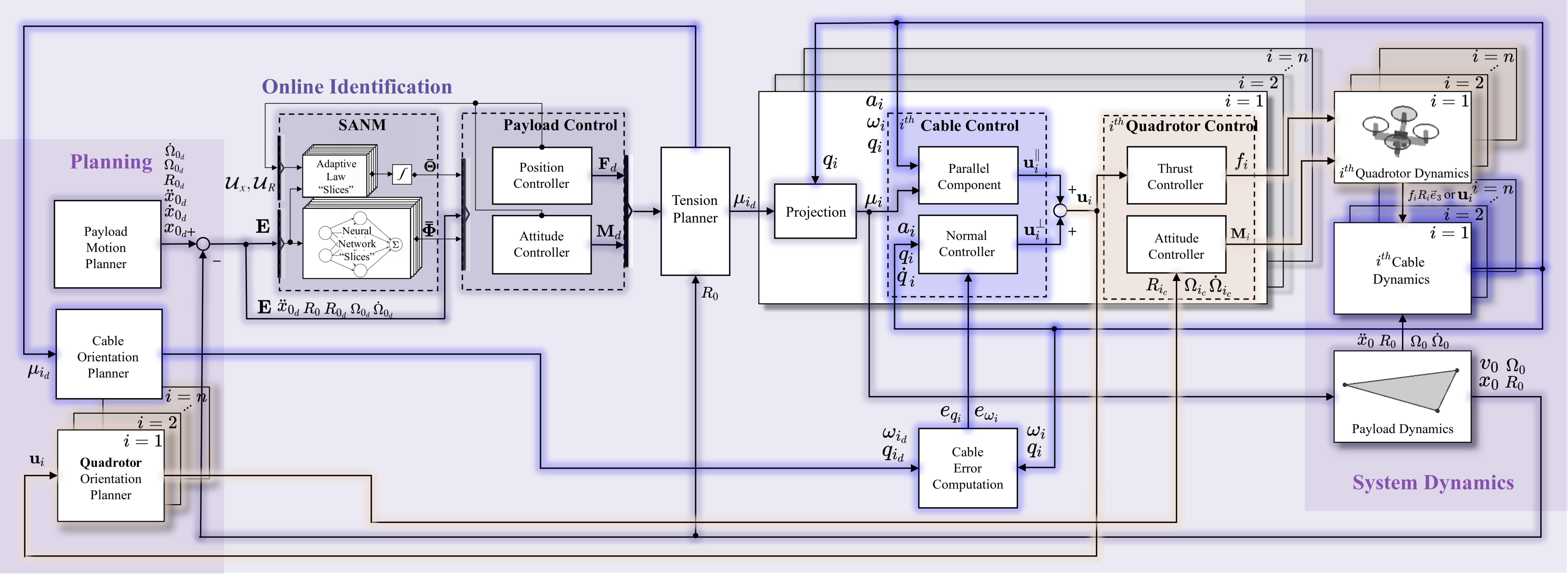}
 \caption{\footnotesize Schematic of simulation block diagram in \textit{MATLAB Simulink}. On the left, the planning part gives the desired commands of the system. Three different colors are used to represent the control loops for the payload, the cables, and the quadrotors, for clearer distinction.} 
      \label{Simulink_Schematic}
\end{figure*}

Given that the optimal approximation error $\lVert\bm{\varpi}_{\textit{x}}\lVert$ and $\lVert\bm{\varpi}_{\textit{R}}\lVert$ are bounded, and the $\lVert\bm{Y}_{\textit{x}}\lVert$ and $\lVert\bm{Y}_{\textit{R}}\lVert$ are also bounded \cite{2018 Geometric Control of Quadrotor UAVs Transporting
a Cable-Suspended Rigid Body}, it follows that $\lVert\bm{\Xi}_{\textit{x}}\lVert$ and $\lVert\bm{\Xi}_{\textit{R}}\lVert$ are likewise bounded. The upper bounds of $\lVert\bm{\Xi}_{\textit{x}}\lVert$ and $\lVert\bm{\Xi}_{\textit{R}}\lVert$ are defined as $\varepsilon_{\bm{\textit{x}}}\in\mathbb{R}^+$ and $\varepsilon_{\bm{\textit{R}}}\in\mathbb{R}^+$, respectively.  Since $\|\bm{e}_{\bm{R}_0}\|<1$ and $\|\bm{\dot{e}}_{\bm{R}_0}\|\leq\|\bm{e}_{\bm{\Omega}_0}\|$ from Eq.~\eqref{rotational error dynamics0}, it holds that:
\begin{equation}
    {\small
    \begin{aligned}
     \bm{\dot{\mathcal{V}}}\leq&\bm{\sum}_{j=1}^3\Bigg{\{}-\frac{1}{2}\lVert\bm{\mathcal{E}}_{\bm{\textit{x}} j}\lVert\Big{\{}\lambda_{\min}\left(\bm{Q}_j\right)\lVert\bm{\mathcal{E}}_{\bm{\textit{x}} j}\lVert-2\varepsilon_{\bm{\textit{x}}}\lambda_{\max}\left(\bm{P}_j\right)  \Big{\}}\Bigg{\}}\\
     &\!-\!k_{R_0}c_{R}\|\bm{e}_{\bm{R}_0}\|^{2}\!-\!\left(k_{\Omega_0}\!-\!c_R\right)\|\bm{e}_{\bm{\Omega}_0}\|^{2}\!+\!k_{\Omega_0}c_{R}\|\bm{e}_{\bm{\Omega}_0}\|\|\bm{e}_{\bm{R}_0}\|\\
     &\!+\!c_R\|\bm{e}_{\bm{\Omega}_0}\|^2+c_R\varepsilon_{\bm{\textit{R}}}\|\bm{e}_{\bm{R}_0}\|+\varepsilon_{\bm{\textit{R}}}\|\bm{e}_{\bm{\Omega}_0}\|-\bm{\sum}_{i=1}^n\bm{z}_{\bm{q}_i}^{\top}\bm{\mathcal{Z}}\bm{z}_{\bm{q}_i}.
    \end{aligned}
    }\label{dot_V_bound}
\end{equation}
By choosing $k_{\Omega_0} > c_R$, we can apply Young’s inequality to yield:
\begin{equation}
    {\small
    \begin{aligned}    
&c_R\varepsilon_{\bm{\textit{R}}}\|\bm{e}_{\bm{R}_0}\|\leq\frac{c_R^2\varepsilon_{\bm{\textit{R}}}^2}{2k_{R_0}c_{R}}+\frac{k_{R_0}c_{R}}{2}\|\bm{e}_{\bm{R}_0}\|^{2},\\
    &\!\varepsilon_{\bm{\textit{R}}}\|\bm{e}_{\bm{\Omega}_0}\|\!\leq\!\frac{\varepsilon_{\bm{\textit{R}}}^2}{2\left(k_{\Omega_0}\!\!-\!c_R\right)}\!+\!\frac{k_{\Omega_0}\!-\!c_R\!}{2}\|\bm{e}_{\bm{\Omega}_0}\|^2.
    \end{aligned}
    } 
    \label{young's inequality}
\end{equation}
Substituting Eq.~\eqref{young's inequality} into Eq.~\eqref{dot_V_bound}, we derive:
\begin{equation}
 {\small
    \begin{aligned}    
    \bm{\dot{\mathcal{V}}}\leq&\bm{\sum}_{j=1}^3\Bigg{\{}-\frac{1}{2}\lVert\bm{\mathcal{E}}_{\bm{\textit{x}} j}\lVert\Big{\{}\lambda_{\min}\left(\bm{Q}_j\right)\lVert\bm{\mathcal{E}}_{\bm{\textit{x}} j}\lVert-2\varepsilon_{\bm{\textit{x}}}\lambda_{\max}\left(\bm{P}_j\right)  \Big{\}}\Bigg{\}}\\
    &-\bm{z}^{\top}_{\bm{\textit{R}}_0}\!\bm{\mathcal{M}}_{\bm{\textit{R}}_0}\,\bm{z}_{\bm{\textit{R}}_0}+\mathbf{C}_{\bm{\textit{R}}_0}-\bm{\sum}_{i=1}^n\bm{z}_{\bm{q}_i}^{\top}\bm{\mathcal{Z}}\bm{z}_{\bm{q}_i},
    \end{aligned}
    }
    \label{dot_V_quadratic}
\end{equation}
where $\bm{z}_{\bm{\textit{R}}_0}=\left(\|\bm{e}_{\bm{R}_0}\|,\|\bm{e}_{\bm{\Omega}_0}\|\right)^{\top}\!\!\in\mathbb{R}^{2}$ and the matrix $\bm{\mathcal{M}}_{\bm{\textit{R}}_0}\in\mathbb{R}^{2\times2}$ is given by:
\begin{equation}
    \begin{aligned}    
       \bm{\mathcal{M}}_{\bm{\textit{R}}_0}={\renewcommand{\arraycolsep}{5pt}\renewcommand{\arraystretch}{1.5}\begin{bmatrix}
\frac{k_{R_0}c_R}{2}&\frac{-k_{\Omega_0}c_R}{2}\\
       \frac{-k_{\Omega_0}c_R}{2} &\frac{k_{\Omega_0}-c_R}{2}\\
    \end{bmatrix}}
    \end{aligned}.
    \label{Matrix_R}
\end{equation}
The constant term $\mathbf{C}_{\bm{\textit{R}}_0}>0$ is expressed as:
\begin{equation}
{\small
    \begin{aligned}    
\mathbf{C}_{\bm{\textit{R}}_0}=\frac{c_R\varepsilon_{\bm{\textit{R}}}^2}{2k_{R}}+\frac{\varepsilon_{\bm{\textit{R}}}^2}{2\left(k_{\Omega_0}\!-\!c_R\!\right)}.
    \end{aligned}
    }
    \label{C_R}
\end{equation}
If the positive constant $c_R$ is chosen sufficiently small:
\begin{equation}
{\small
    \begin{aligned}
    c_R\!< \min\left\{\frac{k_{R_0}k_{\Omega_0}}{k_{\Omega_0}^2+k_{R_0}},k_{\Omega_0}\right\},\label{cR bound}
\end{aligned}
}
\end{equation}
it follows that matrix $\bm{\mathcal{M}}_{\bm{\textit{R}}_0}$ is positive-definite. Design the eigenvalue $\lambda_{\min}\left(\bm{Q}_j\right)\geq\frac{2\varepsilon_{\bm{\textit{x}}}\lambda_{\max}\left(\bm{P}_j\right)}{\lVert\bm{\mathcal{E}}_{\bm{\textit{x}} j}\lVert}$ and choose $\textit{c}_{q}$ to be sufficiently small such that the matrix $\bm{\mathcal{Z}}$ is positive-definite. The Lyapunov derivative then satisfies $\bm{\dot{\mathcal{V}}}\leq0$ whenever the state error vector $(\|\bm{e}_{\bm{x}_0}\|,\|\bm{e}_{\bm{v}_0}\|,\|\bm{e}_{\bm{R}_0}\|,\|\bm{e}_{\bm{\Omega}_0}\|,\lVert\bm{e}_{\bm{q}_i}\lVert, \lVert\bm{e}_{\bm{\omega}_i}\lVert)$ lies outside a certain bounded ball, which guarantees semi-global practical stability. Therefore, the claim in \textbf{\textit{Proposition 2}} is verified. This implies that the full-state error vector of the payload $\|\mathbf{E}(t)\|$ converges to a bounded error ball, and there exists a compact set $\bm{\mathcal{C}}$ such that $\mathbf{E}(t) \in \bm{\mathcal{C}}$ for $\forall t \geq 0$. Consequently, \textbf{\textit{Proposition 3}} holds at all times.

\vspace{1em} 
\noindent\textbf{5. Numerical Simulation}

\begin{figure*}[htbp] 
\vspace{-0em} 
  \centering
  \begin{subfigure}[b]{0.32\textwidth}
    \centering
    \includegraphics[width=\linewidth]{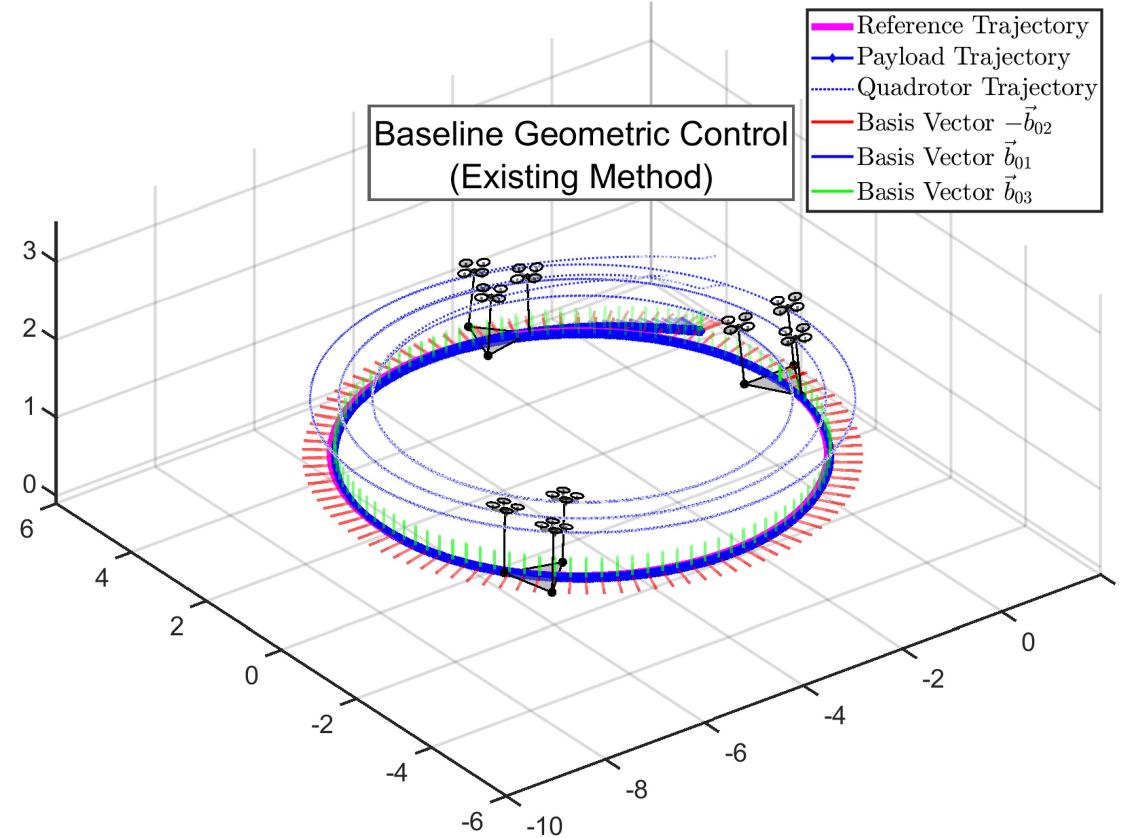}
    \captionsetup{labelformat=empty, labelsep=none} 
    \caption*{}
  \end{subfigure}
  \hspace{2mm}
  \begin{subfigure}[b]{0.32\textwidth}
    \centering
    \includegraphics[width=\linewidth]{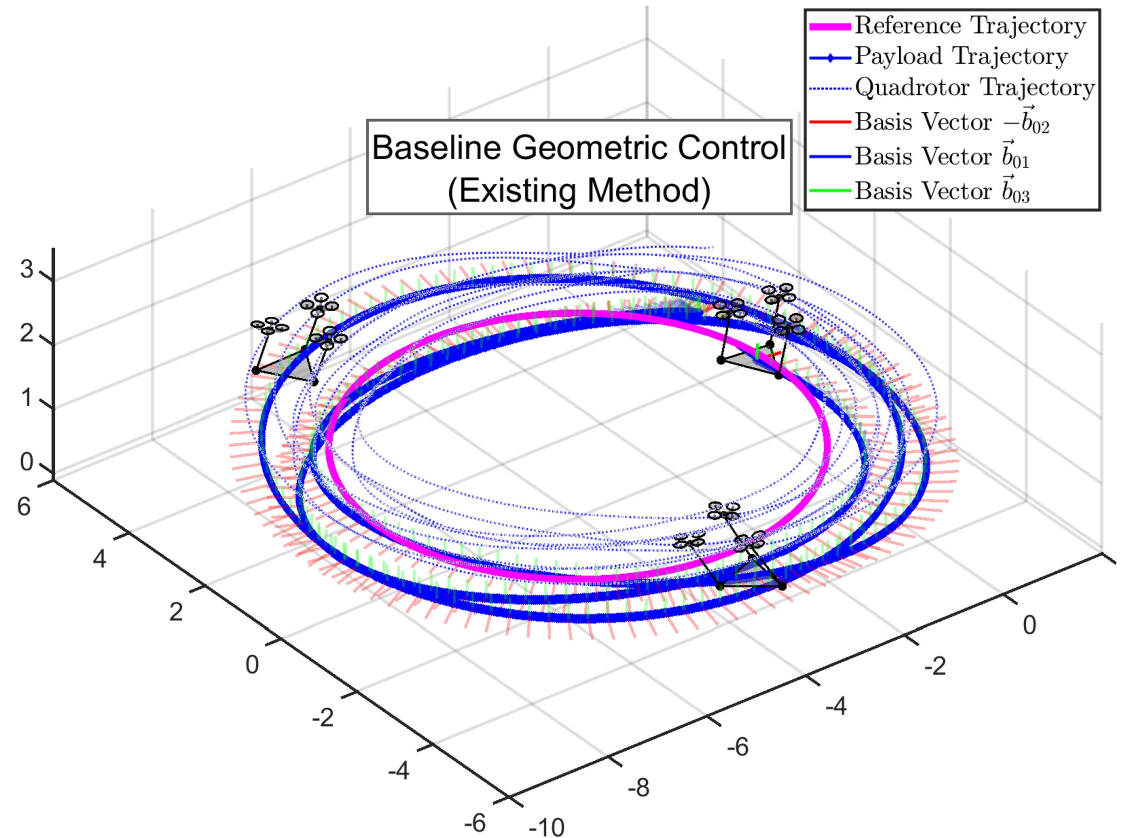}
    \captionsetup{labelformat=empty, labelsep=none} 
    \caption*{}
  \end{subfigure}
  \hspace{2mm}
  \begin{subfigure}[b]{0.32\textwidth}
    \centering
    \includegraphics[width=\linewidth]{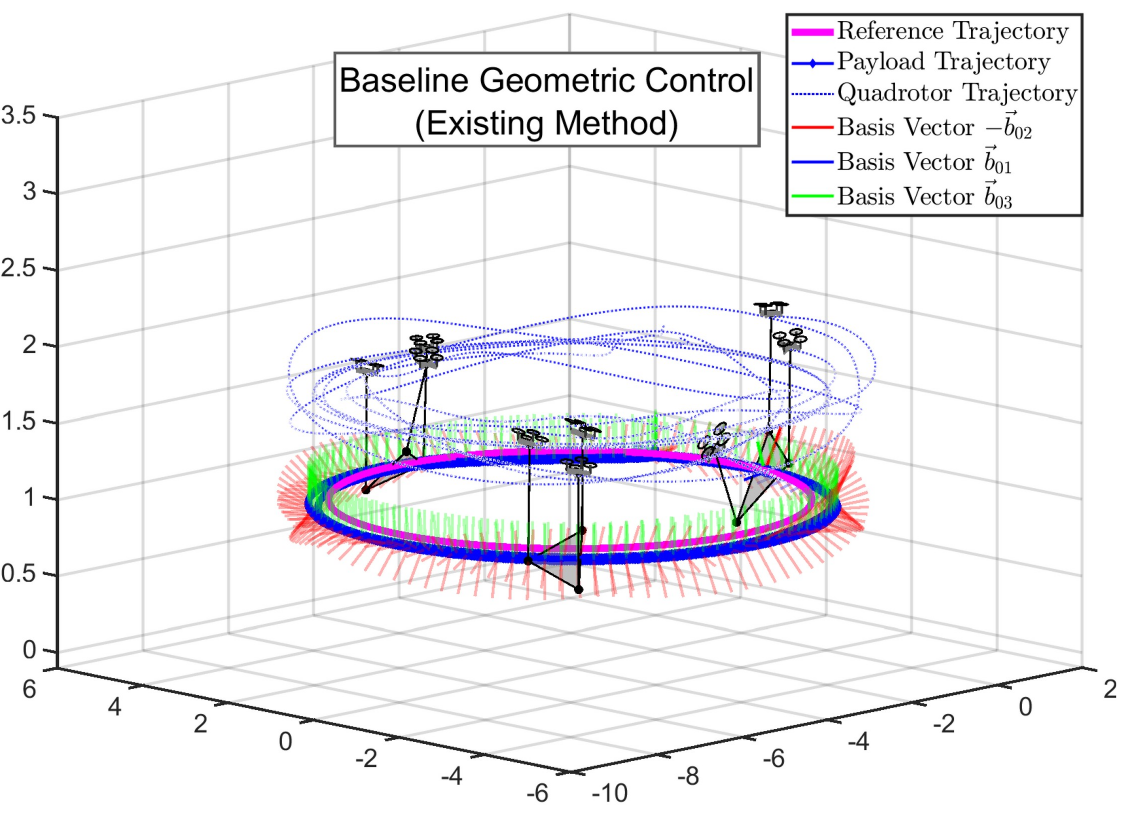}
    \captionsetup{labelformat=empty, labelsep=none} 
    \caption*{}
  \end{subfigure}
    \vspace{-1em} 

  \captionsetup{labelformat=default, labelsep=colon}
  \begin{subfigure}[b]{0.32\textwidth}
    \centering
    \includegraphics[width=\linewidth]{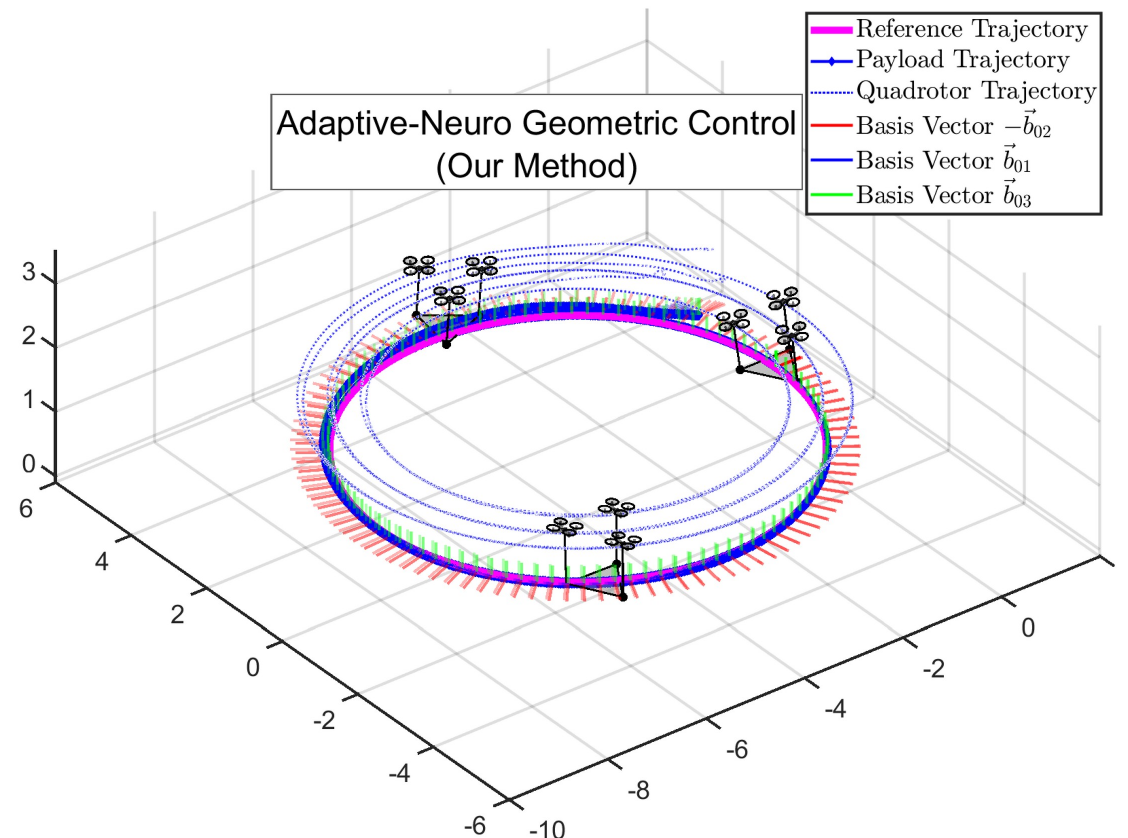}
    \caption{\footnotesize Comparison Group A  }
    \label{Comparison group A}
  \end{subfigure}
  \hspace{2mm}
  \begin{subfigure}[b]{0.32\textwidth}
    \centering
    \includegraphics[width=\linewidth]{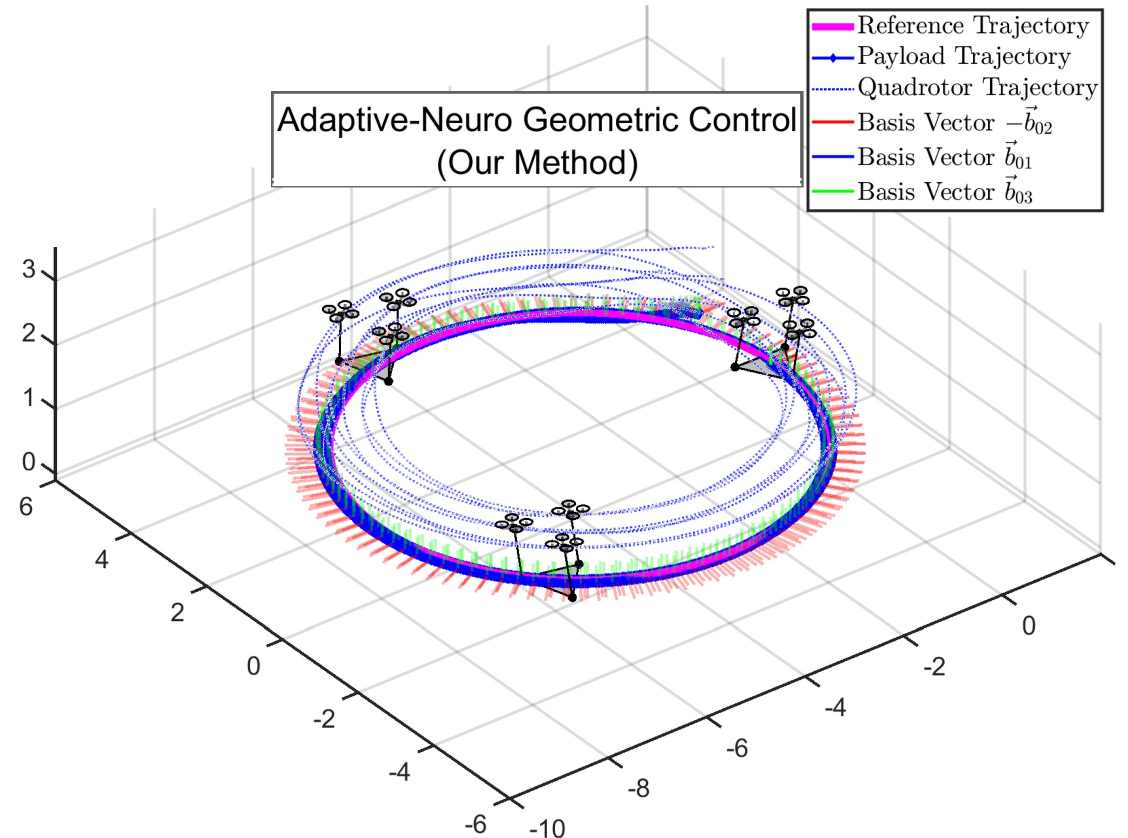}
    \caption{\footnotesize  Comparison Group B}
    \label{Comparison group B}
  \end{subfigure}
  \hspace{2mm}
  \begin{subfigure}[b]{0.32\textwidth}
    \centering
    \includegraphics[width=\linewidth]{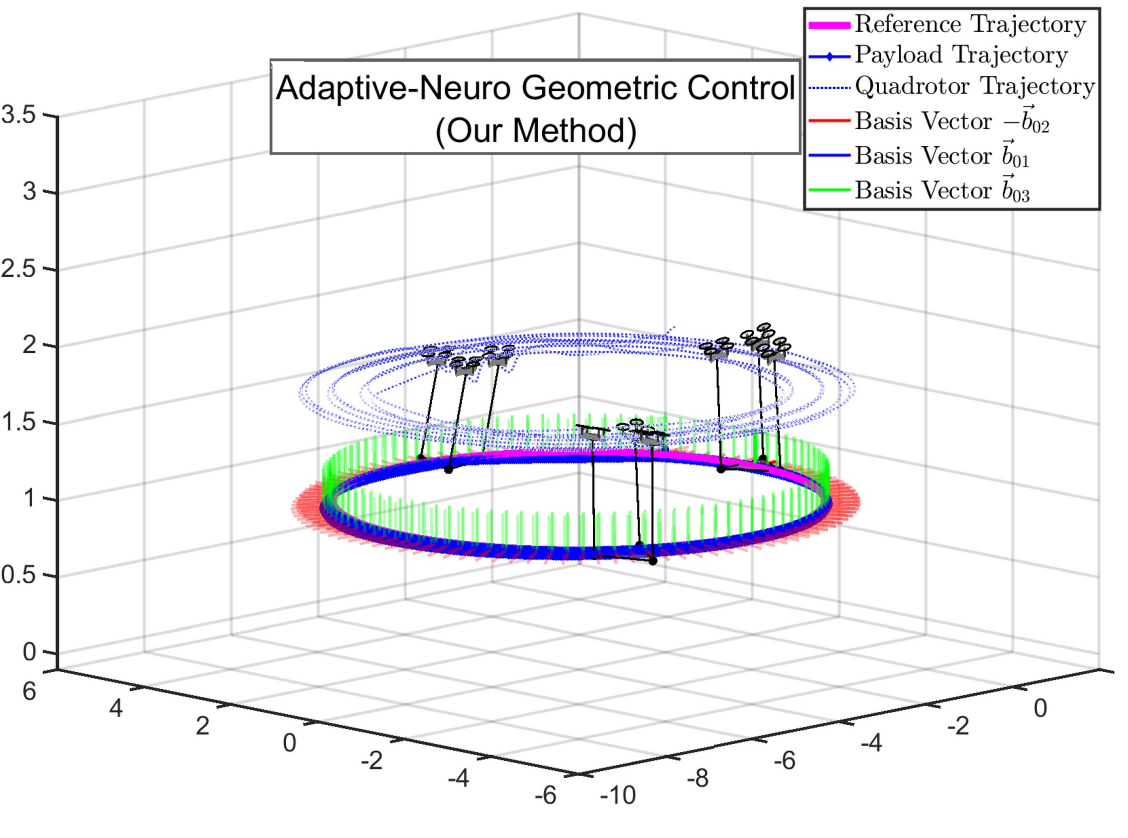}
    \caption{\footnotesize   Comparison Group C}
    \label{Comparison group C}
  \end{subfigure}
  \caption{\footnotesize Comparison results presented in \textit{MATLAB Simulink (ODE3 Bogacki–Shampine Solver)}. For simulation video, refer to \url{https://staff.aist.go.jp/kamimura.a/ES2025/video.mp4}.}
   \vspace{-1em} 
  \label{Comparision results}
\end{figure*}
 \vspace{-0em} 

\begin{figure*}[htbp] 
\vspace{-0em} 
  \captionsetup{labelformat=default, labelsep=colon}
  \begin{subfigure}[b]{0.24\textwidth}
    \centering
    \includegraphics[width=\linewidth]{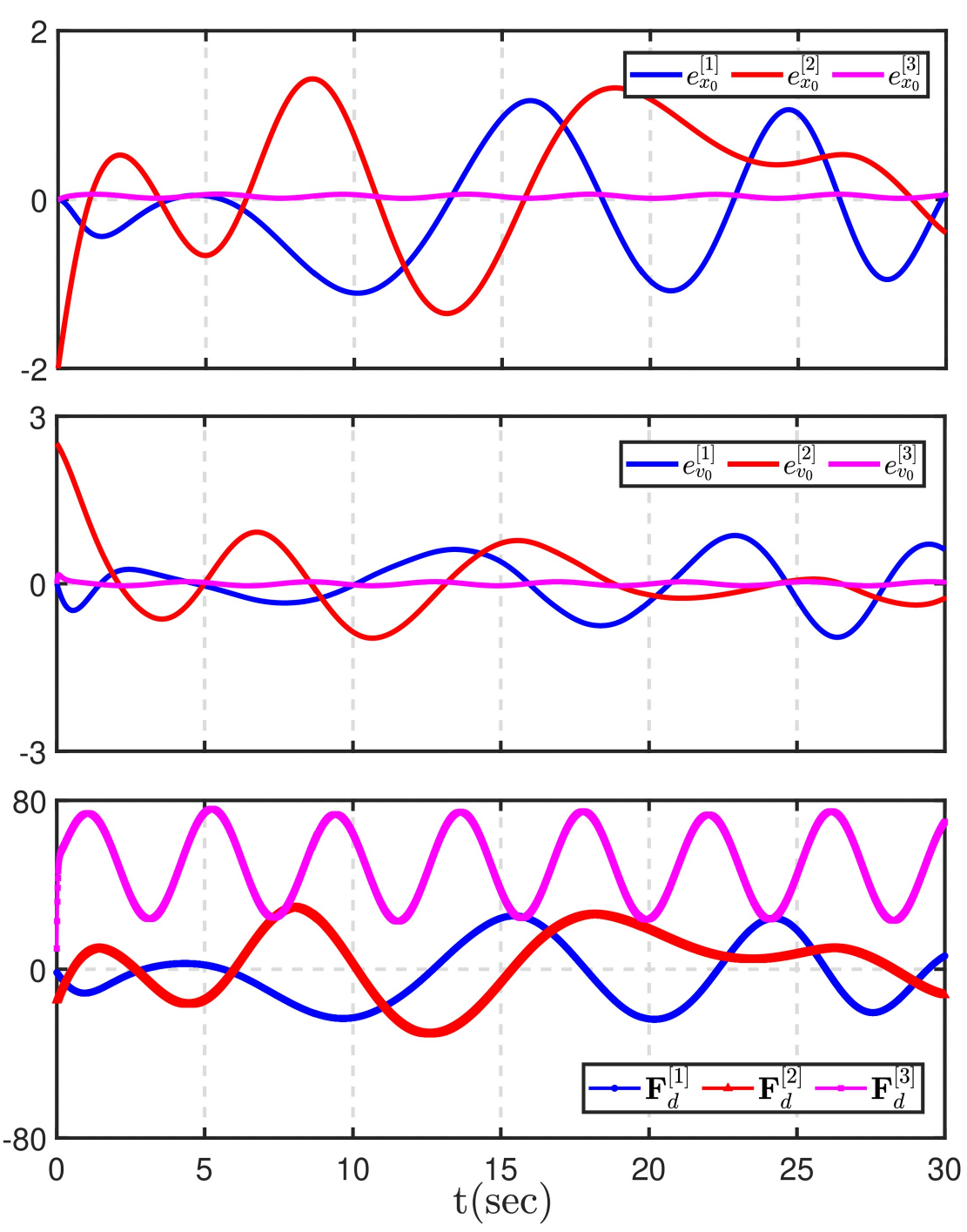}
    \caption{Group B (existing method)}
  \end{subfigure}
  \hspace{1mm}
  \begin{subfigure}[b]{0.24\textwidth}
    \centering
    \includegraphics[width=\linewidth]{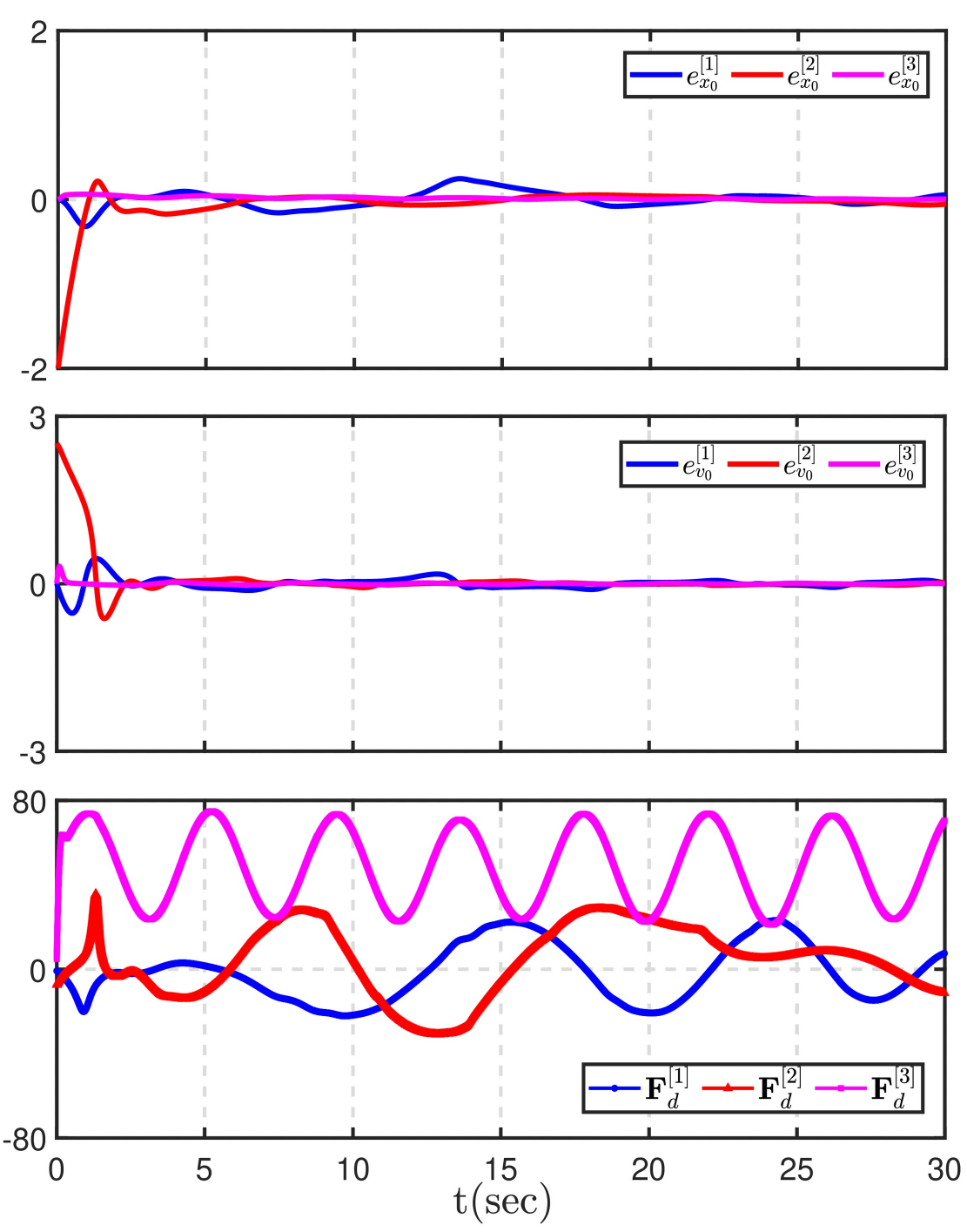}
    \caption{Group B (our method)}
  \end{subfigure}
  \hspace{0mm}
  \vspace{-1.5em} %
  \begin{subfigure}[b]{0.24\textwidth}
    \centering
    \includegraphics[width=\linewidth]{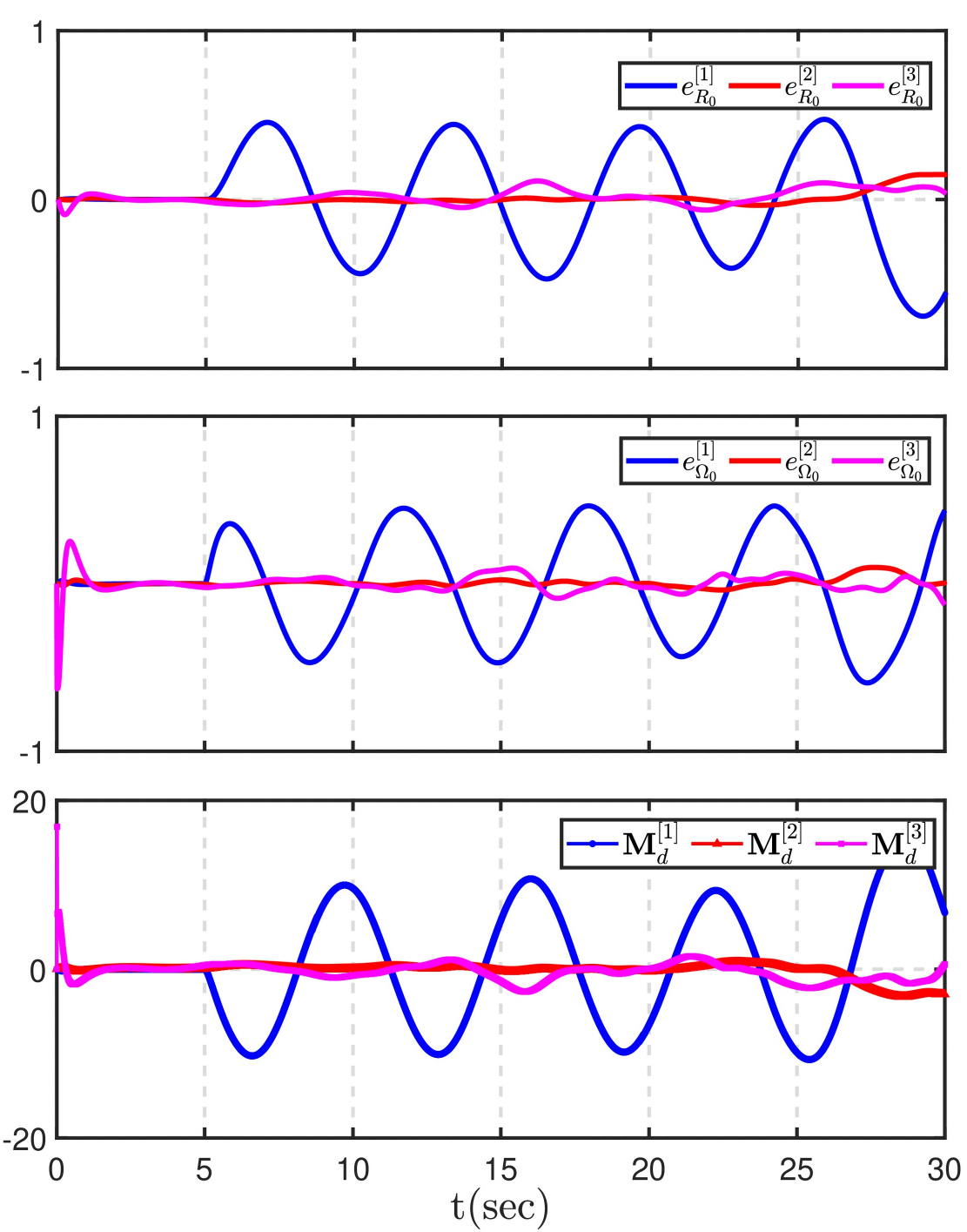}
     \caption{Group C (existing method)}
  \end{subfigure}
  \hspace{1mm}
  \begin{subfigure}[b]{0.24\textwidth}
    \centering
    \includegraphics[width=\linewidth]{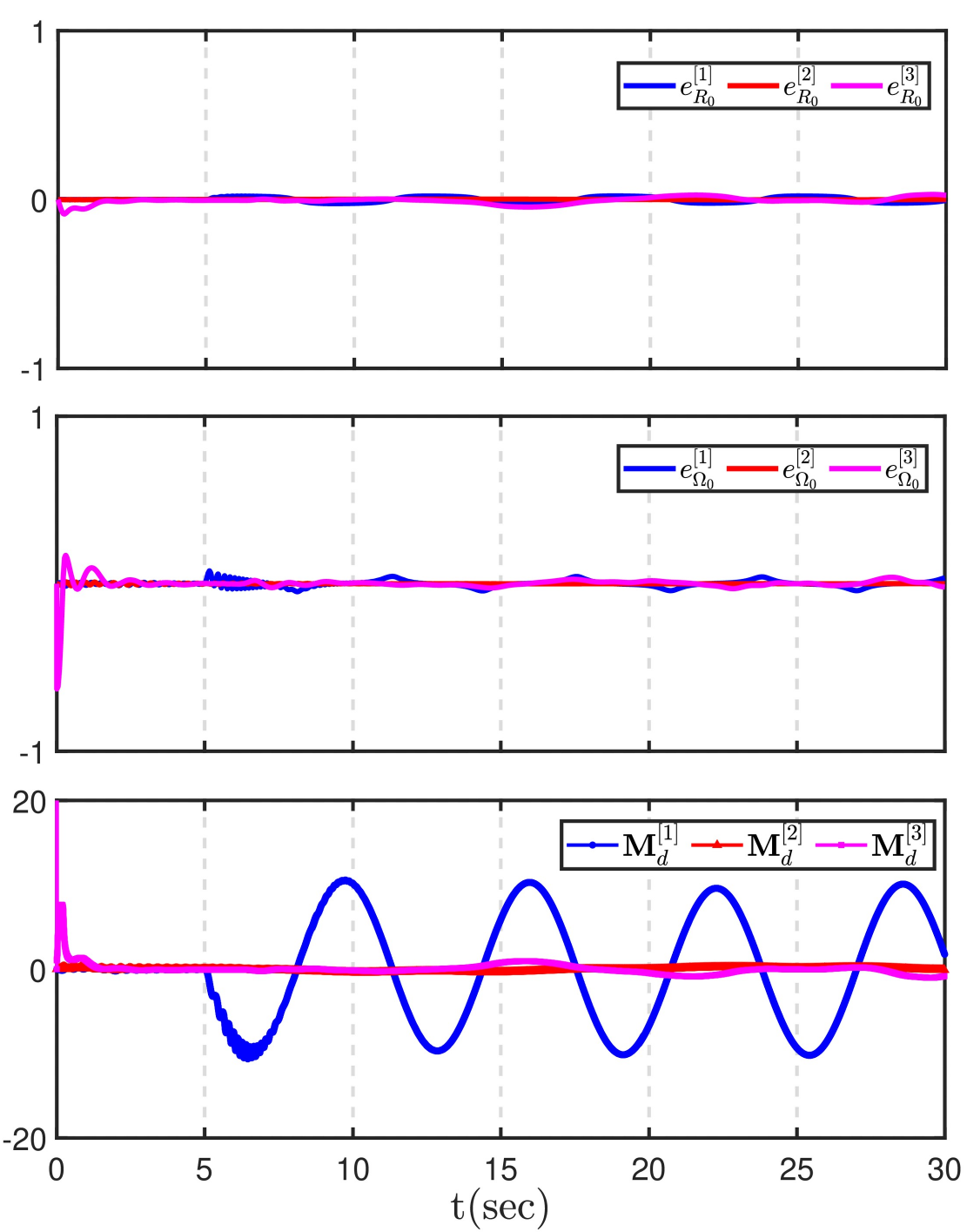}
     \caption{Group C (our method)}
  \end{subfigure}
  \hspace{0mm}
    \vspace{-0em} %
    
  \caption{Numerical comparison results of Groups B and C.}
   \vspace{-0em} %
  \label{Numerical results}
\end{figure*}

\begin{figure*}[htbp] 
\vspace{-0em} 
  \captionsetup{labelformat=default, labelsep=colon}
  \begin{subfigure}[b]{0.49\textwidth}
    \centering
    \includegraphics[width=\linewidth]{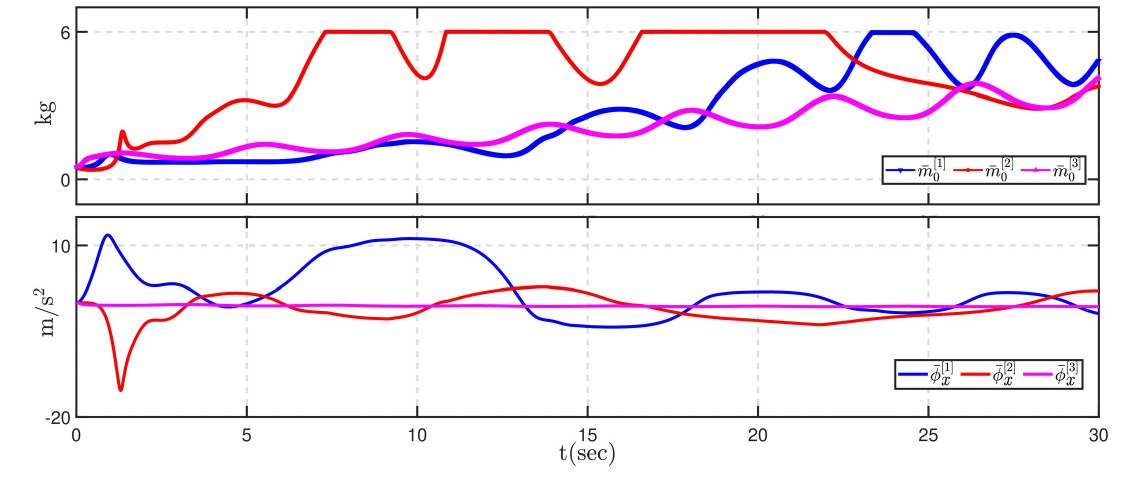}
    \caption{\footnotesize Values of $\bm{\bar{m}}_0^{[j]}$ and $\bm{\bar{\phi}}_{\textit{x}}^{[j]}$ in Group B }
  \end{subfigure}
  \hspace{1mm}
  \begin{subfigure}[b]{0.49\textwidth}
    \centering
    \includegraphics[width=\linewidth]{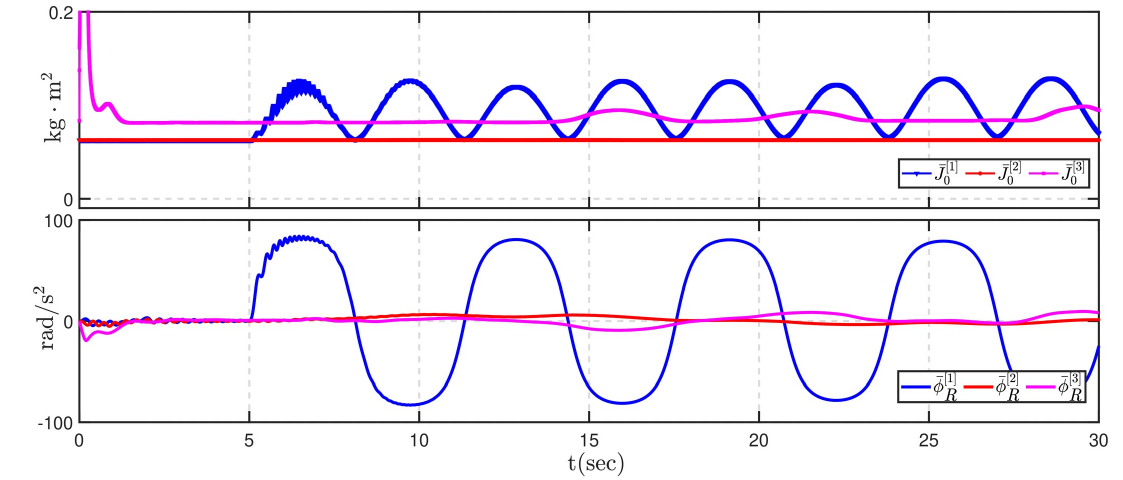}
 \caption{\footnotesize Values of $\bm{\bar{J}}_0^{[j]}$ and $\bm{\bar{\phi}}_{\textit{R}}^{[j]}$ in Group C  }
  \end{subfigure}
  \hspace{0mm}
  \vspace{-0em} %
    
  \caption{ \footnotesize Outputs of SANM module. The values of estimated model parameters $\bm{\bar{m}}_0^{[j]}$, $\bm{\bar{J}}_0^{[j]}$  were updated by adaptive law ``slices". The values of the estimated disturbance features $\bm{\bar{\phi}}_{\textit{x}}^{[j]}$, $\bm{\bar{\phi}}_{\textit{R}}^{[j]}$ were updated by neural network ``slices". }
   \vspace{-0em} %
  \label{SANM_outputs}
\end{figure*}

\begin{figure*}[htbp] 
\vspace{-0em} 
  \centering
  \begin{subfigure}[b]{0.32\textwidth}
    \centering
    \includegraphics[width=\linewidth]{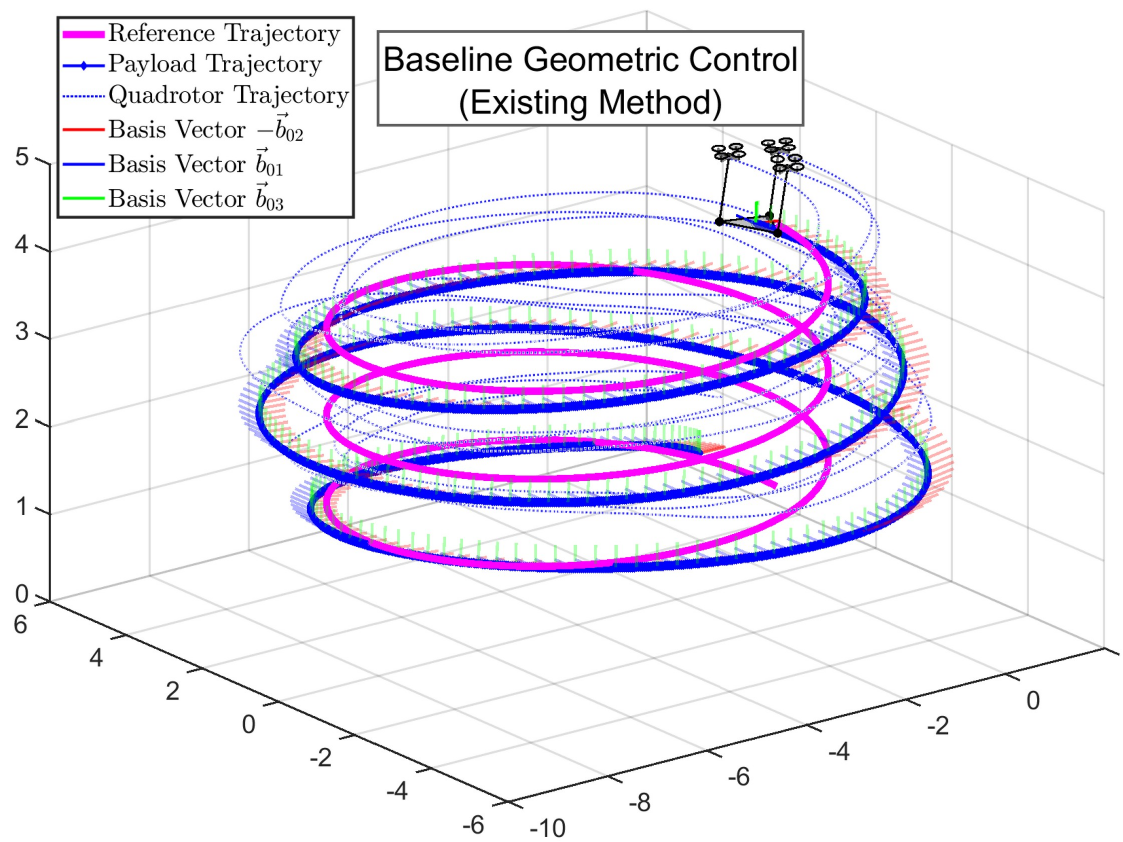}
    \captionsetup{labelformat=empty, labelsep=none} 
    \caption*{}
  \end{subfigure}
  \hspace{2mm}
  \begin{subfigure}[b]{0.32\textwidth}
    \centering
    \includegraphics[width=\linewidth]{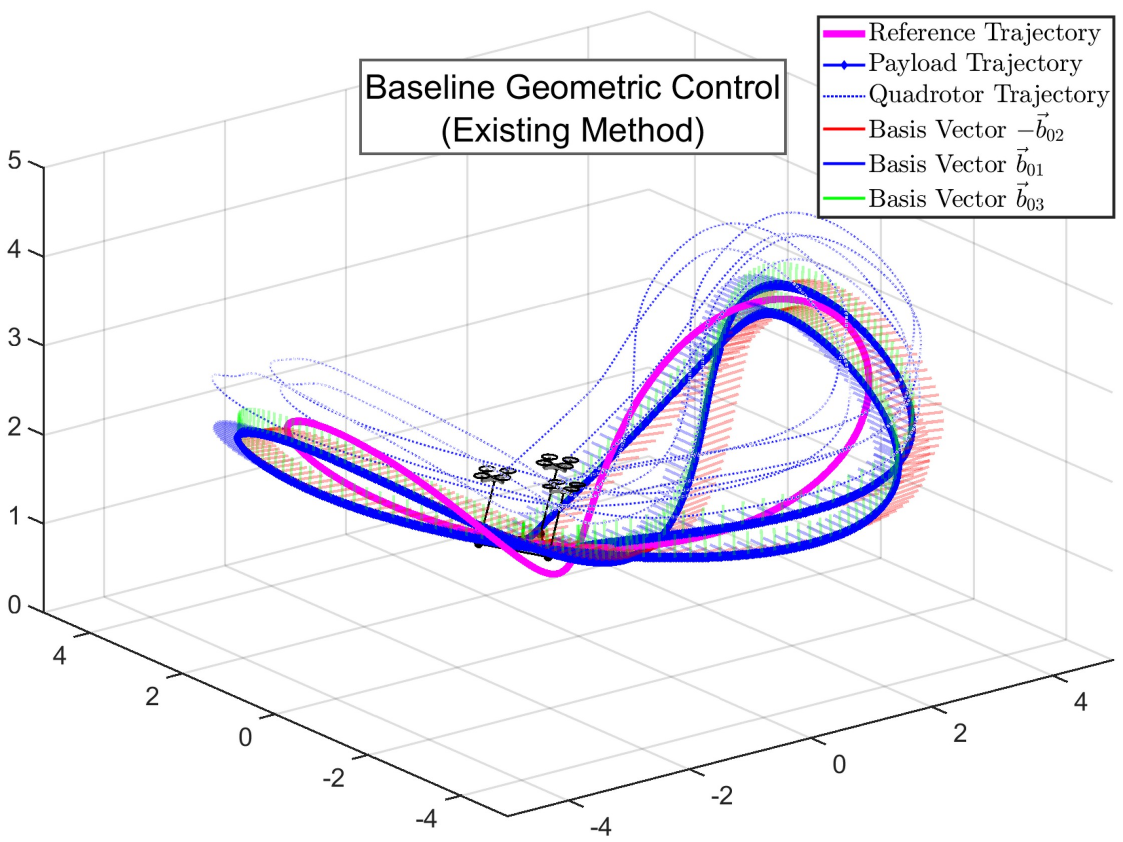}
    \captionsetup{labelformat=empty, labelsep=none} 
    \caption*{}
  \end{subfigure}
  \hspace{2mm}
  \begin{subfigure}[b]{0.32\textwidth}
    \centering
    \includegraphics[width=\linewidth]{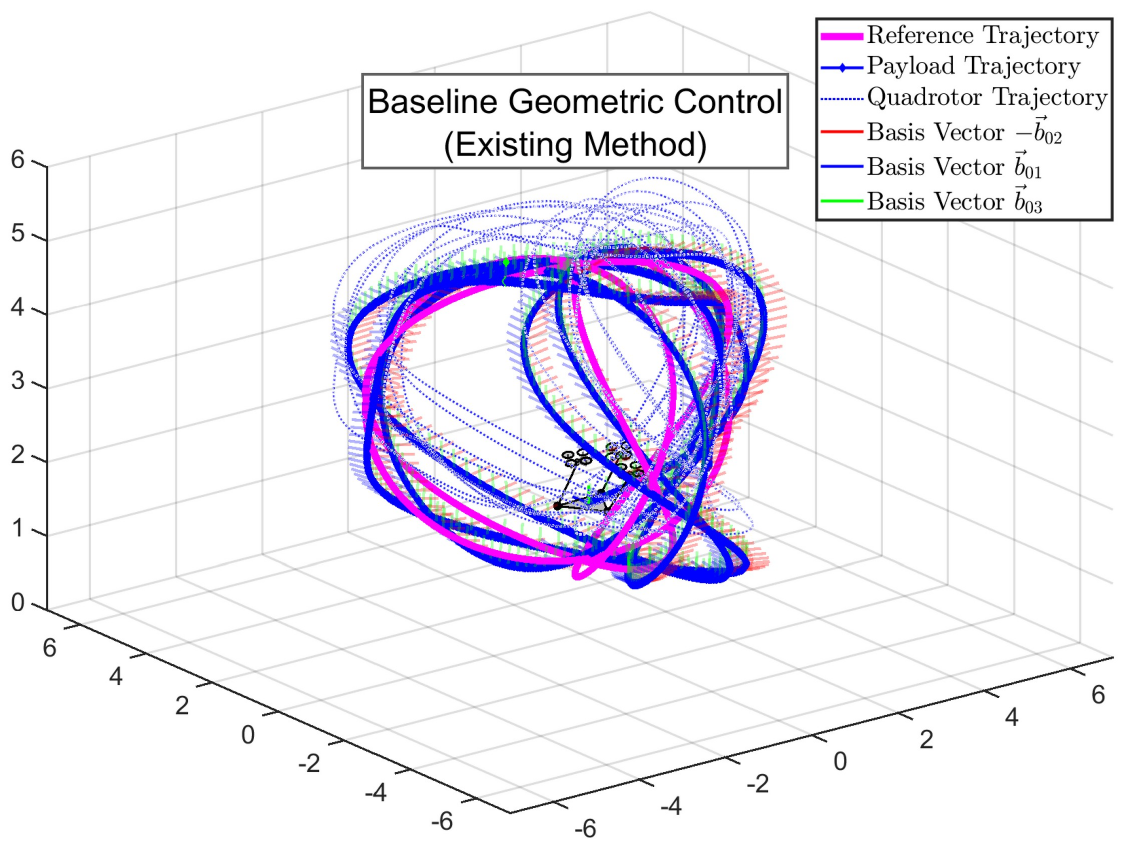}
    \captionsetup{labelformat=empty, labelsep=none} 
    \caption*{}
  \end{subfigure}
    \vspace{-1em} 

  \captionsetup{labelformat=default, labelsep=colon}
  \begin{subfigure}[b]{0.32\textwidth}
    \centering
    \includegraphics[width=\linewidth]{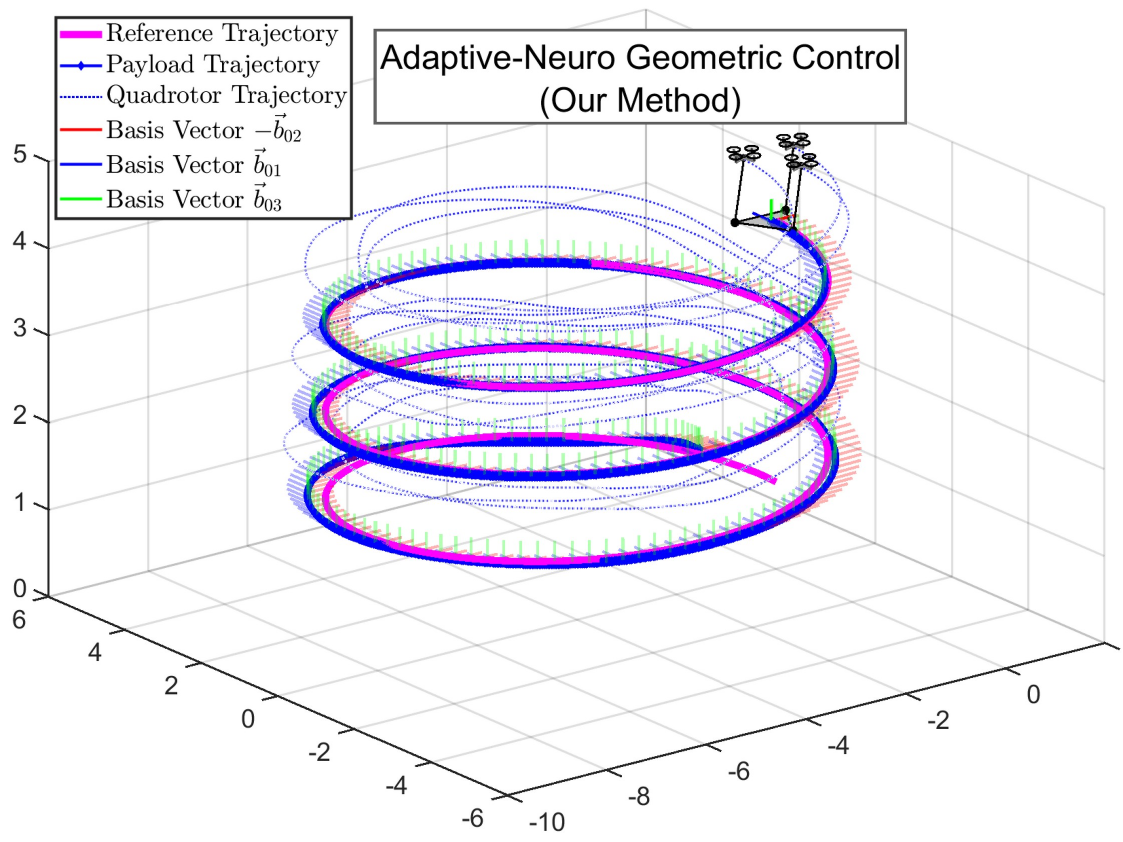}
    \caption{\footnotesize Comparison Group D  }
    \label{Comparison group D}
  \end{subfigure}
  \hspace{2mm}
  \begin{subfigure}[b]{0.32\textwidth}
    \centering
    \includegraphics[width=\linewidth]{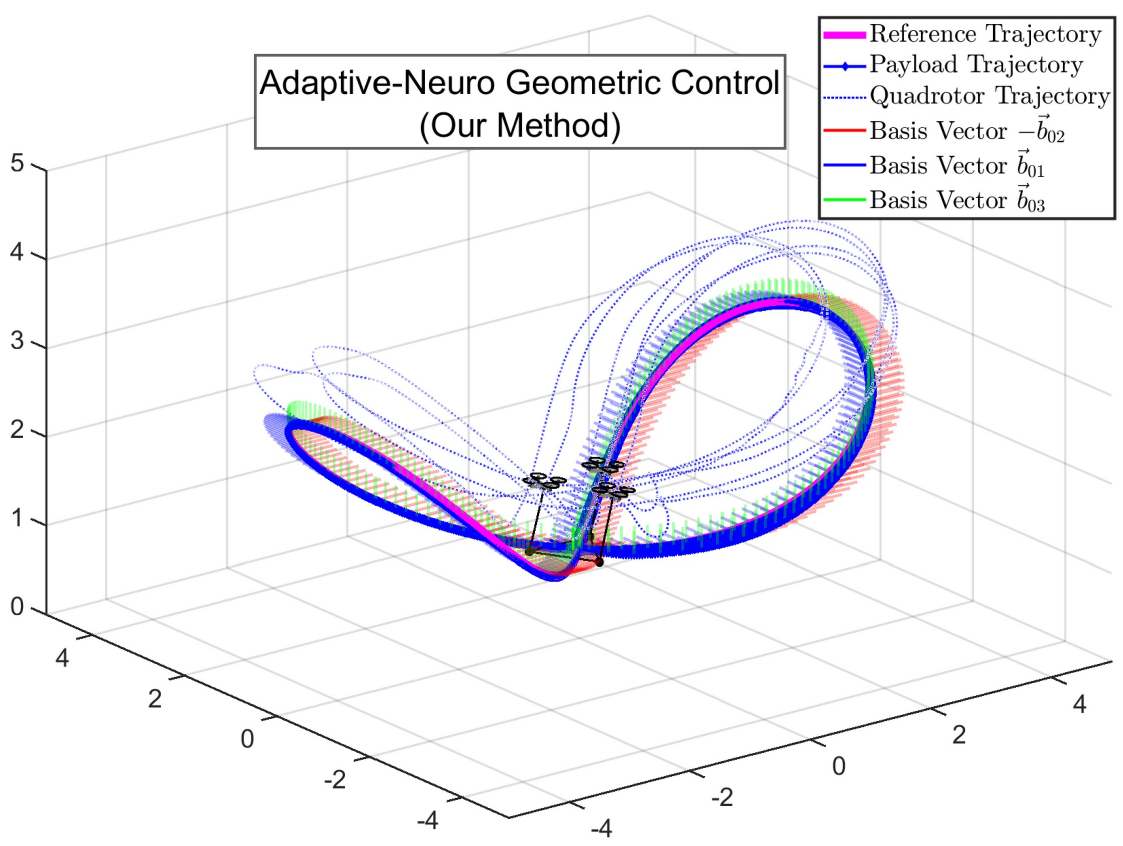}
    \caption{\footnotesize  Comparison Group E}
    \label{Comparison group E}
  \end{subfigure}
  \hspace{2mm}
  \begin{subfigure}[b]{0.32\textwidth}
    \centering
    \includegraphics[width=\linewidth]{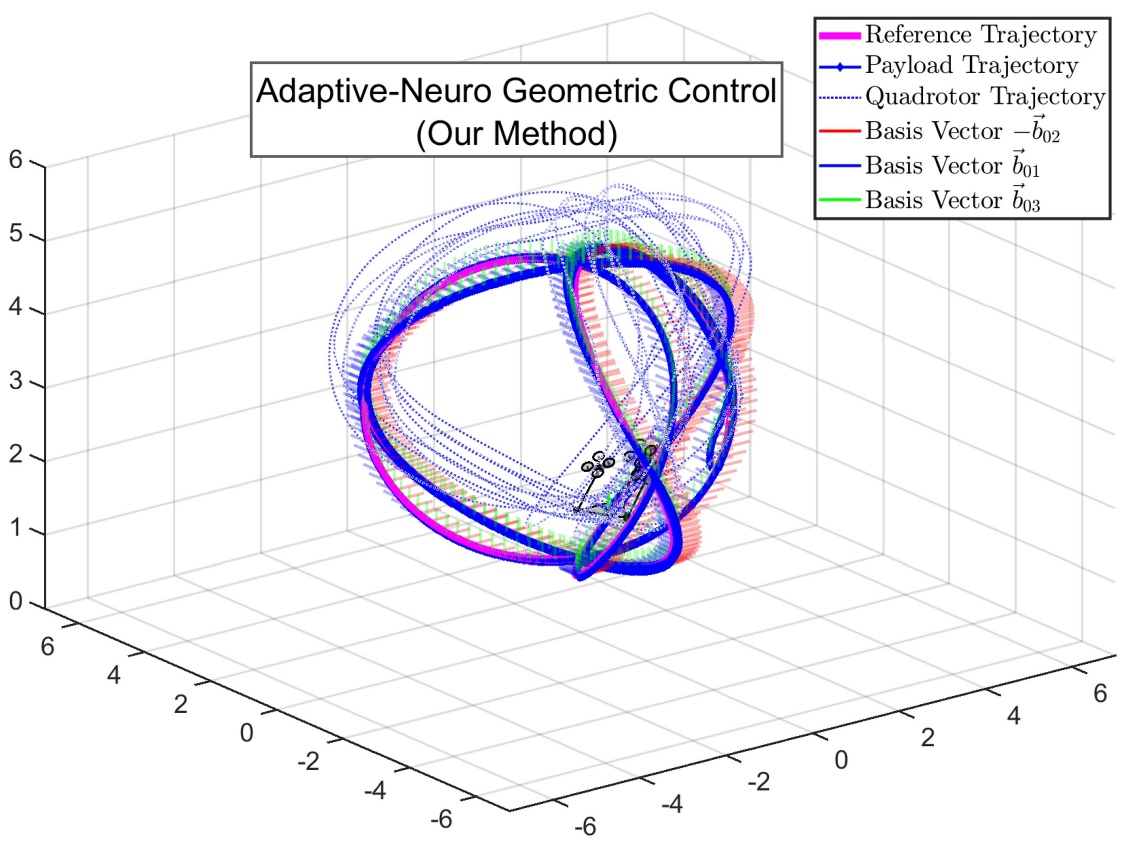}
    \caption{\footnotesize   Comparison Group F}
    \label{Comparison group F}
  \end{subfigure}
  \caption{\footnotesize Comparison results under complex reference tracking tasks with fixed payload attitude.}
   \vspace{-1em} 
  \label{Comparison results2}
\end{figure*}

This section presents the results of the simulation experiments conducted in the \textit{MATLAB Simulink} environment using the \textit{ODE3 Bogacki–Shampine Solver}. A schematic of the simulation block diagram is shown in Fig.~\ref{Simulink_Schematic}. To demonstrate the improved robustness of our method over the existing method \cite{2018 Geometric Control of Quadrotor UAVs Transporting
a Cable-Suspended Rigid Body}, 6 comparison groups (Gourps A, B, C, D, E and F)  were conducted, as shown in Figs.~\ref{Comparison group A}, \ref{Comparison group B}, \ref{Comparison group C}, \ref{Comparison group D}, \ref{Comparison group E}, \ref{Comparison group F}. Throughout the simulations, we ensured that both the adaptive-neuro geometric control (our method) and the existing geometric control were tested under the same setup. The reference model parameters are chosen as
\begin{equation}
{
   \begin{array}{cc}
   &m'_0 = 1~\mathrm{kg}, \,m'_i= 1~\mathrm{kg}, \\
   &\bm{J}_0'=\mathbf{diag}[\frac{1}{8},\frac{1}{8},\frac{1}{6}]~\mathrm{kg}\mathrm{m}^{2},\\
    &\bm{J}_i'=10^{-2}\mathbf{diag}[4,4,8]~\mathrm{kg}\mathrm{m}^{2}.
   \notag
\end{array} 
}
\label{Preset reference parameters}
\end{equation}
The gains of the PD controller and the integral compensation of payload control and quadrotor control are set to
\begin{equation}
{
   \begin{array}{cc}
   &\bm{k_{\text{p}}}=[20,20,1000]^{\top}, \bm{k_{\text{d}}}=[10,10,200]^{\top}, \\
   &k_{R_0}=20, k_{\Omega_0}=10, k_{R_i}=1,k_{\Omega_i}=1.\notag
\end{array} 
}
\label{PD controller gains}
\end{equation}
The gains of the integral compensation terms are chosen as 
\begin{equation}
{
   \begin{array}{cc}
   &\textit{c}_{q}=0.01, \textit{h}_{x_0}=1, \textit{h}_{R_0}=\textit{h}_{x_i}=0.1.\notag
\end{array} 
}
\label{Integral compensation gains}
\end{equation}
The update rates and parameters of the adaptive law ``slices" in the SANM module were designed as
\begin{equation}
{
   \begin{array}{cc}
   &\eta_{\textit{m}_j}=\eta_{\textit{J}_j}=0.01, 
 c_R=0.1,\\
&\mathfrak{s}_{\textit{m}_1}=\mathfrak{s}_{\textit{m}_2}= \mathfrak{s}_{J_1}=\mathfrak{s}_{J_2}=0.01, \mathfrak{s}_{\textit{m}_3}=\mathfrak{s}_{J_3}=0.1,\\
   &\overset{\tiny \text{max}}{m_0}{\rule{0pt}{1.5ex}}=6 \mathrm{kg},\overset{\tiny \text{max}}{\bm{J}_0}{\rule{0pt}{1.5ex}}=\mathbf{diag}[\frac{3}{4},\frac{3}{4}, \frac{1}{3}]~\mathrm{kg}\mathrm{m}^{2},\\
   &\bm{P}_j:=\mathbf{Lyap}(\bm{\Lambda}_{\bm{\textit{x}} j}, \bm{Q}_j)\\
   &\bm{Q}_1= \bm{Q}_2= \mathbf{diag}[\frac{1}{20},\frac{1}{20}], \bm{Q}_3=\mathbf{diag}[1,1]
   \notag
\end{array} 
}
\label{Parameters of adaptive law  slices}
\end{equation}
where the $\mathbf{Lyap}(\bullet,\bullet)$ denotes a function in \textit{MATLAB Control System Toolbox} that solves the continuous-time Lyapunov equation given in Eq.~\eqref{Lyapunov equations}.  

For each neural network ``slice" with index $\circ\in\{\bm{\textit{x}}\}, j\in\{1,2,3\}$ (i.e., ``slices" for translational dynamics), we employed a hidden layer with $l=5$ neurons and their parameters were designed as
\begin{equation}
\begin{array}{cc}
&\begin{bmatrix}\textbf{c}_{11},\textbf{c}_{21},\textbf{c}_{31},\textbf{c}_{41},\textbf{c}_{51}\end{bmatrix}=\begin{bmatrix}-1&-0.5&0&0.5&1\\ -1&-0.5&0&0.5&1\\\end{bmatrix},\\
&\begin{bmatrix}\textbf{c}_{12},\textbf{c}_{22},\textbf{c}_{32},\textbf{c}_{42},\textbf{c}_{52}\end{bmatrix}=\begin{bmatrix}-1&-0.5&0&0.5&1\\ -1&-0.5&0&0.5&1\\\end{bmatrix},\\
&\begin{bmatrix}\textbf{c}_{13},\textbf{c}_{23},\textbf{c}_{33},\textbf{c}_{43},\textbf{c}_{53}\end{bmatrix}=\begin{bmatrix}-1&-0.5&0&0.5&1\\ -1&-0.5&0&0.5&1\\\end{bmatrix},\\
    &b_{11}=b_{21}=b_{31}=b_{41}=b_{51}=1,\\
   &b_{12}=b_{22}=b_{32}=b_{42}=b_{52}=1,\\
   &b_{13}=b_{23}=b_{33}=b_{43}=b_{53}=2,\\
&\gamma_{\bm{\textit{x}}1}=5000,\,\gamma_{\bm{\textit{x}}2}=5000,\,\gamma_{\bm{\textit{x}}3}=1000.
   \notag
   \end{array}
\label{Parameters of neural networks slices1}
\end{equation}
For each neural network ``slice" with index $\circ\in\{\bm{\textit{R}}\}, j\in\{1,2,3\}$ (i.e., ``slices" for rotational dynamics), we also employed a hidden layer with $l=5$ neurons and their parameters were designed as
\begin{equation}
\begin{array}{cc}
&\begin{bmatrix}\textbf{c}_{11},\textbf{c}_{21},\textbf{c}_{31},\textbf{c}_{41},\textbf{c}_{51}\end{bmatrix}=\begin{bmatrix}-1&-0.5&0&0.5&1\\ -1&-0.5&0&0.5&1\\\end{bmatrix},\\
&\begin{bmatrix}\textbf{c}_{12},\textbf{c}_{22},\textbf{c}_{32},\textbf{c}_{42},\textbf{c}_{52}\end{bmatrix}=\begin{bmatrix}-1&-0.5&0&0.5&1\\ -1&-0.5&0&0.5&1\\\end{bmatrix},\\
&\begin{bmatrix}\textbf{c}_{13},\textbf{c}_{23},\textbf{c}_{33},\textbf{c}_{43},\textbf{c}_{53}\end{bmatrix}=\begin{bmatrix}-1&-0.5&0&0.5&1\\ -1&-0.5&0&0.5&1\\\end{bmatrix},\\
    &b_{11}=b_{21}=b_{31}=b_{41}=b_{51}=1,\\
   &b_{12}=b_{22}=b_{32}=b_{42}=b_{52}=1,\\
   &b_{13}=b_{23}=b_{33}=b_{43}=b_{53}=1,\\
&\gamma_{\bm{\textit{R}}1}=1500,\,\gamma_{\bm{\textit{R}}2}=1500,\,\gamma_{\bm{\textit{R}}3}=100.
   \notag
   \end{array}
\label{Parameters of neural networks slices1}
\end{equation}

For comparison groups A, B, and C, the reference trajectory of the payload is chosen as
\begin{equation}
    \bm{x}_{0_{\bm{d}}}(t)=(-4+4\cos(\frac{\pi}{5}t), 4\sin(\frac{\pi}{5}t),1)^{\top}.
\end{equation}
To ensure that the payload is facing the direction of motion, the desired attitude of the payload is specified as
\begin{equation}
\bm{R}_{0_{\bm{d}}}(t)=\begin{bmatrix}
    \frac{\bm{\dot{x}}_{0_{\bm{d}}}}{\lVert\bm{\dot{x}}_{0_{\bm{d}}}\lVert} &\frac{[\bm{\vec{e}}_3]_{\times}\bm{\dot{x}}_{0_{\bm{d}}}}{\lVert[\bm{\vec{e}}_3]_{\times}\bm{\dot{x}}_{0_{\bm{d}}}\lVert}&\bm{\vec{e}}_3
\end{bmatrix}.
\end{equation}

In Group A, both the existing and our methods were tested under the model-matched plant. This group aims to guarantee that the control parameters for both are appropriately configured for a fair comparison. Accordingly, the inertia and mass parameters of the real plant and the reference model are matched, i.e., $m_0=m'_0$, $\bm{J}_0=\bm{J}'_0$, $m_i=m'_i$, $\bm{J}_i=\bm{J}'_i$ and no external disturbances are applied in this group. As shown in Fig.~\eqref{Comparison group A}, both methods achieve satisfactory tracking performance under the model-matched condition. The tracking errors of the payload trajectory and attitude remain close to zero, which confirms that the baseline control gains and parameters are properly tuned. This result also provides a fair starting point for subsequent comparisons in Groups B, C, D, E and F, where model uncertainties and external disturbances are introduced to evaluate the robustness of the proposed method further. 

In Groups B and C,  the preset reference values remained unchanged, while the real-plant values increased to $\bm{J}_0$=$\mathbf{diag}[0.688,0.594,0.783]\mathrm{kg}\mathrm{m}^2$ and $m_0$=$5\mathrm{kg}$. 
For disturbance setups, all the comparison groups experienced all types of disturbances, including $\Delta_{\bm{x}_{i}}$, $\Delta_{\bm{R}_i}$,  $\Delta_{\bm{x}_{i}}^{\parallel}$, $\Delta_{\bm{x}_{i}}^{\bot}$,  $\Delta_{\bm{x}_0}$ and $\Delta_{\bm{R}_0}$. 
To evaluate the disturbance rejection capability, two types of strong and time-varying disturbances were applied to the payload.
In Group B, a disturbance force was imposed:
\begin{equation}
{\small
\begin{aligned}
\Delta_{\bm{x}_0}=&\big{(}15\sin(\sin(0.02t)t)+\cos(0.5t),15\sin(\cos(0.04t+\pi)t)\\
        &+5\cos(0.5t),-25\sin(1.5t)+\cos(0.5t)\big{)}^{\top}\mathrm{N}.
\end{aligned}
}
\notag
\end{equation}
In Group C, a disturbance moment $\Delta_{\bm{R}_0}$ was exerted along the $\bm{\vec{b}}_{01}$ axis starting from $t=5s$:
\begin{equation}
{\small
\begin{aligned}
\Delta_{\bm{R}_0}= (10\sin(t-5),0,0)^{\top}\mathrm{Nm}.
\end{aligned}
}
\notag
\end{equation}

In Groups D, E, and F, the disturbance magnitudes were inherited from Group B, while the reference trajectories were designed to be more complex by introducing variations in altitude. Moreover, these three comparison groups were conducted under the scenario of centralized payload transportation with a fixed payload attitude, i.e., the desired attitude of the payload $\bm{R}_{0_{\bm{d}}}$ remained constant throughout the tasks. 

From the results in Figs.~\ref{Comparision results} and \ref{Comparison results2}, the enhanced robustness of the proposed method was validated, including its adaptivity to parametric uncertainties and its rejection capability against disturbance forces and moments. In Group A, both methods exhibited comparable performance under the model-matched condition, confirming that the control gains were fairly tuned. In Group B, when subjected to strong time-varying disturbance forces, the proposed method maintained smaller tracking errors than the existing method. In Group C, under a strong disturbance moment applied to the payload, the proposed method still preserved stable tracking with faster recovery. The tracking errors in Groups B and C are further illustrated in Fig.~\ref{Numerical results}, which clearly highlight the improved robustness of the proposed method. The results of Groups D, E and F further confirmed this trend, where the proposed method consistently achieved stable tracking performance even under more complex reference trajectories with altitude variations. Additionally, Fig.~\ref{SANM_outputs} presents the outputs of the SANM module, which show the disturbance and model uncertainty features identified online.  Together, these results demonstrate that the proposed controller not only performs reliably under nominal conditions but also provides superior robustness against model uncertainties and time-varying external disturbances.

\vspace{1em} 
\noindent\textbf{6. Conclusions} 

In this paper, we provide an online identification solution for the complex coupled multi-quadrotor centralized transportation system. First, we introduced an online learning perspective, called Dimension-Decomposed Learning (DiD-L). This perspective enables multiple parallel low-dimensional learning processes on the subspaces of the state error domain of the control objective (where the control objective is specified as the payload in this work).
To realize this idea, we introduced a module named Sliced Adaptive-Neuro Mapping (SANM). By constructing several adaptive law and neural network ``slices", the SANM module achieves online identification of payload model parameters and compensation for time-varying disturbances. 

\textbf{\textit{Discussion:}}
The dimension decomposition strategy was initially adopted in our previous conference work \cite{2025 Robustness Enhancement for Multi-Quadrotor Centralized Transportation System via Online Tuning and Learning}, while this work and our current work 
\cite{2025 Dimension-Decomposed Learning for Quadrotor Geometric Attitude Control with Exponential Convergence on SO(3)} formally introduce the concept of Dimension-Decomposed Learning (DiD-L) as a generalized online learning paradigm.  At present, all realizations of DiD-L are based on the proposed Sliced Adaptive-Neuro Mapping (SANM) module. In \cite{2025 Dimension-Decomposed Learning for Quadrotor Geometric Attitude Control with Exponential Convergence on SO(3)}, this approach has been experimentally validated and demonstrated to be applicable to real-world rigid-body motion control. The main advantage of this method lies in the efficient representational capability of neural network ``slices" under the decomposition strategy, which endows the online learning process with a lightweight nature. We have verified that using only five neurons is sufficient to achieve online learning of real-world disturbances (after Kalman filtering). Moreover, the entire online learning process can be executed at 400 Hz on microcontroller units (MCUs), such as \textit{STM32}. Together, these findings suggest that the training process for robotic kinematics could potentially be decomposed into multiple ``subspace learning" processes that can be executed in parallel.

\textbf{\textit{Future Work:}}
We plan to extend the dimension decomposition strategy to both offline training and online inference, thereby broadening its applicability beyond the current setting. In the future, we will further explore its potential in a wider range of robotic applications, such as humanoid robots and bio-inspired robots, where complex high-dimensional dynamics may also benefit from parallelized Dimension-Decomposed learning (DiD-L) across subspaces. Moreover, since the proposed framework integrates online learning with robotics control, it also shows promise in advancing research in artificial intelligence (AI) \& robotics, particularly in the domain of embodied AI \cite{2025 Aligning Cyber Space With Physical World: A Comprehensive Survey on Embodied AI}, where perception, learning, and action are coupled in physical environments.

 \vspace{1em} 

\noindent\textbf{Acknowledgment}

\noindent This work was partially supported by JST SPRING, Japan Grant Number JPMJSP2124.

\vspace{1em} 
 
\noindent\textbf{Conflict of Interest}

\noindent There is no conflict of interest.

\vspace{1em} 

\noindent\textbf{Supporting Information}

\noindent Not applicable.

\addtolength{\textheight}{0.5cm}   




\end{document}